\documentclass[a4paper,11pt]{article}
\usepackage{jcappub} 
\usepackage{lineno}
\newcommand{\angstrom}{\text{\normalfont\AA}}

\usepackage{aas_macros}
\usepackage[dvipsnames]{xcolor}

\newcommand{\eqn}[1]{equation~(\ref{#1})}
\newcommand{\eqns}[2]{equations~(\ref{#1}) and (\ref{#2})}
\newcommand{\eqnsthree}[3]{equations~(\ref{#1}), (\ref{#2}) and (\ref{#3})}
\newcommand{\secn}[1]{Section~\ref{#1}}
\newcommand{\fig}[1]{Figure~\ref{#1}}

\newcommand{\tab}[1]{Table~\ref{#1}}
\newcommand{\app}[1]{Appendix~\ref{#1}}

\arxivnumber{2404.02879} 
\title{Modelling the star-formation activity and ionizing properties of high-redshift galaxies }

\author[1]{Anirban Chakraborty\note{Corresponding author.}}
\author{and Tirthankar Roy Choudhury}
\affiliation{National Centre for Radio Astrophysics, Tata Institute of Fundamental Research, \\Pune University Campus, Ganeshkhind, Pune 411007, India.}

\emailAdd{anirban@ncra.tifr.res.in; anirban.chakraborty096@gmail.com}

\abstract{Early results from the JWST observations have reported a surprisingly high number of UV-bright galaxies at $z \geq 10$, which appears to challenge the theoretical predictions from standard galaxy formation models in the $\Lambda$CDM framework at these redshifts. To alleviate this tension, several cosmological and astrophysical interpretations have been advanced. However, all of these proposed scenarios carry noteworthy consequences for other large-scale processes in the early Universe, particularly cosmic reionization, since high-redshift galaxies are believed to be the primary ionizing sources during the Epoch of Reionization (EoR). To investigate this, we introduce a semi-analytical model of galaxy formation and evolution that explains the evolving galaxy UV luminosity function (UVLF) over $6 \lesssim z \lesssim 15$, and also jointly tracks the time evolution of the globally averaged neutral hydrogen fraction in the intergalactic medium. The model self-consistently accounts for the suppression of star formation in low-mass galaxies due to reionization feedback and is constrained by comparing the model predictions with various observational probes like the UVLF data from HST and JWST, recent measurements of the neutral hydrogen fraction, and the CMB scattering optical depth. Our analysis confirms that a rapid enhancement in the star-formation rate efficiency and/or UV luminosity per stellar mass formed is necessary for consistency with the JWST UVLF estimates at $z \geq 10$. We further find that it is possible to jointly satisfy the current reionization constraints when the escape fraction is assumed to be halo-mass dependent, requiring higher Lyman-continuum leakage from low-mass galaxies. We also examine the relative contribution of galaxies with different UV luminosities towards the ionizing photon budget for the EoR and investigate the large-scale bias of high-$z$ galaxies.}


\begin{document}
\maketitle
\flushbottom

\section{Introduction}
\label{sec:intro}

Over the past few decades, our understanding of the statistical and astrophysical properties of galaxies has seen remarkable improvement. This has been largely facilitated by confronting theoretical models of hierarchical structure formation with high-quality observational data accrued from deep galaxy surveys \cite{Sommerville2015, Dayal&Ferrara2018}. Despite advancements in understanding the properties of dark matter (DM) halos as a function of mass and redshift from N-body simulations \cite{Springel2005, BoylanKolchin2009, Klypin2011, Klypin2016} and analytical formalisms \cite{ST99, SMT2001}, modeling the various physical processes, such as gas accretion and cooling, star formation rate, feedback mechanisms, chemical enrichment of the interstellar gas, that regulate the formation and evolution of galaxies inside these virialized DM halos continues to be a rather challenging task. In the literature, considerable amount of effort has been devoted to studying this galaxy-halo connection with the help of hydrodynamical simulations \cite{Vogelsberger2014, Schaye2015, Dave2019} as well as semi-analytical models \cite{Choudhury2005, Samui2007, Mitra2011, Dayal14, Mason2015, Lacey2016, Mutch2016, Furlanetto2017, Mirocha2017, Behroozi19}. Such studies have yielded crucial insights into various facets of galaxies across different cosmic epochs; for example, their mass assembly histories, the properties of their constituent stellar and black-hole populations, their sizes and structural morphology, their gas and metal contents, and more. These frameworks have even been employed for studying how the formation of the first stars and galaxies initiated a major phase transition of the Universe \cite{Dayal2017, Park++2019, Maity2022, Munoz2022_21cmFAST}, known as the \textit{Epoch of Reionization} \cite{Barkana&Loeb2001, Dayal&Ferrara2018, Wise2019, TRC2022Review}. However, our limited understanding of the properties of these earliest luminous sources makes it difficult to accurately reconstruct the ionization and thermal history of the intergalactic medium (IGM) during this transition. Consequently, one of the main objectives of many current and upcoming telescopes, e.g., the James Webb Telescope (JWST) \cite{JWST_highzgalaxies, JWST_DeepSurveys}, Thirty Meter Telescope (TMT) \cite{TMTScienceCase_2015}, Giant Magellan Telescope (GMT) \cite{GMT_TMT_WhitePaper2019}, European Extremely Large Telescope (E-ELT), is to conduct a comprehensive study of the galaxy population at high redshifts and understand their contribution to the process of reionization more precisely. 

Over the past decade, the commissioning of large galaxy surveys using space-based and ground-based telescopes like the Hubble Space Telescope (HST), the Spitzer Space Telescope, the Subaru Telescope \cite{Bouwens2015, Bouwens2017, Atek2018, Ono2018, Bowler2020, harikane2022} has resulted in a growing sample of high redshift galaxies that have been identified using the Lyman-break technique. An important statistical quantity that is usually measured by such surveys is the galaxy luminosity function (LF), defined as the comoving number density of galaxies per unit luminosity interval. These LFs, measured at a rest-frame wavelength of about 1500 $\angstrom$ and out to redshifts as high as $z$ = 10, carry vital information about the rate and efficiency of cosmic star formation, characteristics of the stellar populations, the prevalence of dust and heavy elements in the early Universe, among other things. More recently, the launch of the James Webb Space Telescope (JWST) opened up a new window to peer into the early phases of cosmic evolution, offering an unprecedented glimpse into the earliest generation of galaxies at $z \gtrsim 10$ \cite{Naidu2022, Castellano2022, Finkelstein2022, Atek2023, Adams2023, Bradley2023,  Whitler2023}. Using these early JWST observations, several independent studies have published measurements of the galaxy UV luminosity function (UVLF) at high redshifts ($ 9 \lesssim z \lesssim 17$) \cite{Castellano2023_UVLF, Finkelstein2023_UVLF, Donnan2023, Harikane2023, Bouwens2023, McLeod2024}. All of these works found a surprisingly large abundance of UV bright galaxies, which is in sharp contrast to the expectations from traditional galaxy formation models in the $\Lambda$CDM framework \cite{Haslbauer2022, Lovell2023, Mirocha2023, Munoz2023, Mason2023}. Moreover, the stellar mass estimates for some candidate galaxies identified by JWST at such early times seem to be in tension with predictions from standard models as well \cite{Boylan2023, Labbe2023, Castellano2023_UVLF}.

This discrepancy between observations and theoretical expectations calls for revisiting the existing astrophysical and cosmological models that serve as building blocks for simulating the formation and growth of galaxies. To ease this tension with the observations, a plethora of cosmological and astrophysical hypotheses were proposed. These ranged from seeking possible modifications to the baseline $\Lambda$CDM cosmology \cite{Liu2022, Padmanabhan2023, Biagetti2023, Hutsi2023, Parashari2023, Gong2023, Maio2023} to invoking various astrophysical factors like boosted efficiency of star formation \cite{Dekel2023, Renzini2023, Sipple2023}, stochasticity in the star-formation activity \cite{Mirocha2023, Shen2023, Pallottini2023}, the presence of Pop-III stars with a top-heavy initial mass function \cite{Inayoshi2022, Harikane2023} and negligible dust attenuation beyond $z \gtrsim 11$ \cite{Ferrara2023}. However, as these high-$z$ JWST surveys observed a relatively small volume of the sky, the possibility that their measured UVLFs may have been greatly affected by the effects of cosmic variance \cite{Ucci2021, Jespersen2024}, besides other survey systematics, cannot be ignored either. 

Since existing observations favor a scenario where reionization is primarily driven by star-forming galaxies \cite{Wise2014, Bouwens2015_sources, Dayal2020, Atek2024_Spectroscopy}, all of these proposed possibilities, in principle, can have significant implications for the growth of ionized regions in the IGM. For instance, if the efficiency of star formation is higher at early redshifts, then this would also increase the total number density of ionizing photons in the IGM available for reionization, thereby affecting its onset and progress. Moreover, as the IGM gets progressively reionized, the associated ultraviolet background (UVB) would also photo-heat the gas to higher temperatures within ionized regions \cite{Miralda1994}. This increase in temperature may lead to the photo-evaporation of gas from relatively low-mass halos present inside these ionized regions, thereby quenching their subsequent star formation abilities. This effect, known as radiative feedback, in principle, should impact the observations of the galaxy UV luminosity function at the faint end and also the subsequent progress and morphology of reionization \cite{Finlator2011, Finlator2012, Maity2022, Hutter2021}. Given this intricate connection between galaxy formation and cosmic reionization, it is imperative to self-consistently couple these two processes in theoretical studies that aim to probe the physics of the high-redshift Universe. 

In this paper, we introduce a semi-analytical model of galaxy formation and evolution to compute the rest-frame galaxy UV luminosity function (UVLF) across a wide range of redshift ($6 \lesssim z \lesssim 15$). We self-consistently account for the suppression of star formation in low-mass galaxies due to reionization feedback by tracking the time evolution of the globally averaged ionization fraction in the IGM. It is worthwhile to emphasize that while the ionization history is sensitive to the product of the star-formation efficiency $(f_*)$ and the escape fraction $(f_{\rm esc})$ of ionizing photons, the UVLF for faint and intermediate magnitudes can be determined from the knowledge of $f_*$. Therefore, combining LF observations with reionization observations holds great promise in unraveling the characteristics of high-redshift galaxies, especially in constraining the unknown fraction of ionizing photons that escape from the interstellar medium of these galaxies into the IGM \cite{Mitra2015, Park++2019, Maity2022_ParameterConstraints}. In this work, our main aim is to use this flexible semi-analytical model for computing different observables and place constraints on the astrophysical properties of high-redshift galaxies by comparing the model predictions against a wide variety of recent observations \emph{simultaneously}. Throughout this work, the cosmological parameters are taken to be $\Omega_m$ = 0.308, $\Omega_{\Lambda}$ = 0.692, $\Omega_b$ = 0.0482, $h$ = 0.678, $\sigma_8$ = 0.829 and $n_s$ = 0.961 \cite{Planck2014}.

The paper is organized as follows: In \secn{sec:theory_model}, we describe the details of the theoretical model used to compute the UV luminosity function and the global reionization history. \secn{sec:datasets} describes the various observational constraints used in this work and the Bayesian formalism used for parameter estimation. We discuss the results obtained and their implications in \secn{sec:results}. Finally, we conclude with a summary of our main results in \secn{sec:conclusion}.  \\

\section{Theoretical Model}
\label{sec:theory_model}

In this section, we describe the theoretical framework for modelling the star formation and ionizing properties of galaxies at high redshifts ($z \geq 6$) and calculating the different global observables.

\subsection{The UV Luminosity Function Model}
\label{sec:galform_model}

The first step in modelling the UV luminosity function is to assign galaxies to dark matter halos.  We assume that each dark matter halo contains only one galaxy and that the luminosity of the galaxy will primarily be determined by the mass of the associated DM halo. The stellar mass of a galaxy, $M_*$, can be related to the mass of the host halo, $M_h$, through the relation \cite{Park++2019, Behroozi&Silk2015, Sun&Furlanetto2016, Dayal14, Mitra2015}-
\begin{equation}
\label{eq:stellar_mass}
M_*(M_h) = f_*(M_h) \bigg(\dfrac{\Omega_b}{\Omega_m}\bigg) M_h
\end{equation}
where the star formation efficiency parameter $f_*(M_h)$ is taken to have a \textit{power-law} dependence on $M_h$ as 
\begin{equation}
\label{eq:fstar}
f_*(M_h) = f_{*,10} \bigg(\dfrac{M_h}{10^{10} M_\odot}\bigg)^{\alpha_*}
\end{equation}
The star formation rate (SFR) of the galaxy is given by
\begin{equation}
\label{eq:SFR}
\dot{M}_*(M_h,z) = \dfrac{M_*(M_h)}{t_{*}(z)},
\end{equation}
where $t_{*}(z)$ is the mean characteristic star formation time scale. We assume that the mean characteristic star formation time scale is proportional to the halo dynamical time scale which, in turn, is proportional to the Hubble time scale at redshifts of our interest.

This allows us to write
\begin{equation}
\label{eq:tstar}
t_{*}(z)  = c_{*}~t_H(z) = c_{*}~\dfrac{1}{H(z)}
\end{equation}
with $c_*$ being a dimensionless proportionality constant. Combining all the relevant equations, we can rewrite the SFR for a galaxy hosted by a halo of mass $M_h$ as 
\begin{equation}
\label{eq:SFR_expanded}
\dot{M}_*(M_h,z) = \dfrac{f_{*,10}}{c_{*}}~H(z)\bigg(\dfrac{M_h}{10^{10} M_\odot}\bigg)^{\alpha_*} \bigg(\dfrac{\Omega_b}{\Omega_m}\bigg) M_h.
\end{equation}

Having calculated the SFR of the galaxy, we can use it to determine its intrinsic rest-frame UV continuum luminosity $L_\mathrm{UV}$ at 1500 $\angstrom$ through the conversion factor $\mathcal {K}_{{\rm UV}}$, which is defined as
\begin{equation}
\mathcal {K}_{{\rm UV}} = \dfrac{\dot{M}_*(M_h,z) }{L_{{\rm UV}}(z)}.
\end{equation}

The value of $\mathcal {K}_{{\rm UV}}$ is usually sensitive to the recent star formation history, metallicity of the stellar populations, dust content as well as the choice of initial mass function (IMF) \cite{Madau&Dickinson2014, Sun&Furlanetto2016}. The $1500~\angstrom$  UV luminosity of a galaxy can be written as 
\begin{equation}
\label{eq:LUV_nofb_init}
L^{\rm nofb}_{{\rm UV}} = \dfrac{\dot{M_*}(M_h,z) }{\mathcal {K}_{{\rm UV}}} = \dfrac{1}{\mathcal {K}_{{\rm UV}}} \bigg[\dfrac{f_{*,10}}{c_{*}}~H(z)\bigg(\dfrac{M_h}{10^{10} M_\odot}\bigg)^{\alpha_*} \bigg(\dfrac{\Omega_b}{\Omega_m}\bigg) M_h \bigg],
\end{equation}
where the superscript ``nofb'' on the left-hand side indicates that the above relation holds for galaxies in the absence of radiative feedback; we will discuss the modifications arising from feedback later.

Given that the uncertain conversion factor $\mathcal {K}_{{\rm UV}}$ is completely degenerate with combination $\dfrac{f_{*,10}}{c_{*}}$, we define a new dimensionless quantity $\varepsilon_{{\rm *10,UV}}$ by combining it with the other unknown quantities as shown below and refer to it as the ``UV efficiency'' parameter.
\begin{equation}
\label{eq:epsilonUVstar}
\varepsilon_{{\rm *10,UV}} \equiv \dfrac{f_{*,10}}{c_{*}} \dfrac{1}{\mathcal {K}_{{\rm UV}}/{\mathcal {K}_{{\rm UV,fid}}}}.
\end{equation}
We adopt a constant fiducial conversion factor  $\mathcal {K}_{{\rm UV,fid}} = 1.15485 \times 10^{-28} \ {\rm \mathrm{M}_{{\odot }}}\ {\rm yr}^{-1}/\ {\rm ergs}\ {\rm s}^{-1}\ {\rm Hz}^{-1}$, that has been evaluated for continuous mode star formation (at time scales of $\gtrsim$ 100 Myr) with a 0.1 - 100 $M_\odot$ Salpeter IMF and metallicity $Z = 0.001 ( = 0.05~Z_\odot)$ using {\tt{STARTBURST99 v7.0.1}}\footnote{https://www.stsci.edu/science/starburst99/docs/default.htm}\cite{Starburst99}.
Therefore,  \eqn{eq:LUV_nofb_init} becomes 
\begin{equation}
\label{eq:LUV_nofb}
L^{\rm nofb}_{{\rm UV}} = \dfrac{\varepsilon_{{\rm *10,UV}}}{\mathcal {K}_{{\rm UV,fid}}}~H(z)\bigg(\dfrac{M_h}{10^{10} M_\odot}\bigg)^{\alpha_*} \bigg(\dfrac{\Omega_b}{\Omega_m}\bigg) M_h
\end{equation}

The relations so far do not account for radiative feedback arising from photo-heating, which can be important in ionized regions. Similar to previous studies \cite{Choudhury&Dayal2019, Sobacchi&Mesinger2013}, we model the radiative feedback mechanism through a simplistic approach wherein a “critical" mass scale $M_{{\rm crit}}$ is defined to characterize the effect of UV feedback in suppressing the gas fraction inside low-mass halos residing in ionized regions. The $1500~\angstrom$ UV luminosity of a galaxy affected by radiative feedback  can be written as 
\begin{equation}
\label{eq:LUV_fb}
L^{\rm fb}_{{\rm UV}} = f_g(M_h)~\dfrac{\varepsilon_{{\rm *10,UV}}}{\mathcal {K}_{{\rm UV,fid}}}~H(z)\bigg(\dfrac{M_h}{10^{10} M_\odot}\bigg)^{\alpha_*} \bigg(\dfrac{\Omega_b}{\Omega_m}\bigg) M_h,
\end{equation} 
where the function $f_g(M_h)$ quantifies the fraction of gas mass that is retained inside a halo of mass $M_h$ after photo-heating due to the ultraviolet background. For this work, we assume that the UV radiative feedback acts gradually, being pivoted about $M_{{\rm crit}}(z)$ where the gas fraction is taken to be 50 percent \cite[e.g.,][]{Sobacchi&Mesinger2013, Dayal2015, Choudhury&Dayal2019, Hutter2021, Maity2022}.
\begin{equation}
\label{eq:fgas}
f_g(M_h) = 2^{-M_{{\rm crit}}/M_h} = \exp \bigg[- \left(\ln 2\right) \dfrac{~M_{{\rm crit}}}{M_h} \bigg].
\end{equation}

Finally, we can relate the UV luminosity at 1500 $\angstrom$ to the absolute UV magnitude using the standard AB magnitude relation \cite{Oke_ABmag, Oke&Gunn_ABmag}.
\begin{equation}
{\rm log_{10}}\left(\frac{L_{\rm UV}}{{\rm ergs \ s^{-1} \ Hz^{-1}}} \right) = 0.4 \times (51.6 - M_{\rm UV}). 
\end{equation} 

Having found the relation between $M_h$ and $L_\mathrm{UV}$ (or $M_\mathrm{UV}$) for both neutral and ionized regions, we now discuss the model for UVLF. As radiative feedback will affect the gas content of only those galaxies that are present in already ionized regions, the globally averaged UVLF (${\rm \Phi^{total}_{UV}}$) can be written as a weighted combination of the feedback-affected UVLF (${\rm \Phi^{fb}_{UV}}$) from ionized regions and the feedback-unaffected UVLF (${\rm \Phi^{nofb}_{UV}}$) from neutral regions, i.e.,
\begin{align}
\label{eqn:lumfunc_full}
\Phi^{\rm total}_{\rm UV} &= Q_{\rm II}(z)~{ \Phi^{\rm fb}_{\rm UV}} + [1-Q_{\rm II}(z)]~{\rm \Phi^{\rm nofb}_{\rm UV}}
\nonumber \\
&= Q_{\rm II}(z) \frac{{\rm d}n}{{\rm d}M_h} \left|\frac{{\rm d}M_h}{{\rm d}{L^{\rm fb}_{\rm UV}}}\right|~\left|\frac{{\rm d}{ L^{\rm fb}_{\rm UV}}}{{\rm d}{\rm M_{UV}}} \right| 
+ \big[1 - Q_{\rm II}(z)\big] \frac{{\rm d}n}{{\rm d}M_h} \left|\frac{{\rm d}M_h}{{\rm d}{L^{\rm nofb}_{\rm UV}}}\right|~\left|\frac{{\rm d}{ L^{\rm nofb}_{\rm UV}}}{{\rm d}{M_{\rm UV}}}\right|,
\end{align}
where $Q_{\rm II}(z)$ is the global ionization fraction at redshift $z$ and ${\rm d}n / {\rm d}M_h$ is dark matter halo mass function. Strictly speaking, as galaxies are highly biased objects, the fraction of them residing in ionized regions is likely to be larger than the global ionization fraction, as has been assumed in \eqn{eqn:lumfunc_full}. However, accounting for this effect of galaxy clustering in our semi-analytical UVLF calculations is challenging. This can be more easily done when working with numerical simulations, which we plan to take up in a future work.

We use the fitting function of Jenkins \cite{Jenkins2001} for ${\rm d}n / {\rm d}M_h$ as that provides a better match to the mass function obtained from numerical simulations. For calculating the halo mass function, we use the publically available package {\tt{hmf}}\footnote{https://github.com/halomod/hmf} \cite{HMFcodePaper}.

In order to allow for the possibility that the galaxy properties evolve with redshift across the range $6 \lesssim z \lesssim 15$, we model the parameters $\varepsilon_{\rm *10,UV}$ and $\alpha_*$ as $z$-dependent, using a $\textit{tanh}$ parameterization, as follows
\begin{equation}
\log_{10}(\varepsilon_{\rm *10,UV}) = \ell_{\varepsilon,0} + \dfrac{\ell_{\varepsilon, \mathrm{jump}}}{2} \tanh\left(\dfrac{z-z_\varepsilon}{\Delta z_\varepsilon}\right),
\end{equation}
and
\begin{equation}
\alpha_\ast = \alpha_0 + \dfrac{\alpha_\mathrm{jump}}{2} \tanh\left(\dfrac{z-z_\alpha}{\Delta z_\alpha}\right).
\end{equation}
In the above formulation, the parameter $\log_{10} (\varepsilon_{*10,\mathrm{UV}})$ asymptotes to $\ell_{\varepsilon,0} - \ell_{\varepsilon, \mathrm{jump}} / 2$ at low redshifts and to $\ell_{\varepsilon,0} + \ell_{\varepsilon, \mathrm{jump}} / 2$ at high redshifts, with the transitions occurring at a characteristic redshift $z_\varepsilon$ over a range $\Delta z_\varepsilon$. Thus, $\ell_{\varepsilon,0}$ can be interpreted as the mean value of $\log_{10} (\varepsilon_{*10,\mathrm{UV}})$ while $\ell_{\varepsilon, \mathrm{jump}}$ characterizes the jump in the parameter. The parameters for the slope $\alpha_*$ can also be interpreted in an identical manner. The motivation behind choosing such a parameterization is two-fold (please refer to \app{appendix:tanhParameterisation} for more details) - firstly, the asymptotic behavior of the $\textit{tanh}$ function ensures that these parameters do not indefinitely increase to extremely high values at the highest redshifts and secondly, it allows for a wide range of possible evolutionary scenarios for the parameters, ranging from cases where they transition rapidly between their two asymptotic limits to those where they evolve more gradually, with ample flexibility in their corresponding asymptotic values and their transition redshifts ($z_\alpha~\text{or}~z_\varepsilon)$.

\subsection{The Reionization Model}
\label{sec:reion_model}

In order to calculate the evolving UVLFs using \eqn{eqn:lumfunc_full}, one needs to know the evolution of the global ionization fraction $Q_{II}(z)$, which can be obtained by solving the following first-order differential equation \cite[][]{Shapiro1987,Madau1999},
\begin{equation}
\label{eq:reion_eq0}
\dfrac{dQ_{II}}{dt} = \dfrac{\dot{n}_{ion}}{\bar{n}_H} - \chi_{\mathrm{He}}(z)~\bar{n}_H ~(1+z)^3 ~\alpha_B~\mathcal{C}(z)~Q_{II},
\end{equation}
where, $\dot{n}_{ion}$ is the comoving number density of ionizing photons produced by the galaxies and escaping into the IGM per unit time, $\bar{n}_{\rm H}$ is the mean comoving number density of hydrogen, $\mathcal{C}(z)$ is the IGM clumping factor, $\alpha_B = 2.6 \times 10^{-13}~\mathrm{cm^{3} ~s^{-1}}$ is the Case-B recombination coefficient, $\chi_{\mathrm{He}}(z) = 1 + \eta Y_p/4X_p$  is the number of free electrons per hydrogen atom (which includes the excess contribution from helium, depending on its ionization state $\eta$). 

In our case, $\chi_{\mathrm{He}}(z) = 1.08 (1.16)$ for $z \geq 3$ ($z < 3$, essentially assuming that helium gets singly ionized along with hydrogen and is doubly ionized at $z < 3$). We take the value of the clumping factor $\mathcal{C}$ of ionized gas to be redshift independent and fix its value to  $\mathcal{C}$ = 3. Although the clumping factor may, in principle, vary with redshift owing to its dependence on the density, ionization, and temperature distribution in the IGM; for simplicity, we assume it to be constant in this work. This choice is in reasonable agreement with radiative transfer simulations \cite{McQuinn2011, Shull2012, Finlator2012, Pawlik2015, DAloisio2020} that find $\mathcal{C}$ to very gradually increase, after the establishment of the ionizing UV background, towards low redshifts and finally approach this constant value as the gas gets completely relaxed in response to cosmic reionization.

In neutral regions, i.e., the regions unaffected by radiative feedback, we model the ionizing photon production rate in terms of the SFR as
\begin{equation}
\label{eq:niondot_expanded}
\dot{n}^{\rm nofb}_\mathrm{ion} (z) = \eta_{\gamma*} \int_{M_{cool}(z)}^{\infty} f_{\rm esc}(M'_{ h})~ \dot{M}_*(M'_h,z)~\dfrac{dn{(M'_h,z)}}{dM_h} dM'_h,
\end{equation}
where $\eta_{\gamma*}$ is the total number of ionizing photons produced in the DM halo per unit mass of stars, and $f_\mathrm{esc}(M_h)$ is the fraction of ionizing photons that escape from the star-forming halos of mass $M_h$ into the IGM. The integration limit $M_{{\rm cool}}(z)$ denotes the minimum mass of halos for which the gas can cool via atomic transitions and form stars at a redshift $z$. At any given redshift $z$, we parameterize $M_{{\rm cool}}(z)$ in such a way that it corresponds to the mass threshold appropriate for atomic cooling, having a constant virial temperature  $T_{vir}$ of $10^4$ K \cite{Sobacchi&Mesinger2013}
\begin{equation}
M_{cool} (z) = 10^{8}~ h^{-1} ~\mathrm{M}_{{\odot }} \left(\frac{\mu }{0.6}\right)^{-3/2} \left(\frac{\Omega _{\rm m}}{\Omega _{\rm m}^{\phantom{{\rm }} z}} \frac{\Delta _{\rm c}}{18\pi ^{2}}\right)^{-1/2} \left(\frac{T_{\rm vir}}{1.98\times 10^{4}\,{\rm K}}\right)^{3/2} \left(\frac{1+z}{10}\right)^{-3/2},  
\end{equation}
where $\Omega _{\rm m}^{\phantom{{\rm }} z} = \Omega _{\rm m}\left(1+z\right)^{3}/[\Omega _{\rm m}\left(1+z\right)^{3}+\Omega _{\Lambda }]$, $\Delta _{\rm c} = 18\pi ^{2} + 82d - 39d^{2}$,  $d=\Omega _{\rm m}^{\phantom{{\rm }} z}-1$ and $\mu$ is the mean molecular weight. 

We model the ionizing UV escape fraction $f_{esc}(M_h)$ from the galaxy using a \textit{power-law} parameterization \cite{Park++2019,Qin2021}
\begin{equation}
\label{eq:fescape}
f_{\rm esc}(M_{\rm h}) = f_{{\rm esc, 10}}\left(\frac{M_{\rm h}}{10^{10}{M}_{\odot }}\right)^{\alpha _{\rm esc}}.
\end{equation}
Then using \eqn{eq:SFR_expanded} for the SFR, we obtain the photon production rate as
\begin{equation}
\dot{n}^{\rm nofb}_{\rm ion} (z) =  \dfrac{f_{*,10}}{c_{*}}~f_{{\rm esc, 10}}~\eta_{\gamma*} \bigg(\dfrac{\Omega_b}{\Omega_m}\bigg) ~H(z) ~ \int_{M_{cool}(z)}^{\infty} \left(\frac{M'_{\rm h}}{10^{10}{M}_{\odot }}\right)^{\alpha _{\rm esc}+\alpha_*}  M'_h ~\dfrac{dn{(M'_h,z)}}{dM_h}~dM'_h.
\end{equation}
We can simplify this expression further by writing in terms of $\varepsilon_{*10, \mathrm{UV}}(z)$, see \eqn{eq:epsilonUVstar}
\begin{equation}
\label{eq:niondot_neutral}
\dot{n}^{\rm nofb}_{\rm ion} (z) =  \varepsilon_{{\rm *10,UV}}~\varepsilon_{{\rm esc,10}}~\eta_{\gamma*,{\rm fid}}  \bigg(\dfrac{\Omega_b}{\Omega_m}\bigg) ~H(z) ~ \int_{M_{cool}(z)}^{\infty} \left(\frac{M'_{\rm h}}{10^{10}{ M}_{\odot }}\right)^{\alpha _{\rm esc}+\alpha_*}  M'_h ~\dfrac{dn{(M'_h,z)}}{dM_h}~dM'_h,
\end{equation}
where
\begin{equation}
\varepsilon_{{\rm esc,10}} \equiv \dfrac{{\mathcal {K}_{{\rm UV}}}}{{{\mathcal {K}_{{\rm UV,fid}}}}} ~ \dfrac{\eta_{\gamma*}}{\eta_{\gamma*,{\rm fid}}}~f_{{\rm esc, 10}}.
\end{equation}
We take $\eta_{\gamma*,{\rm fid}}$ = $4.62175 \times 10^{60}$ photons per M$_\odot$, as calculated for continuous mode star formation at an age of 100 Myr with a 0.1 - 100 M$_\odot$ Salpeter IMF and metallicity $Z = 0.001 ( = 0.05~Z_\odot)$ using {\tt{STARTBURST99 v7.0.1}} \cite{Starburst99}.

At this point, let us relate the above quantities to another parameter widely used in the literature, namely, the rate of ionizing photons produced per unit UV luminosity at 1500${\angstrom}$, denoted by $\xi_{ion}$. It is straightforward to show that
\begin{equation}
\xi_{\rm ion} = \mathcal {K}_{\rm UV} ~ \eta_{\gamma*}.  
\end{equation}
This immediately implies that
\begin{equation}
\varepsilon_{{\rm esc,10}} \equiv \dfrac{\xi_{{\rm ion}}}{\xi_{{\rm ion, fid}}}~f_{{\rm esc, 10}},
\label{eq:new_xi_escape_fraction}
\end{equation}
where $\xi_{{\rm ion, fid}} \equiv \mathcal{K}_{{\rm UV,fid}}~\eta_{\gamma*,{\rm fid}}$ corresponds to the fiducial values used in the paper. For the fiducial values mentioned above, we obtain $\log_{10} \big[\xi_{\rm ion,fid}/({\rm ergs}^{-1}\ {\rm Hz}) \big] \approx 25.23$, similar to that assumed in other works \cite{Robertson2015}.

For ionized regions, we take into account radiative feedback by incorporating the retained gas fraction $f_g(M_h)$ as
\begin{align}
\label{eq:niondot_ion}
\dot{n}^{\rm fb}_{\rm ion} (z) &=  \varepsilon_{{\rm *10,UV}}~\varepsilon_{{\rm esc,10}}~\eta_{\gamma*,{\rm fid}}  \bigg(\dfrac{\Omega_b}{\Omega_m}\bigg) ~H(z)
\nonumber \\
& \quad \times \int_{M_{cool}(z)}^{\infty} f_g(M'_h) \left(\frac{M'_{\rm h}}{10^{10}{ M}_{\odot }}\right)^{\alpha _{\rm esc}+\alpha_*}  M'_h ~\dfrac{dn{(M'_h,z)}}{dM_h}~dM'_h.
\end{align}
The globally averaged number of hydrogen ionizing photons per unit time per unit comoving volume contributing to reionization is given by \cite{Dayal2017} -
\begin{equation}
\label{eq:niondot_total}
\dot{n}_{\rm ion}(z) = Q_{II}(z)~\dot{n}^{\rm fb}_{\rm ion}(z) + [1- Q_{II}(z)]~\dot{n}^{\rm nofb}_{\rm ion}(z) 
\end{equation}

Using \eqnsthree{eq:niondot_neutral}{eq:niondot_ion}{eq:niondot_total}, we solve the first-order differential equation, outlined in \eqn{eq:reion_eq0}, numerically through Euler's backward scheme to obtain the solution curve $Q_{II}(z)$, beginning with redshift $z_{\rm start} = 20$ where $Q_{II}(z_{\rm start})$ = 0.

Once the global reionization history $Q_{II}(z)$ is obtained, we can compute the Thomson scattering optical depth of the CMB photons for that particular model as
\begin{equation}
\label{eq:tauCMB}
\tau_{el} \equiv \tau (z_{\rm LSS}) = \sigma _T \bar{n}_{H}c \int _0^{z_{\rm LSS}} \frac{\mathrm{d}z^{\prime }}{H(z^{\prime })} ~ (1 + z^{\prime })^2 ~ \chi _{\mathrm{He}}(z^{\prime }) ~ Q_{\mathrm{II}}(z^{\prime }),
\end{equation}
where $z_{\rm LSS}$ is the redshift of last scattering,  $\bar{n}_{\rm H} $ is the mean comoving number density of hydrogen, and $\sigma_T$ is the Thomson cross-section. In practice, the contribution to the integral above is limited to the start of reionization $z_{\text{start}}$, before which $Q_{\mathrm{II}}(z > z_{\text{start}}) = 0$.

In total, our model of high-redshift star-forming galaxies has \emph{eleven} free parameters, which we summarize below:
\begin{itemize}
    \item  $\ell_{\varepsilon,0} - \ell_{\varepsilon,\mathrm{jump}} / 2$, $\ell_{\varepsilon,0} + \ell_{\varepsilon,\mathrm{jump}} / 2$, $z_\varepsilon$, $\Delta z_\varepsilon$: The parameters of the $\textit{tanh}$ function that models the redshift evolution in the normalization of the UV efficiency of high-$z$ galaxies, evaluated for halos of mass $10^{10} M_\odot$. In this parameterization, the value of $\log_{10}(\varepsilon_{{\rm *10,UV}})$ transitions from $\ell_{\varepsilon,0} - \ell_{\varepsilon,\mathrm{jump}} / 2$ to $\ell_{\varepsilon,0} + \ell_{\varepsilon,\mathrm{jump}} / 2$ over a redshift interval of $\Delta z_\varepsilon$ centered around $z = z_\varepsilon$. 

    \item  $\alpha_0 - \alpha_\mathrm{jump} / 2$, $\alpha_0 + \alpha_\mathrm{jump} / 2$, $z_\alpha$, $\Delta z_\alpha$: The parameters of the $\textit{tanh}$ function that models the redshift variation in the power-law scaling of the star-formation efficiency $f_{\rm *}$ with halo mass. In this parameterization, the value of $\alpha_*$ changes from  $\alpha_0 - \alpha_\mathrm{jump} / 2$ to $\alpha_0 + \alpha_\mathrm{jump} / 2$, with the transition occurring over a redshift interval of $\Delta z_\alpha$ centered at $z = z_\alpha$. 
  
    \item  $\log_{10} (M_\mathrm{crit} / M_\odot)$: The critical halo mass (in $M_\odot$) below which the gas fraction is suppressed due to radiative feedback. 
    
    \item  $\log_{10} (\varepsilon_{\rm esc,10})$: The normalization of the fraction of ionizing UV radiation escaping from high-$z$ galaxies, $\varepsilon_{\rm esc}$, evaluated for halos of mass $10^{10} M_\odot$. 
    
    \item  $\alpha_{\rm esc}$: The power-law scaling of the ionizing UV escape fraction $f_{\rm esc}$ with halo mass. 
\end{itemize}

\section{Observational Datasets and Likelihood Analysis}
\label{sec:datasets}
We utilize several available data sets to constrain the model parameters and the corresponding UV luminosity function and reionization histories through a Bayesian analysis. In this section, we list the different observational constraints that are used in this study and also describe the Bayesian formalism used to constrain the free parameters of our model.

\begin{enumerate}
    \item An important integrated constraint on the reionization history comes from the Thomson scattering optical depth ($\tau_{el}$) of CMB photons by the free electrons produced due to reionization. For our analysis, we use the latest Planck measurement \citep{Planck2020} of $\tau_{el} = 0.054 \pm 0.007$.
    
    \item  We use the galaxy UV luminosity functions at six redshift bins spanning a wide redshift range (6 $\leq z \leq$ 15), that are available from various surveys conducted with the Hubble Space Telescope  \cite{Bouwens2021} and the James Webb Space Telescope \cite{Donnan2023, Harikane2023, Bouwens2023, McLeod2024}. The UVLF measurements from the HST compiled in \cite{Bouwens2021} are the most comprehensive determinations at $z < 9$ available to date. The UVLF measurements from the different JWST studies listed above were deduced by analyzing the imaging data obtained from the Early Release Observations (ERO) of the SMACS J0723 cluster and the Stephan’s Quintet, as well as from the Early Release Science programs such as CEERS and GLASS. The results reported by \cite{McLeod2024} additionally utilized publicly available data from JWST Cycle-1 programs.
    
    From these studies, we consider only the relatively faint galaxies, i.e., observational data points with $M_{\rm UV} \geq -21$. This is because our model does not account for feedback due to active galactic nucleus (AGN) activity or the severe dust attenuation inside bright galaxies \cite{Mauerhofer&Dayal2023}.
    
    \item  We also utilize measurements of the global neutral hydrogen fraction ($Q_\mathrm{HI}$) in the IGM at different cosmic epochs. At relatively lower redshifts ($5.4 \leq z \leq 6$), we use the recent findings from a study of a large sample of quasar absorption spectra taken with the X-Shooter and ESI instruments \cite{Gaikwad2023}, combined with a radiative transfer code to model the Lyman-$\alpha$ opacities. For higher redshifts ($z \geq 6$), we use the constraints on $Q_\mathrm{HI}$ derived from Lyman-$\alpha$ damping wing observations of individual or stacked samples of high redshift quasars \cite{Davies2018, Greig2022, Durovcikova2024}, and a collection of UV-bright galaxies observed with the JWST \cite{Umeda2023}. It is worth mentioning that all the constraints on $Q_\mathrm{HI}$ used in this work are model-dependent, and hence one must be careful in interpreting the results. We will investigate this aspect in \secn{subsec:onlyJ23_case}.

\end{enumerate}

Our \textbf{default} analysis includes \emph{all} the data points mentioned above. However, to understand the relative importance of some of these observations, we study other variants of the analysis retaining only a subset of the data.

We use a Bayesian analysis to constrain the free parameters of our model by comparing the model predictions with the above observational constraints. Our objective is to compute the conditional probability distribution or the posterior $\mathcal {P}(\boldsymbol \theta \vert \mathcal{D})$ of the model parameters $\boldsymbol \theta$ given the observed data sets $\mathcal{D}$ mentioned above. This is calculated using the Bayes theorem
\begin{equation}
    \mathcal{P}(\boldsymbol \theta \vert \mathcal{D})=\frac{\mathcal {L}(\mathcal {D} \vert \boldsymbol \theta) ~\pi (\boldsymbol \theta)}{\mathcal{P}(\mathcal{D})},
\end{equation}
where $\mathcal {L}(\mathcal{D} \vert \boldsymbol \theta)$ is the likelihood i.e. the conditional probability distribution of the data $\mathcal {D}$ given the model parameters $\boldsymbol \theta$, $\pi(\boldsymbol \theta)$ is the prior distribution of the parameters of the model, and $\mathcal {P}(\mathcal{D})$ is the model evidence which is redundant in our work. The likelihood is assumed to be multidimensional Gaussian.
\begin{equation}
\mathcal {L}(\mathcal {D} \vert \boldsymbol \theta) = \exp \left(-\frac{1}{2} \sum _{i}\left[\frac{\mathcal {D}_i-\mathcal {M}_i(\boldsymbol \theta)}{\sigma _i}\right]^2 \right) 
= \prod _i \exp \left(-\frac{1}{2} \left[\frac{\mathcal {D}_i-\mathcal {M}_i(\boldsymbol \theta)}{\sigma _i}\right]^2 \right),
\end{equation}
where $\mathcal{D}_i$ are the values of the measured data points, $\mathcal{M}_i(\boldsymbol \theta)$ are the values predicted by the model for the parameters $\boldsymbol \theta$ and $\sigma_i$ are the observational error bars on the data points. The summation index $i$ runs over all data points used.

We obtain the posterior distribution of the model's free parameters using the Monte Carlo Markov Chain (MCMC) method. We make use of the publicly available package {\tt{COBAYA}}\footnote{https://cobaya.readthedocs.io/en/latest/} \cite{cobaya} for this purpose. The samples are drawn using 8 parallel chains. The chains are assumed to have converged when the Gelman–Rubin $R-1$ statistic becomes less than a threshold 0.01. We discard the first  30$\%$  of the steps in the chains as ‘burn-in’ and use the rest for our analysis.

The parameter priors and the constraints on them obtained from the different MCMC runs are summarized in \tab{tab:mcmc_results}. We have assumed flat priors for all the free parameters. The priors on the parameter set $\ell_{\varepsilon,0} - \ell_{\varepsilon,\mathrm{jump}} / 2, \ell_{\varepsilon,0} + \ell_{\varepsilon,\mathrm{jump}} / 2, z_\varepsilon$, $\Delta z_\varepsilon$ and $\alpha_0 - \alpha_\mathrm{jump}/ 2, \alpha_0 + \alpha_\mathrm{jump}/ 2, z_\alpha, \Delta z_\alpha$ have been chosen to encompass a broad spectrum of scenarios for the redshift evolution of the corresponding model quantities, $\log_{10}(\varepsilon_{{\rm *10,UV}})$ and $\alpha_*$, centered at $z \gtrsim 8$. The prior on the redshift-independent critical mass scale $M_{\rm{crit}}$ affected by radiative feedback is assumed to lie in the interval [$10^9$, $10^{11}$] $M_\odot$. The lower bound of $M_{\rm{crit}}$ is motivated by the typical Jeans mass scale at the virial overdensity in ionized regions at $z \sim 6$. The prior range of $\varepsilon_{esc,10}$ and $\alpha_{esc}$ too have been kept sufficiently wide.

\section{Results and Discussion}
\label{sec:results}

In this section, we present our constraints on the free parameters of the model and discuss the implications of these results. In this work, we execute three variants of MCMC runs using different combinations of observational datasets, as summarised below - 

\begin{itemize}
    \item \textbf{default}: For this case, we use all the data sets outlined in \secn{sec:datasets}. This constitutes our default run. 

    \item \textbf{only-HST}: Here, we use \textbf{only} the UVLF measurements made with the HST at $z \leq 9$ as the observational data for the UVLFs. We keep the other two sets of observations ($Q_{HI}$ and $\tau_{el}$) unchanged, as compared to the default case. The main objective of this run is to understand the significance of the inclusion of the JWST datasets ($z \gtrsim 9$).

    \item \textbf{only-J23}: In this case, we use the most recent \textbf{model-independent} constraints on the neutral hydrogen fraction at $z \leq 6.5$ in the intergalactic medium based on the fraction of dark pixels identified in Lyman-$\alpha$ and Lyman-$\beta$ forests \cite{Jin2023}. Furthermore, we allow only those reionization histories where it gets completed (i.e. $Q_{II} = 1 $) at $ z \geq 5.3$. This is motivated by the recent studies of large-scale fluctuations of the effective Lyman-$\alpha$ optical depth measured from high redshift quasar spectra \cite{Becker2015, Bosman2018, Kulkarni2019, Choudhury2021, Bosman2022}. We keep the other two sets of observations (UVLFs and $\tau_{el}$) unchanged, as compared to the default case. The main objective of this run is to understand the effect of altering the reionization observational constraints used.
\end{itemize}

\begin{table*}
 \caption{Parameter constraints obtained from the MCMC-based analysis for the presently available data. The first eleven rows correspond to the free parameters of the model while the others are the derived parameters. The free parameters are assumed to have uniform priors in the range mentioned in the second column. The other numbers show the mean value with 1$\sigma$ errors for different parameters of the model, as obtained from the three MCMC runs (see \secn{sec:results}).\\} 
 \label{tab:mcmc_results}
 \begin{tabular*}{\columnwidth}{l@{\hspace*{20pt}}l@{\hspace*{20pt}}l@{\hspace*{20pt}}l@{\hspace*{20pt}}l@{\hspace*{20pt}}}
  \hline   \hline  \\
    Parameters & Priors &  default & only-HST & only-J23 \\  \\
  \hline   \hline \\
{\boldmath$\ell_{\varepsilon,0} + \ell_{\varepsilon,\mathrm{jump}} / 2 $} & [-2.0, 2.0] & $0.45^{+0.67}_{-0.98}$   & $< -0.972$  & $0.45^{+0.61}_{-0.98} $  \\  \\

{\boldmath$\ell_{\varepsilon,0} - \ell_{\varepsilon,\mathrm{jump}} / 2$} & [-2.0, 1.0] & $-0.924\pm 0.062$  &  $-0.874^{+0.057}_{-0.099}  $  & $-0.930\pm 0.064 $ \\ \\

{\boldmath$z_\varepsilon $} & [8.0, 18.0] & $14.2^{+2.3}_{-1.8}$ & $> 11.6$ & $14.2^{+2.4}_{-1.8} $\\ \\

{\boldmath$\Delta z_\varepsilon $} & [1.5, 6.0] & $< 3.78$  & ---  & $< 3.79 $\\ \\

{\boldmath$\alpha_0 + \alpha_\mathrm{jump} / 2 $} & [0.0, 7.0] & $4.2^{+2.2}_{-1.4}$  &  $< 3.68$ & $3.9\pm 1.6$  \\ \\

{\boldmath$\alpha_0 - \alpha_\mathrm{jump} / 2 $} & [0.0, 1.0]  & $0.313\pm 0.057$ & $0.293^{+0.096}_{-0.060}$ & $0.318\pm 0.060$ \\ \\

{\boldmath $z_\alpha$} & [8.0, 18.0] &  $13.71^{+0.81}_{-1.7}$  & $> 14.0$ & $13.82^{+0.93}_{-1.9}      $ \\ \\

{\boldmath$\Delta z_\alpha$} & [1.5, 6.0] & $< 2.48$ & ---  & $< 2.56 $  \\ \\

{\boldmath$\log_{10}~(\varepsilon_{\mathrm{esc,10}})$} & [-3.0, 1.0]  & $-0.957^{+0.13}_{-0.060} $ & $-0.973^{+0.15}_{-0.056}  $ &  $-0.81^{+0.30}_{-0.24} $ \\ \\

{\boldmath$\alpha_{esc}   $} & [-3.0, 1.0] & $-0.38^{+0.19}_{-0.15}$  & $-0.43^{+0.24}_{-0.16} $  & $> 0.0112  $ \\ \\

{\boldmath$\log_{10}(M_{\mathrm{crit}}/M_\odot)$} & [9.0, 11.0] &$< 9.70 $ & $< 9.75$ & $< 9.58 $\\ \\

\hline \\

$\tau_{el}                 $ & - &  $0.0558^{+0.0020}_{-0.0030}$ & $0.0554^{+0.0024}_{-0.0036}$ & $0.0575^{+0.0058}_{-0.0076}$ \\ \\

$\ell_{\varepsilon,\mathrm{jump}}$ & -  &$1.37^{+0.68}_{-0.98}$ & $-0.16^{+0.20}_{-0.96}$ & $1.38^{+0.62}_{-0.99}      $\\ \\

$\alpha_\mathrm{jump}  $ & - &$3.9^{+2.3}_{-1.3} $ &  $2.5^{+1.1}_{-2.7} $  & $3.6\pm 1.6$ \\ \\

$\ell_{\varepsilon,0}      $ & - &$-0.24^{+0.32}_{-0.50} $ &  $-0.082^{+0.099}_{-0.48}$  & $-0.24^{+0.30}_{-0.50}     $ \\ \\

$\alpha_0                  $ & - & $2.25^{+1.1}_{-0.66} $  &   $1.23^{+0.54}_{-1.3}$ &$2.11\pm 0.80$ \\ \\

\hline
\end{tabular*}
\end{table*}

\subsection{Results from the {\tt{default}} case}
\label{subsec:results_default_case}

\begin{figure}[htbp]
\centering
\includegraphics[width=\columnwidth]{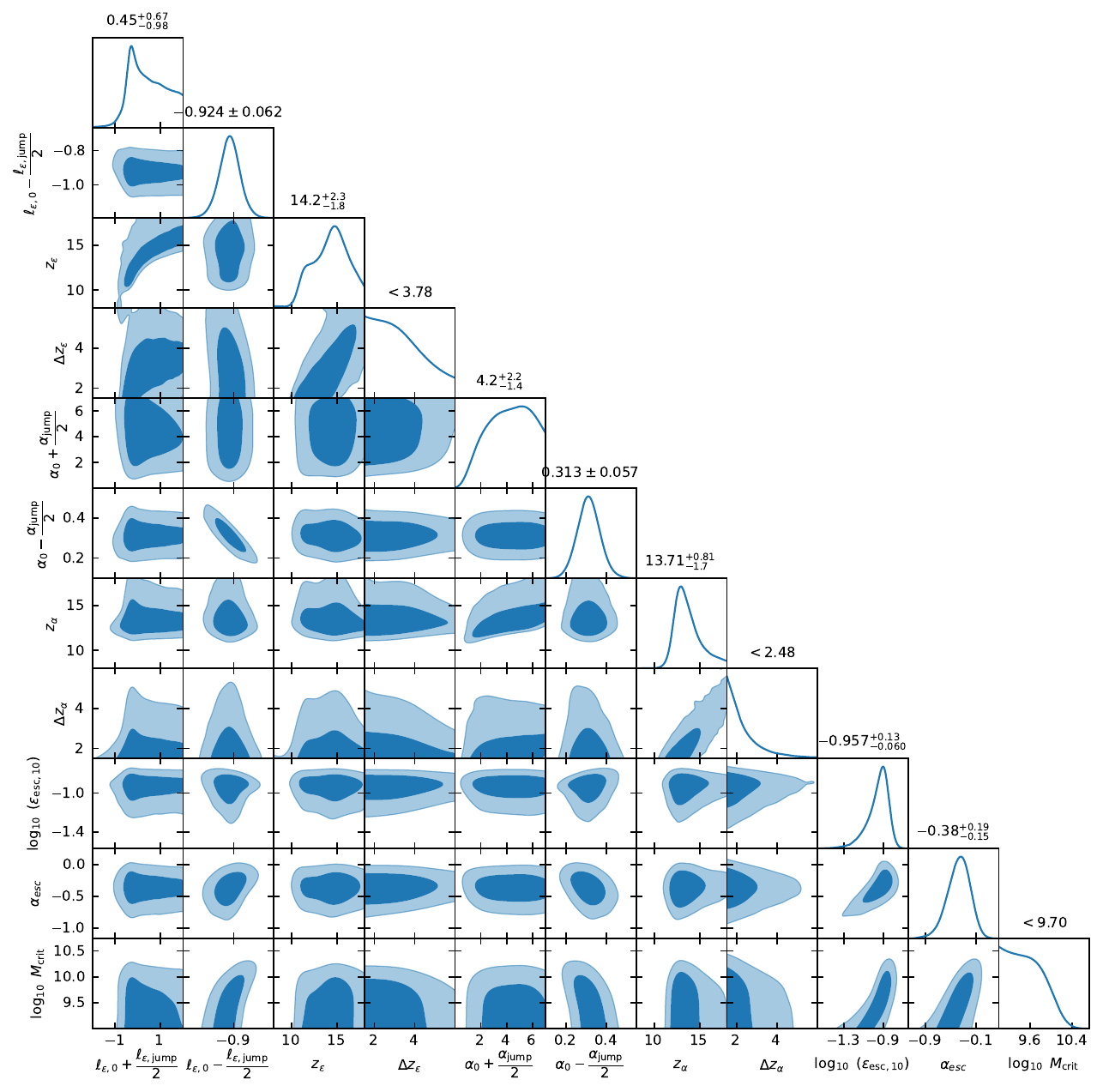}
\caption{Posterior distributions of the free parameters for the \textbf{default} case. The diagonal panels show the one-dimensional distribution, while the contour plots in the off-diagonal panels represent the two-dimensional joint distribution. The contour levels represent 68\% and 95\% confidence levels. The mean and 68\% confidence intervals are denoted above the one-dimensional posterior distributions of the respective parameters.}
\label{fig:allQHI_corner}
\end{figure}

We begin the discussion with the results obtained from the MCMC runs of the  \textbf{default} case. The posterior distributions of the free and derived parameters are shown in \fig{fig:allQHI_corner}. The corresponding constraints on the free and derived parameters can be found in \tab{tab:mcmc_results}. 

\begin{figure}[htbp]
\centering
\includegraphics[width=\columnwidth]{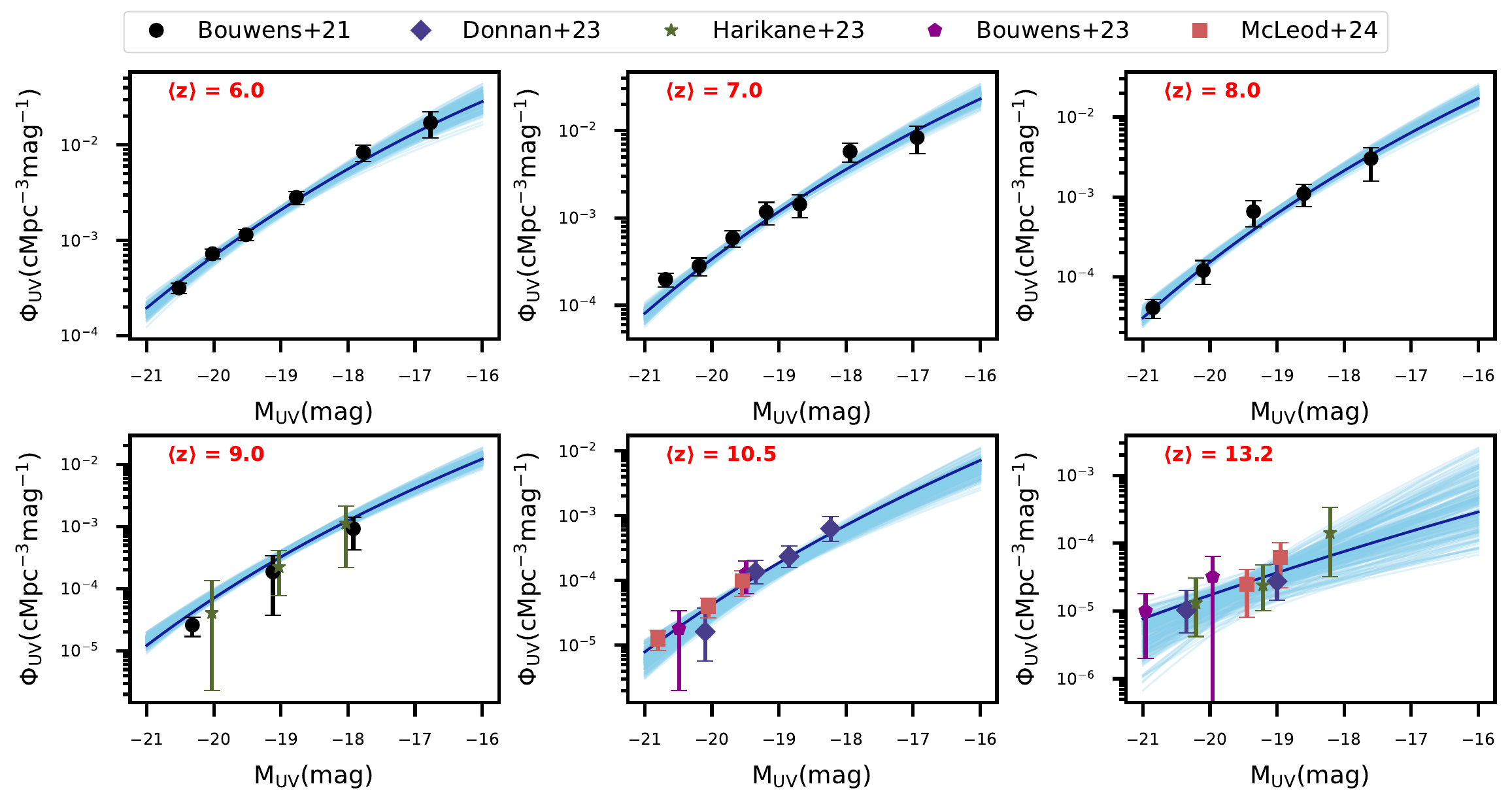}
\caption{The galaxy UV luminosity functions at six different redshift bins (with their respective mean values $\langle z \rangle$ mentioned in the upper left corner) for 200 random samples drawn from the MCMC chains for the
\textbf{default} case. In each panel, the solid dark-blue line corresponds to the best-fit model, while the colored data points show the different observational constraints \cite{Bouwens2021, Donnan2023, Harikane2023, Bouwens2023, McLeod2024} used in the likelihood analysis.}
\label{fig:allQHI_UVLF}
\end{figure}

\begin{figure}[htbp]
\centering
\includegraphics[width=\columnwidth]{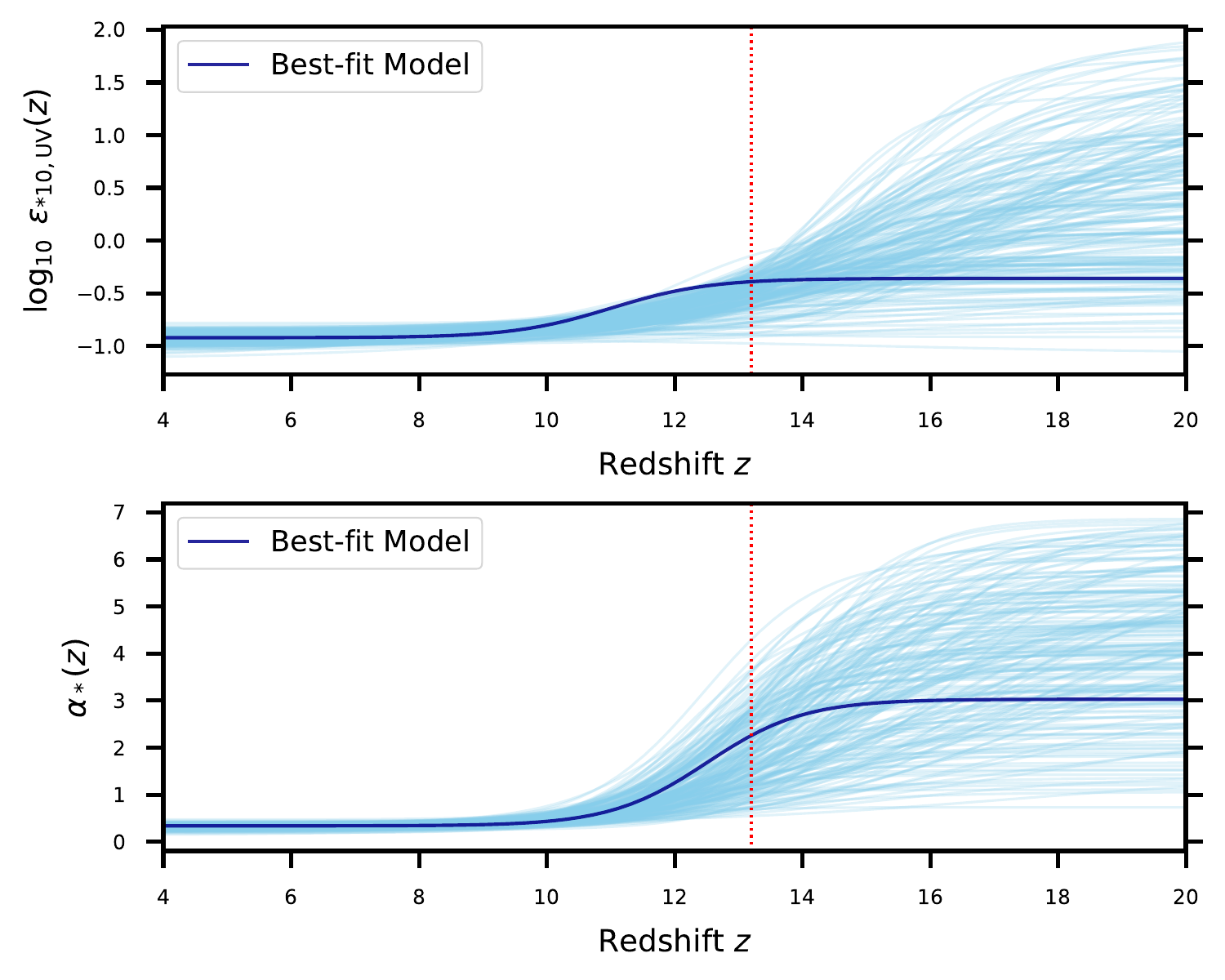}
\caption{The redshift evolution of the normalization (top panel) and power-law (bottom panel) scaling of the UV/star-formation efficiency with halo mass for 200 random samples drawn from the MCMC chains for the \textbf{default} case. The vertical red dotted line shows the highest mean redshift ($\langle z \rangle$ = 13.2) of the UVLF measurements used in this work.}
\label{fig:allQHI_sfe_params}
\end{figure}

From \tab{tab:mcmc_results} and \fig{fig:allQHI_corner}, we notice that the individual constraints on the various free parameters characterizing the star formation and UV efficiency towards the high redshift regime ($z \gtrsim 10$) are not that stringent. Despite this, it is clear that we require the UV efficiency parameter, $\varepsilon_{{\rm *10,UV}}$, to assume higher values at $z \geq 10$ to match the evolving UVLF observations from the JWST. This is in agreement with many recent studies \cite{Dekel2023, Mason2023, Qin2023, Mirocha2023, Inayoshi2022, Munoz2023, Yung2024} and could be indicative of several possible scenarios (or their combination) occurring at  $z \gtrsim 10$. The most straightforward explanation would be an increasing star formation efficiency $f_\ast$ at early times. Recent analytical studies \cite{Dekel2023} provide support for this scenario, suggesting that if the gas clouds within massive early galaxies have densities exceeding a few times $10^{-3} \mathrm{cm^{-3}}$ and are devoid of metals, then their free-fall timescale would be shorter than the duration needed by stellar feedback mechanisms, such as winds, supernova explosions etc, from low metallicity stars to become effective. This could lead to the occurrence of \emph{feedback-free} starbursts in halos that are more massive than a threshold of $10^{10.54}~\mathrm{M}_\odot$ at $z \sim 10$, with this mass threshold decreasing further at higher redshifts. An alternate possibility is that these early galaxies probably formed their stars in a very short amount of time (i.e. the typical value of $c_\ast$ is lower), exhibiting \emph{bursty} star formation histories. Although the stochastic nature of star formation at $z \geq  10$  can in principle account for the overabundance seen by the JWST \cite{Mason2023, Mirocha2023, Shen2023} relying on the Eddington bias, it remains to be seen whether they alone are sufficient to resolve the discrepancy \cite{Sun2023, Pallottini2023}. Essentially, both of these scenarios point towards increased star formation rates at early redshifts as an explanation. 

Another intriguing conjecture is that perhaps the UV light-to-mass ratio for these early galaxies is much larger due to a greater abundance of high-mass stars, suggesting a more top-heavy initial mass function (IMF) \cite{Inayoshi2022, Harikane2023, Finkelstein2023_UVLF, Trinca2023, Yung2024}. At high redshifts, a higher CMB temperature \cite{Jermyn2018, Chon2022} and lower gas-phase metallicity \cite{Chon2021} can collectively increase the Jeans' mass for star-forming clouds, inhibiting the formation of low-mass stars in these systems. Several theoretical works in the literature claim that the first generation of stars (the so-called \textit{Population III} stars) born out of pristine primordial gas follow a top-heavy initial mass function \cite{Susa2014, Hirano2015, Stacy2016}. In such a case, the factor $\kappa_{\rm UV}$ would be $\sim$ 3 – 4 times lower at early times than our adopted fiducial value (see e.g., Table 1 of Ref.\cite{Inayoshi2022}, Fig. 20 of Ref.\cite{Harikane2023}), resulting in higher UV luminosity for the same star formation rate. However, a recent study \cite{Cueto2023} that invokes an evolving IMF argues that this scenario alone is unlikely to explain the observed $z > 10$ UVLFs since a top-heavy IMF would also influence the stellar and chemical feedback mechanisms regulating Pop-II/Pop-III star formation activity.

Our analysis finds the power-law scaling index of the mass-dependent star-forming efficiency $f_*$ to increase with halo mass and steepen from a nearly constant value of $\alpha_\ast$ $\approx$ 0.3 at $z \lesssim$ 9 to $\alpha_\ast$ $\gtrsim$ 1 at higher redshifts. This transition is preferred to be reasonably rapid and occurs somewhere between $z \approx 11 - 13$. At a fixed luminosity, increasing $f_\ast$ (either through $f_{\ast,10}$ or $\alpha_\ast$) means assigning relatively smaller, and thus more abundant, DM host halos to the brighter galaxies. This helps our models to explain the observed flatness and mild evolution of the bright end of the  $z\geq 10$ UVLFs. The value of the power-law index obtained for $z \leq 9$  from our analysis agrees qualitatively with other semi-analytical or semi-numerical studies that used a power-law index of $\sim 0.4 - 0.5$ to match the UVLF observations from the HST \cite{Park++2019, Qin2020_CMB, Qin2021, Maity2022_ParameterConstraints, Sipple2023}.

Whatever the physical reasoning may be, these findings clearly reveal that processes regulating star formation and the production of UV photons in the earliest galaxies may be significantly different from those seen in galaxies at lower redshifts. We show the model-predicted UVLFs for 200 random samples from the MCMC chains in \fig{fig:allQHI_UVLF}, along with the data points of the various observations \cite{Bouwens2021, Donnan2023, Harikane2023, Bouwens2023, McLeod2024} that were used in the MCMC analysis. The evolution in the star-formation activity preferred by the data has been plotted in \fig{fig:allQHI_sfe_params} for the same 200 random samples.

\begin{figure}[htbp]
\centering
\includegraphics[width=\columnwidth]{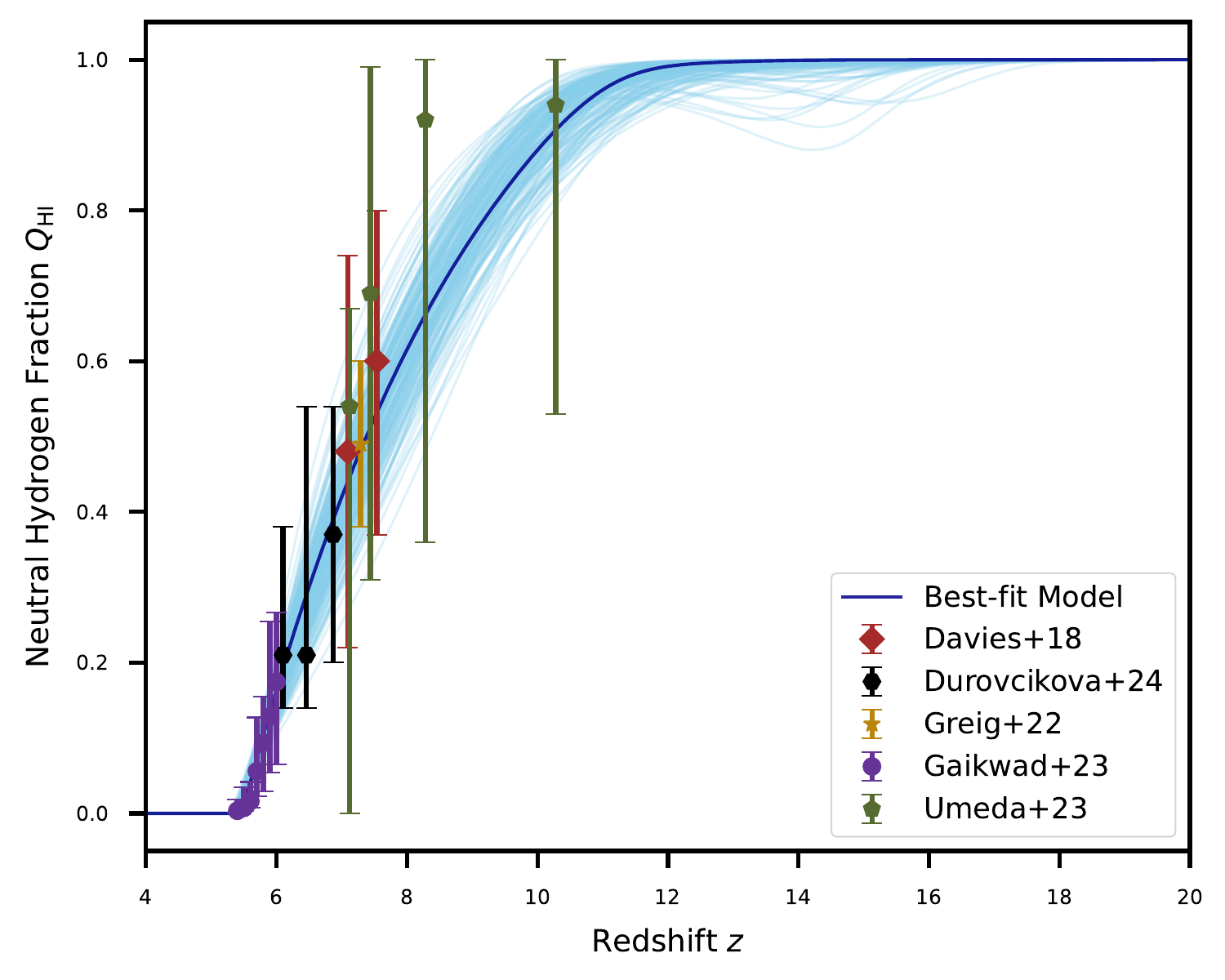}
\caption{The reionization histories for 200 random samples drawn from the MCMC chains for the \textbf{default} case. The data points denote the different observational constraints for $Q_{\rm HI}$($z$) used in the likelihood analysis.}
\label{fig:allQHI_reionhist}
\end{figure}

We are only able to obtain an upper limit for the halo mass scale $M_{\rm crit}$ affected by radiative feedback from the current measurements of the UVLFs, with the corresponding marginalized 1D probability distribution remaining relatively constant below approximately $10^{9.7}M_\odot$ (68\% confidence limit) and falling sharply beyond this value. The inability to constrain $M_{\rm crit}$ robustly from our analysis can possibly be attributed to the large uncertainties at the faint end of the measured UVLFs, where reionization feedback effects are expected to be most pronounced, and/or to the lack of a discernible faint-end turnover in the observational datasets used in this work. 

Our obtained constraints mildly prefer a negative value of $\alpha_{esc}$. This inclination for $\alpha_{esc} < 0$ implies that the ionizing escape fraction is higher for galaxies that are formed in lower-mass halos. From the constraints obtained on $\log_{10}(\varepsilon_{esc,10})$, we find that an escape fraction of $\sim 11 \%$ is required for $10^{10}$ M$_\odot$ halos for the assumed fiducial value of $\log_{10} \big[\xi_{\rm ion,fid}/({\rm ergs}^{-1}\ {\rm Hz}) \big] \approx 25.23$\footnote{It must be emphasized that the constraints on the escape fraction are completely degenerate with $\xi_{ion}$. Recent studies from JWST have inferred high values of $\log_{10} \big[\xi_{\rm ion}/({\rm ergs}^{-1}\ {\rm Hz}) \big]$ in the range of $25.6-26.0$ \cite{CurtisLake2023, Rinaldi2023, Endsley2023, Atek2024_Spectroscopy, Simmonds2024,AlvarezMarquez2024, Calabro2024}, which would imply a much lower escape fraction ($\approx 3\%$) for 10$^{10} M_\odot$ halos.}. Low-mass DM halos typically have shallower gravitational potential wells. This means that the gas within these halos is not as strongly bound as in the more massive halos and thus, is more susceptible to being blown away by supernovae explosions or other outflows. This facilitates the creation of low-column density channels through which ionizing photons can escape. There also exists a strong correlation between $\log_{10}(\varepsilon_{esc,10})$ and $\alpha_{esc}$, as also seen in \fig{fig:allQHI_corner}. Since the escape fraction follows a power-law functional form, it is understandable that, for a constant value of $\varepsilon_{esc,10}$, decreasing $\alpha_{esc}$ will result in more effective escape of ionizing photons from halos with $M_h < 10^{10} M_\odot$. This would need to be counterbalanced by also decreasing the normalization $\varepsilon_{esc,10}$ to remain consistent with data. The mass-dependency of $f_{esc}$ suggested by our analysis is in qualitative agreement with that found in different hydrodynamic simulations \cite{Paardekooper2015, Xu2016, Lewis2020, Kimm2014} and other semi-numerical models \cite{Park++2019, Maity2022_ParameterConstraints, Mutch2023} of reionization. The typical behavior of the favored reionization histories can be understood from \fig{fig:allQHI_reionhist} where we have plotted the evolution of the globally averaged neutral fraction of 200 random samples from the MCMC chains.
We notice that some of the allowed models show double reionization histories in which the globally averaged neutral hydrogen fraction evolves non-monotonically with redshift. In these models, the star formation rate (SFR) is significantly higher at $z \gtrsim 12$, followed by a sharp decline at lower redshifts. This evolution results in an initial phase of partial reionization, followed by recombination as the SFR drops, and a subsequent second stage of complete reionization. Since results from the JWST suggest a higher-than-expected star formation rate density at very early times, ruling out these scenarios of double reionization would require measurements of the neutral hydrogen fraction at $z>10$.

\subsection{Implications from the {\tt{default}} case}
\label{subsec:implications_default_case}

Despite its simplicity, our semi-analytical formalism can be used to compute several other observables and quantities related to the high-redshift galaxy population. Therefore, we now move towards looking at several implications of our model. 

\begin{figure}[htbp]
\centering
\includegraphics[width=0.8\columnwidth]{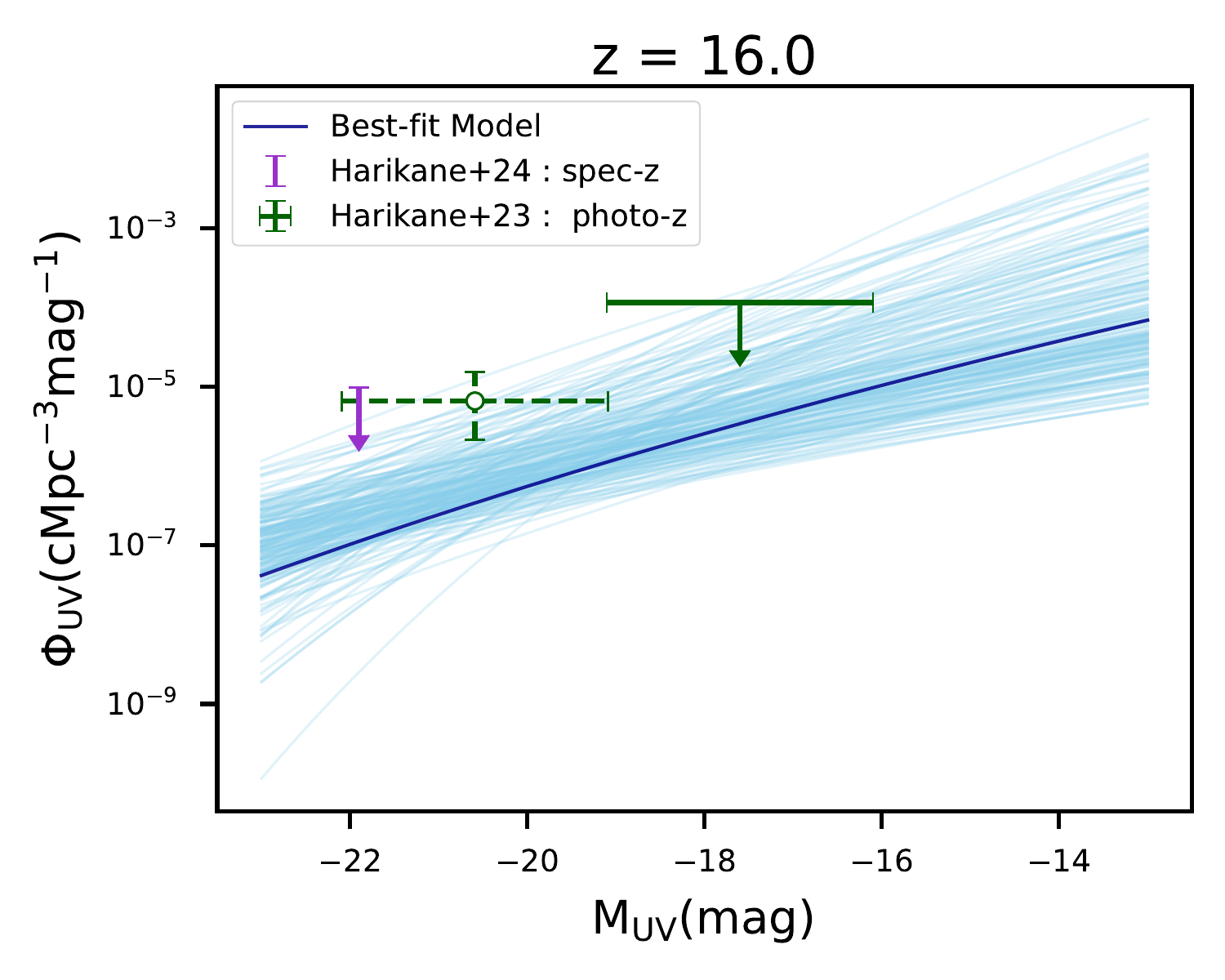}
\caption{The predicted  $z$ = 16 galaxy UV luminosity function for 200 random samples drawn from the MCMC chains for the \textbf{default} case. The dark-blue solid line corresponds to the best-fit model. We represent the Harikane et al. (2023) estimate at $\mathrm{M_{UV}} \approx -20.6$ using open circles as it was based on two photometrically selected galaxies, one of which (CEERS-93316) is now spectroscopically confirmed to be at lower redshifts ($z = 4.9$), and the other candidate from the Stephan’s Quintet field (S5-z16-1) also has a bimodal photometric redshift solution ($z = 16.4 ~\mathrm{or}~ 4.6$).}
\label{fig:UVLF16}
\end{figure}

One major advantage of our formalism lies in its capability to predict the galaxy UV luminosity function at high redshifts and also towards the faint end, which are currently beyond the reach of existing observations. In \fig{fig:UVLF16}, we show the galaxy UV luminosity function at $z$ = 16 predicted by our model, which is otherwise constrained by UVLF observations at $\langle z \rangle \leq$ 13.2, for 200 random samples drawn from the MCMC chains for the \textbf{default} case. Our UVLF predictions are in reasonable agreement with the recent upper limits obtained by the photometric \cite{Harikane2023} as well as spectroscopic study \cite{Harikane2024} of $z \geq$ 15 galaxies with the JWST. In the near future, any robust detection of galaxies at $z>15$ by the JWST  will enable us to test these predictions better.

\begin{figure}[htbp]
\centering
\includegraphics[width=0.8\columnwidth]{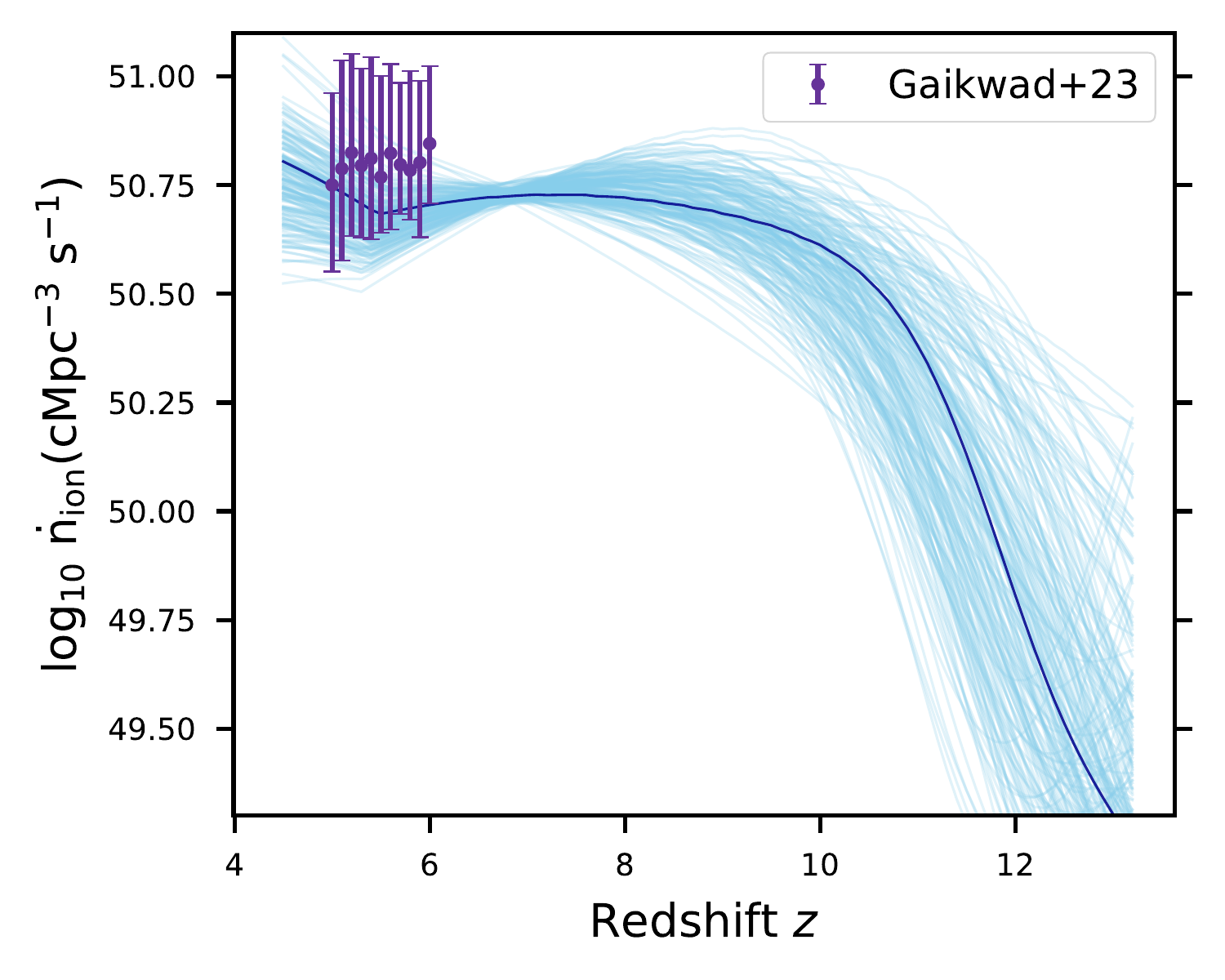}
\caption{The redshift evolution of the emissivity of ionizing photons from star-forming galaxies for 200 random samples drawn from the MCMC chains for the \textbf{default} case. The solid dark-blue line corresponds to the best-fit model, while the colored data points represent the observational measurements \cite{Gaikwad2023}.}
\label{fig:niondot_samples}
\end{figure}

We next proceed to check the evolution of the ionizing emissivity $\dot{n}_{\rm ion}(z)$ as a function of redshift. At lower redshifts ($z \lesssim$ 6), one can infer $\dot{n}_{\rm ion}$ from the measurements of the HI photoionization rate $\Gamma_{\rm HI}$ and the mean free path $\lambda_{\rm mfp}$ of ionizing photons obtained by combining Lyman-$\alpha$ forest observations with detailed hydrodynamical simulations \cite{Bolton2007, Becker2013, DAloisio2018, Becker2021, Gaikwad2023}. From \fig{fig:niondot_samples}, we see that the ionizing emissivity in our models quickly increases with decreasing redshift as new structures form in the Universe and then turns over to steadily decline up to $z \approx 5.5$, after which it again rises.  In our model, the intrinsic production rate of ionizing photons traces the star formation rate at all redshifts. The decrease in $\dot{n}_{\rm ion}$ seen at $z \lesssim 9$ occurs for two main reasons. Firstly, star formation inside low-mass halos, which predominantly contribute to the ionizing radiation at early times, gets increasingly quenched due to radiative feedback as reionization progresses. Secondly, the average escape fraction $\langle f_{esc} \rangle$ also decreases with time as more massive galaxies, which have lower escape fractions in our model and are less affected by photo suppression, start forming at lower redshifts. This slow evolution in the ionizing photon emission makes reionization progress gradually in our models. Nevertheless, our models predict an ionizing emissivity consistent with the recent observations at $5 \lesssim z \lesssim 6$ \cite{Gaikwad2023}, albeit on the lower end of their allowed 1$\sigma$ confidence interval. In our models, an increase in $\dot{n}_{\rm ion}(z)$ is again seen once reionization is complete and the UV ionizing background has been fully established. Interestingly, some hydrodynamical simulations \cite{Gaikwad2023, Keating2020_ATON0, Keating2020_ATON1} also find a similar evolution of the ionizing emissivity, i.e., a rise, followed by a dip and again a rise in  $\dot{n}_{\rm ion}$ as redshift decreases. These authors claim that such an emissivity evolution is needed to match the Lyman-$\alpha$ forest statistics (e.g., the evolution of the observed Ly$\alpha$ transmitted mean flux with redshift) at $z \lesssim 5.5$ \cite{Keating2020_ATON0, Keating2020_ATON1}.

\begin{figure}[htbp]
\centering
\includegraphics[width=\columnwidth]{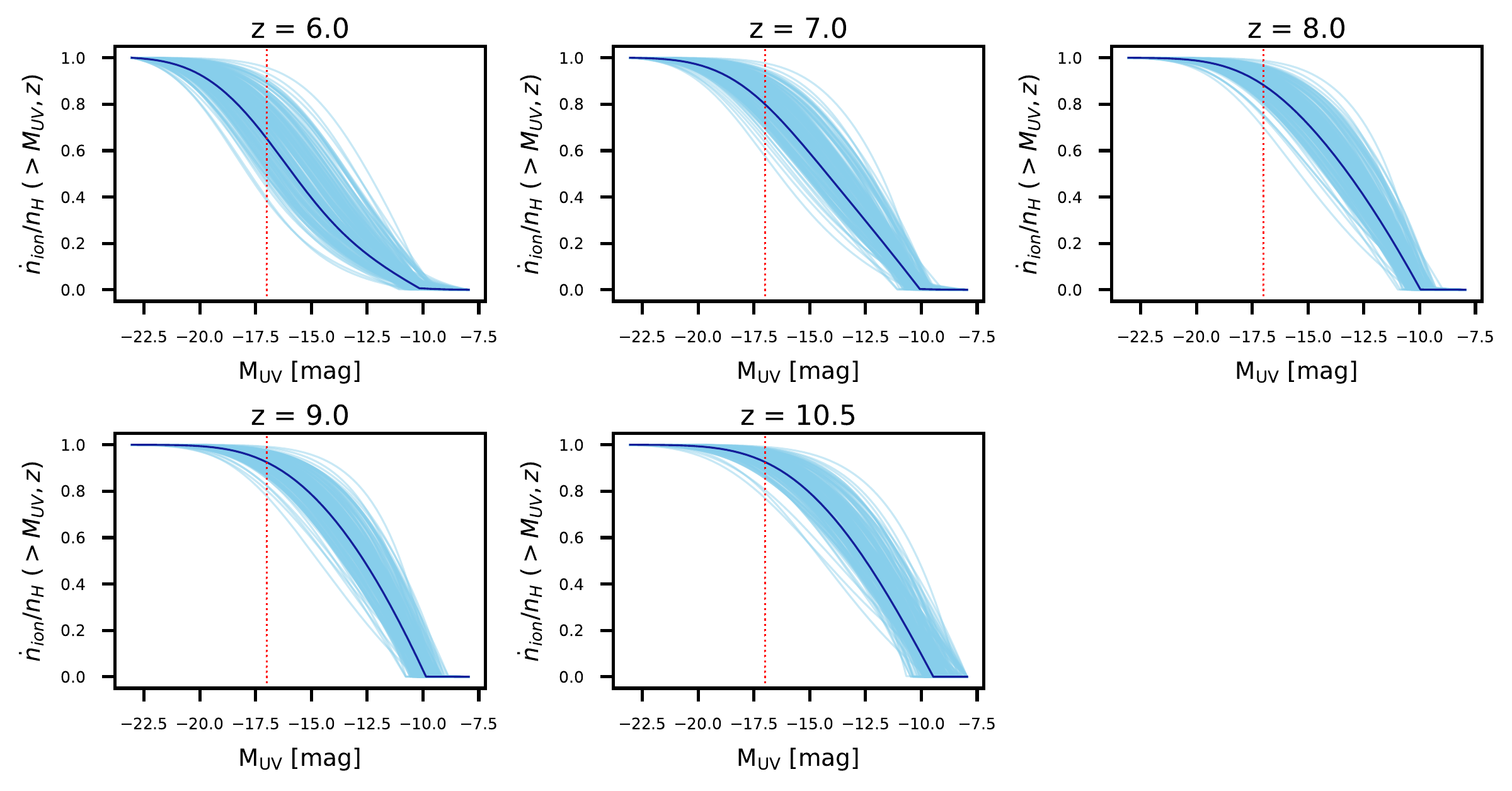}
\caption{The cumulative distribution of the instantaneous ionizing emissivity as a function of galaxy UV luminosity at different redshifts for 200 random samples drawn from the MCMC chains for the \textbf{default} case. In each panel, the solid dark-blue line corresponds to the best-fit model, while the vertical dotted red line denotes the cumulative contribution from galaxies fainter than $\mathrm{M_{UV}}$ = -17.}
\label{fig:cum_nion_samples}
\end{figure}

Another important topic of ongoing debate is the contribution of low-luminosity galaxies to the ionizing photon budget for reionization. Some studies \cite{Sharma2016, Naidu2020, Joshi2024} contend that faint galaxies ($M_{UV} \geq$ -17) make only a negligible contribution, while others \cite{Anderson2017, Lewis2020, Atek2024_Spectroscopy, Simmonds2024} argue that these sources play a dominant role in the reionization of the Universe. In our model, the star-formation rate increases with increasing halo mass, whereas the abundance of halos (governed by the halo mass function) and the escape fraction of UV ionizing photons decreases with increasing halo mass. These competing effects essentially shape the ionizing photon budget for reionization at a given redshift. To understand the relative importance of galaxies with different UV luminosities in reionizing the Universe, we calculate the relative contributions to $\dot{n}_{\rm ion}(z)$ from various UV magnitude bins at different redshifts spanning various stages of the reionization process as follows. 
\begin{align}
\label{eqn:dniondotdMUV_full}
\frac{{\rm d}\dot{n}_{\rm ion}}{{\rm d} M_{\rm UV}} &= Q_{\rm II}(z)~\frac{{\rm d}\dot{n}^{\rm fb}_{\rm ion}}{{\rm d}M_{\rm UV}}+ [1-Q_{\rm II}(z)]~\frac{{\rm d}\dot{n}^{\rm nofb}_{\rm ion}}{{\rm d} M_{\rm UV}}
\nonumber \\
&= Q_{\rm II}(z)~\frac{{\rm d}\dot{n}^{\rm fb}_{\rm ion}}{{\rm d}M_h} \left|\frac{{\rm d}M_h}{{\rm d}M_{\rm UV}}\right|_{\rm fb}
+ \big[1 - Q_{\rm II}(z)\big] \frac{{\rm d}\dot{n}^{\rm nofb}_{\rm ion}}{{\rm d}M_h} \left|\frac{{\rm d}M_h}{{\rm d}M_{\rm UV}}\right|_{\rm nofb}
\end{align}
where ${\rm d}\dot{n}_{\rm ion}/{\rm d} M_{\rm UV}$ denotes the comoving number density of ionizing photons per unit time per unit magnitude contributed by galaxies having UV magnitudes between $M_{\rm UV}$ and $M_{\rm UV} + {\rm d} M_{\rm UV}$, while the terms ${\rm d}\dot{n}^{\rm nofb}_{\rm ion}/{\rm d}M_h$ and ${\rm d}\dot{n}^{\rm fb}_{\rm ion}/{\rm d}M_h$ represent the integrands (with pre-factors) of \eqns{eq:niondot_neutral}{eq:niondot_ion} respectively.

In \fig{fig:cum_nion_samples}, we present the distributions of the cumulative contribution made to the instantaneous ionizing emissivity by galaxies fainter than a certain UV magnitude for 200 random samples drawn from the MCMC chains for the \textbf{default} case. It is evident that galaxies fainter than a fiducial value of $M_{\rm UV} = - 17$ produce the majority of the ionizing photons that drive cosmic reionization, with their relative contribution dropping from $\gtrsim$ 80\% at $z \sim 10.5$  to $\gtrsim$ 40\% at $z \sim 6$. 

\begin{figure}[htbp]
\centering
\includegraphics[width=.45\textwidth]{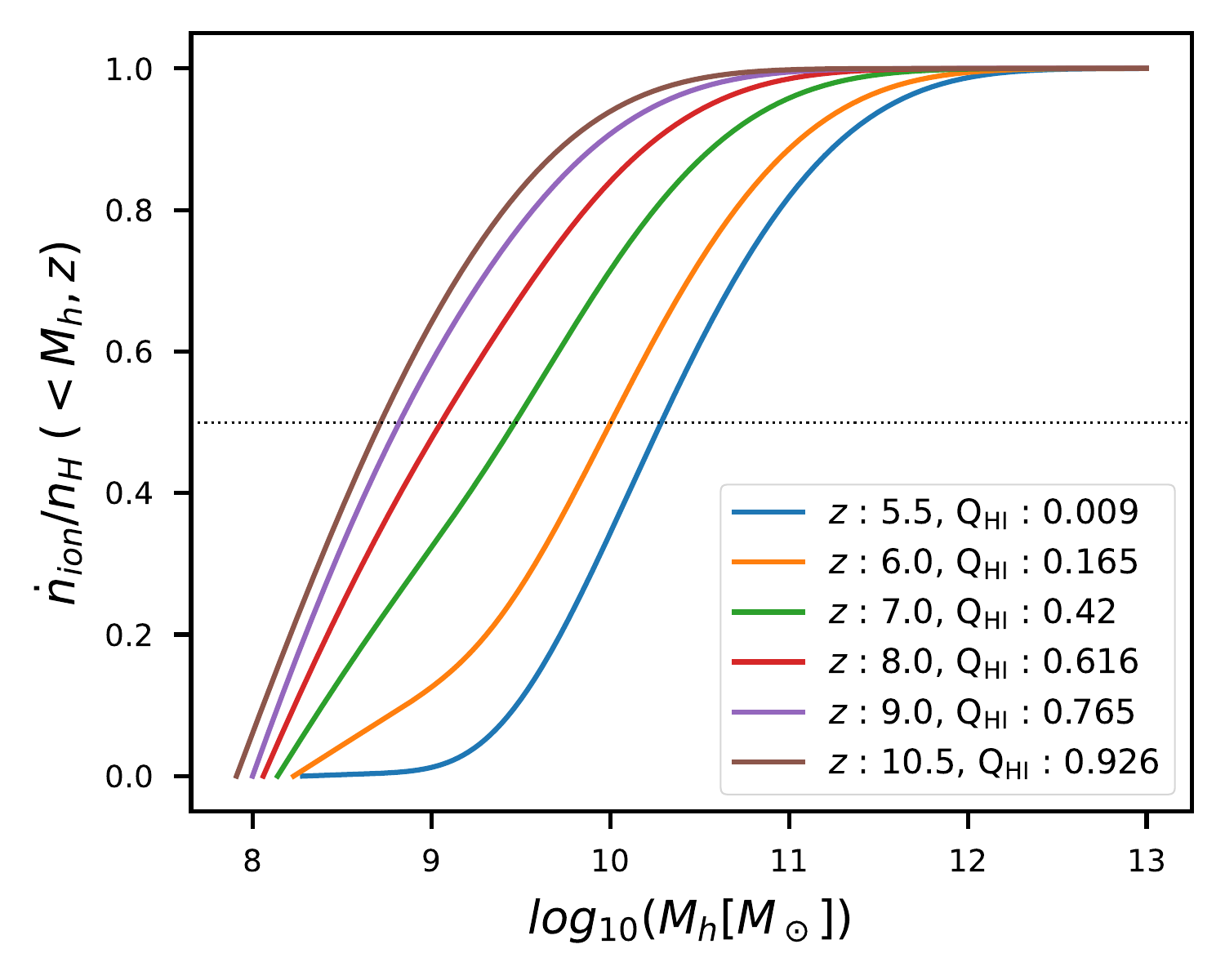}
\qquad
\includegraphics[width=.45\textwidth]{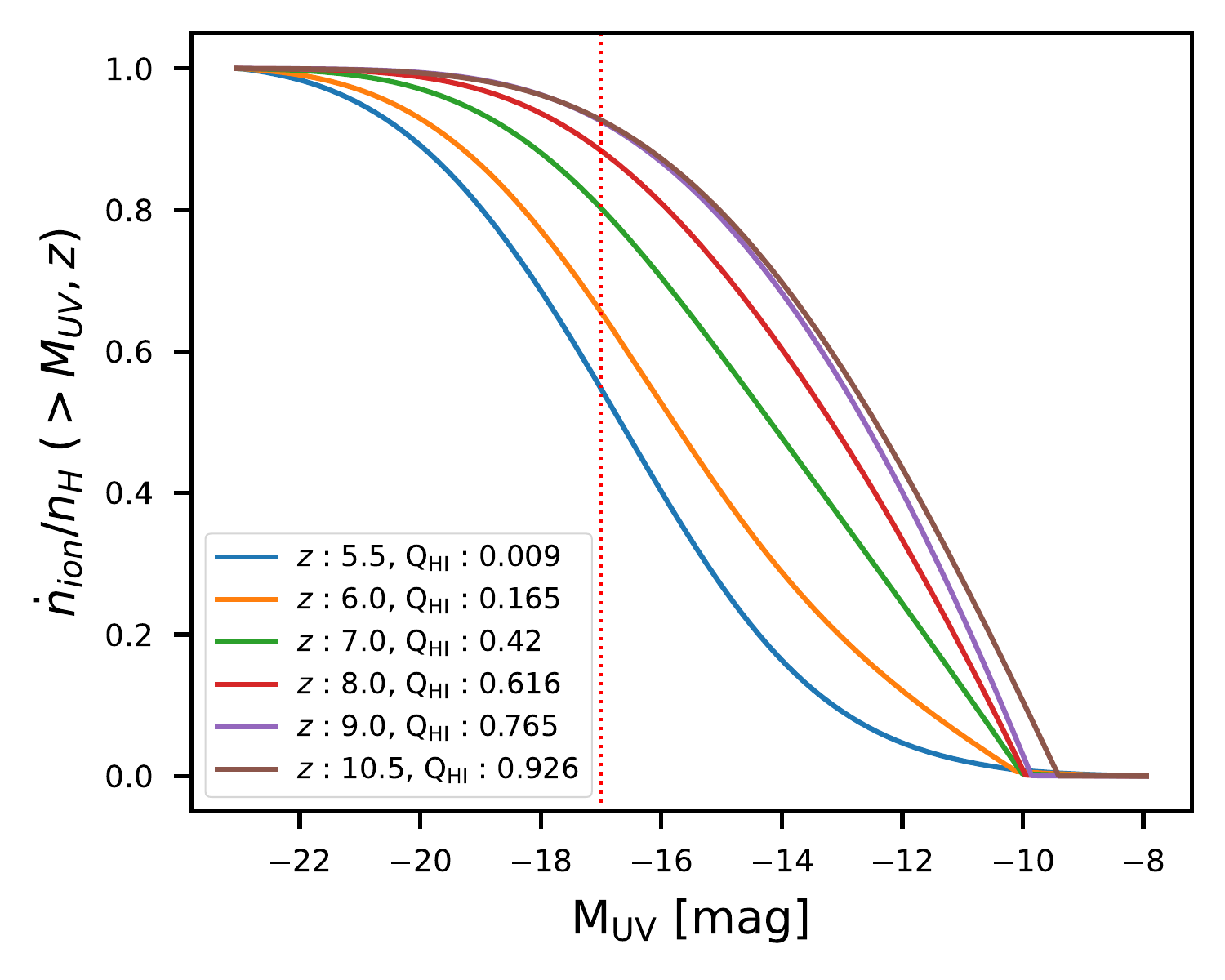}
\caption{The cumulative contribution to the instantaneous ionizing photon
rate as a function of halo mass (left panel) and UV magnitude of the resident galaxies (right panel) at different redshifts resulting from
the best-fit model of the \textbf{default} case. The horizontal black dash-dotted line in the left panel corresponds to a cumulative contribution of 50\% to the photon budget, while the vertical red dotted line in the right panel represents the cumulative contribution from galaxies fainter than $\mathrm{M_{UV}}$ = -17.}
\label{fig:cum_allz_nion_BF}
\end{figure}

We also specifically show the evolution of the instantaneous ionizing emissivity as a function of galaxy UV luminosity and its host DM halo mass with redshift for the best-fit model of the \textbf{default} case in \fig{fig:cum_allz_nion_BF}. The cumulative distribution shifts towards higher halo masses (and, analogously towards brighter UV galaxies) by more than an order of magnitude as reionization progresses from $z$ = 9 to $z$ = 6. This trend mirrors the rapid build-up and evolution of the halo mass function towards higher mass halos. As a result, the galaxies responsible for collectively providing 50\% of the instantaneous ionizing emissivity tend to shift towards more massive host halos as redshift decreases.

In the hierarchical structure formation model, it is well known that the statistical properties of galaxies are dictated by the statistical properties of their host dark matter halos. As a result, we next move to examine the clustering properties of galaxies predicted by our model based on the spatial clustering of DM halos and our constructed galaxy-halo connection. To do so, we focus on the galaxy bias $b_{gal}$, which quantifies the enhanced clustering of galaxies with respect to that of the underlying matter distribution, and check how compatible our predictions are with the bias measurements that have recently become available at $z \geq 6$ \cite{DalmassoHST, DalmassoJWST}.

We calculate the effective number-weighted linear bias of galaxies $b^{\rm eff}_g(z)$ at a given redshift $z$ as
\begin{equation}
\label{eqn:effective_bias_definition}
b^{\rm eff}_{\rm gal}(z) = \dfrac{\int _{M_{\rm UV,min}}^{M_{\rm UV,max}} {\rm dM_{UV}}~ \{Q_{\rm II}(z)~{ b^{\rm fb}_{\rm gal}({\rm M_{UV}},z)~\Phi^{\rm fb}_{\rm UV}} + [1-Q_{\rm II}(z)]~{\rm \Phi^{\rm nofb}_{\rm UV}}~b^{\rm nofb}_{\rm gal}({\rm M_{UV}},z) \} }{\int _{M_{\rm UV,min}}^{M_{\rm UV,max}} {\rm dM_{UV}}~ \{Q_{\rm II}(z)~{ \Phi^{\rm fb}_{\rm UV}} + [1-Q_{\rm II}(z)]~{\rm \Phi^{\rm nofb}_{\rm UV}} \}}
\end{equation}
where $b^{\rm fb}_{\rm gal}({\rm M_{UV}},z)$ and $b^{\rm nofb}_{\rm gal}({\rm M_{UV}},z)$ represents the linear bias of galaxies, with absolute magnitude $\rm M_{UV}$ at redshift $z$, residing in ionized and neutral regions respectively.

In our models, as the mass of a host dark matter halo $M_h$ entirely determines the UV magnitude of its resident galaxy based on the $M_h - M_{UV}$ relation, we can write the galaxy bias $b^{\rm fb}_{\rm gal}$ ($b^{\rm nofb}_{\rm gal}$) in ionized (neutral) regions in terms of the linear halo bias $b_h$, as follows
\begin{align}
\label{eqn:bgal_definition}
b^{\rm fb}_{\rm gal}({\rm M_{UV}},z) &= b_h( M_h ({\rm M_{UV}})|_{\rm fb},z)
\nonumber \\
b^{\rm nofb}_{\rm gal}({\rm M_{UV}},z) &= b_h( M_h ({\rm M_{UV}})|_{\rm nofb},z)
\end{align}

For the bias calculations, we take $M_{\rm UV,min}$ = $-21$  and compute the linear halo bias from the formula provided by Sheth \& Tormen \cite{ST99} based on the peak-background split theory. We use the python module {\tt{halomod}}\footnote{https://github.com/halomod/halomod} accompanying the {\tt{hmf}} codebase for these calculations.

\begin{figure}[htbp]
\centering
\includegraphics[width=0.47\textwidth]{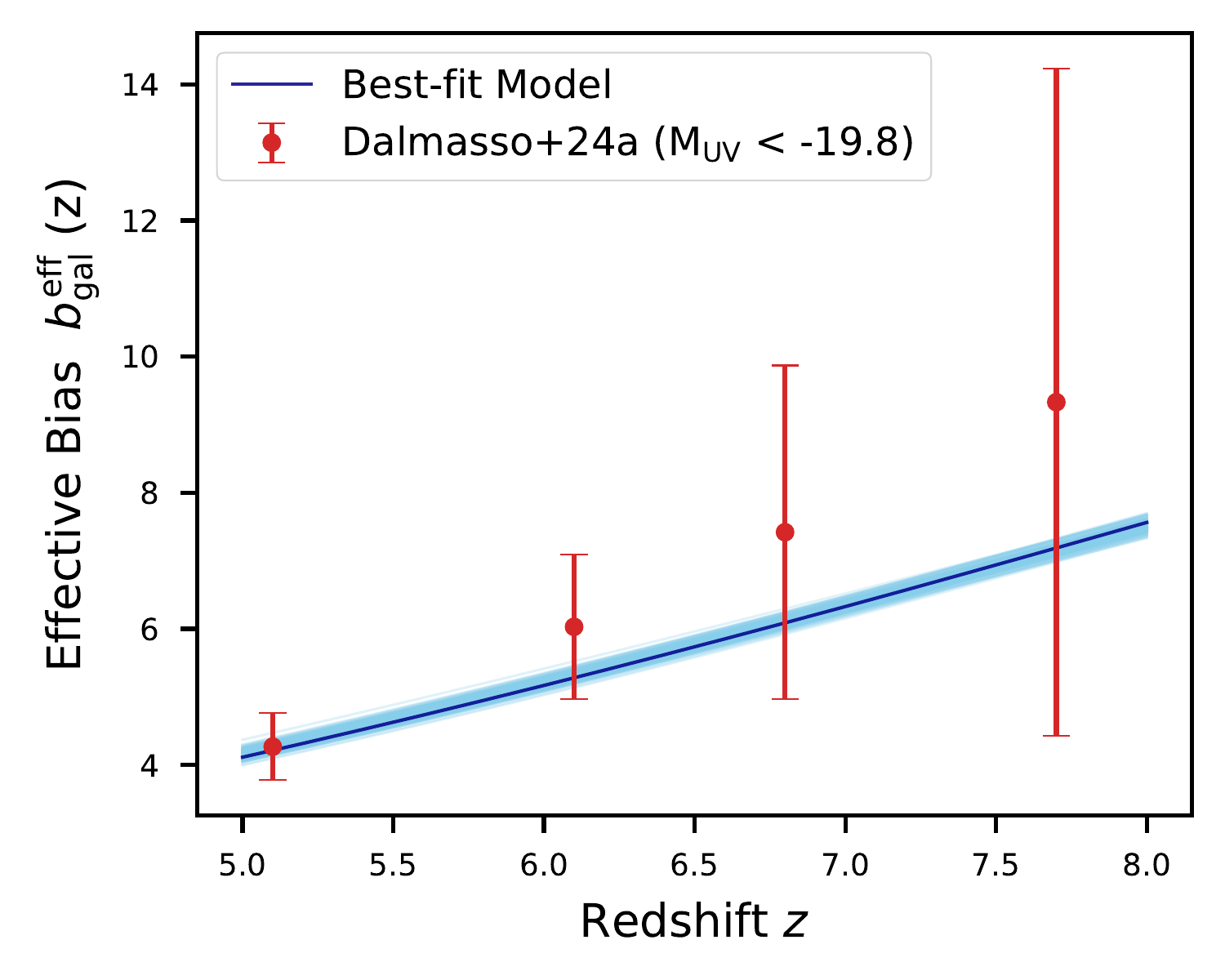}
\qquad
\includegraphics[width=0.47\textwidth]{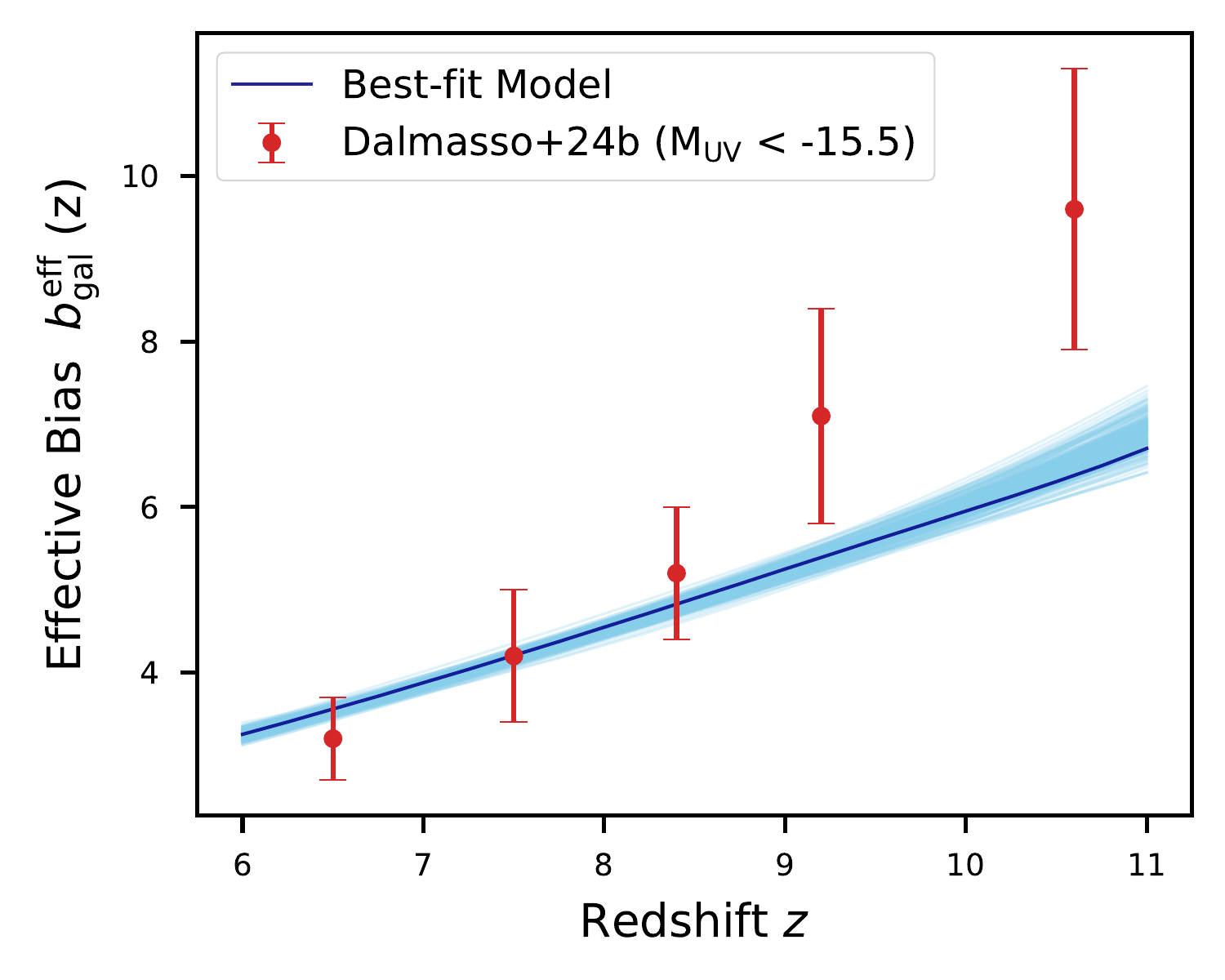}
\caption{The evolution of the effective galaxy bias (\eqn{eqn:effective_bias_definition}) as a function of redshift for two magnitude cuts - $M_{UV} < -19.8$ (left) and $M_{UV} < -15.5$ (right), for 200 random samples drawn from the MCMC chains for the \textbf{default} case. The colored data points denote the recent measurements using Hubble Legacy Fields data \cite{DalmassoHST} and JWST/NIRCam observations \cite{DalmassoJWST}. } 
\label{fig:effective_gal_bias}
\end{figure}

In \fig{fig:effective_gal_bias}, we show the effective galaxy bias predicted by our models as a function of redshift for two different magnitude cuts (i.e. $M_{\rm UV,max}$) along with recent bias measurements from Hubble Legacy Fields data \cite{DalmassoHST} and JWST/NIRCam observations \cite{DalmassoJWST}. We find that our model predicts an increasing trend of the galaxy bias with redshift and agrees well with the observations over $5 \leq z \leq 8$ for both cases, after accounting for the observational errors. However, at $z > 9$, we predict much weaker clustering than that reported by these studies. As the halo bias is known to strongly increase with halo mass at a fixed redshift, this means that our model predicts slightly less massive halo masses for UV-detectable galaxies than that suggested by the clustering observations at these redshifts. In fact, from the redshift evolution of the observed galaxy bias, the authors of these studies \cite{DalmassoJWST} also inferred that characteristic dark matter halo mass $M_{h,c}$ hosting the observed galaxies slightly increases from a value of $10^{10} M_\odot$ over $6 \leq z \leq 8.5$ to $10^{10.5} M_\odot$ at $z>9$ (see Fig. 2 of \cite{DalmassoJWST}). This underscores the importance of supplementing UVLF measurements with clustering measurements in robustly inferring the galaxy-halo connection at high redshifts.

Over the past few years, several observational studies at high redshifts have reported an increasing trend of the galaxy bias with luminosity at a fixed redshift \cite{Barone-Nugent2014, Harikane2016, Qiu2018, harikane2022}. To investigate whether our model can reproduce this trend, we calculate the average bias of a galaxy with UV magnitude $M_{UV}$ 
at a redshift $z$ as follows
\begin{equation}
\label{eq:average_bias_definition}
\langle b_{gal} \rangle = Q_{\rm II}(z)~ b^{\rm fb}_{\rm gal}({\rm M_{UV}},z) + [1-Q_{\rm II}(z)]~b^{\rm nofb}_{\rm gal}({\rm M_{UV}},z)     
\end{equation}
where symbols have their usual meaning.

\begin{figure}[htbp]
\centering
\includegraphics[scale=0.4]{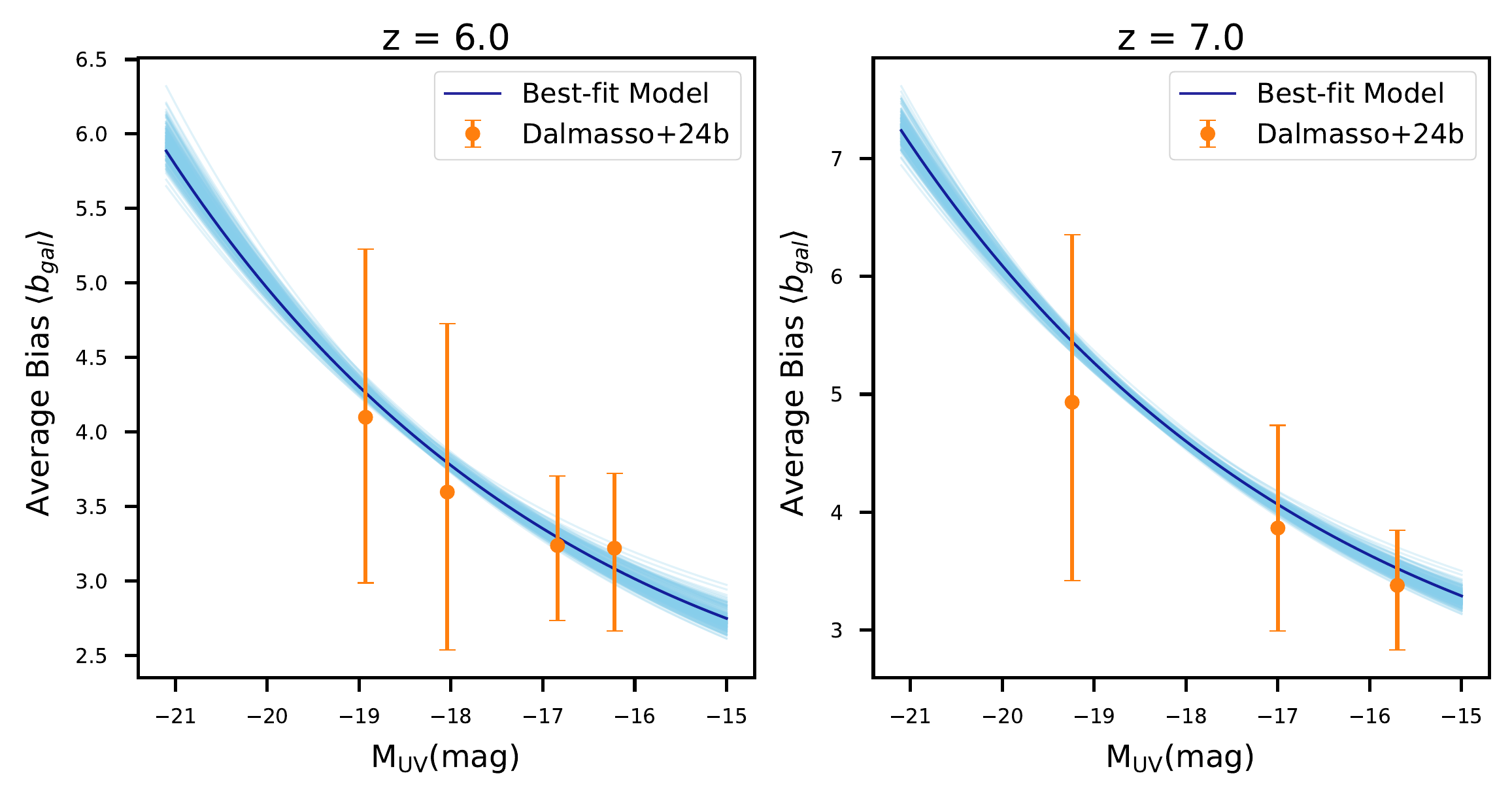}
\caption{The evolution of the average galaxy bias (\eqn{eq:average_bias_definition}) as a function of the absolute UV  magnitude at two redshift bins for 200 random samples drawn from the MCMC chains for the \textbf{default} case. The colored data points denote the recent measurements \cite{DalmassoJWST} using JWST/NIRCam observations of the GOODS South field.}
\label{fig:average_galaxy_bias}
\end{figure}

From \fig{fig:average_galaxy_bias}, we can see that our model predicts the galaxy bias to be increasing with UV luminosity at $z = 6$ and $z = 7$, which is consistent with the recent observations \cite{DalmassoJWST}. This directly follows from the fact that in our models, brighter galaxies reside in more massive dark matter halos which exhibit stronger clustering than less massive ones.

\subsection{Importance of the inclusion of the recent UVLF measurements from JWST}
\label{subsec:onlyHST_case}

\begin{figure}[htbp]
\centering
\includegraphics[width=\columnwidth]{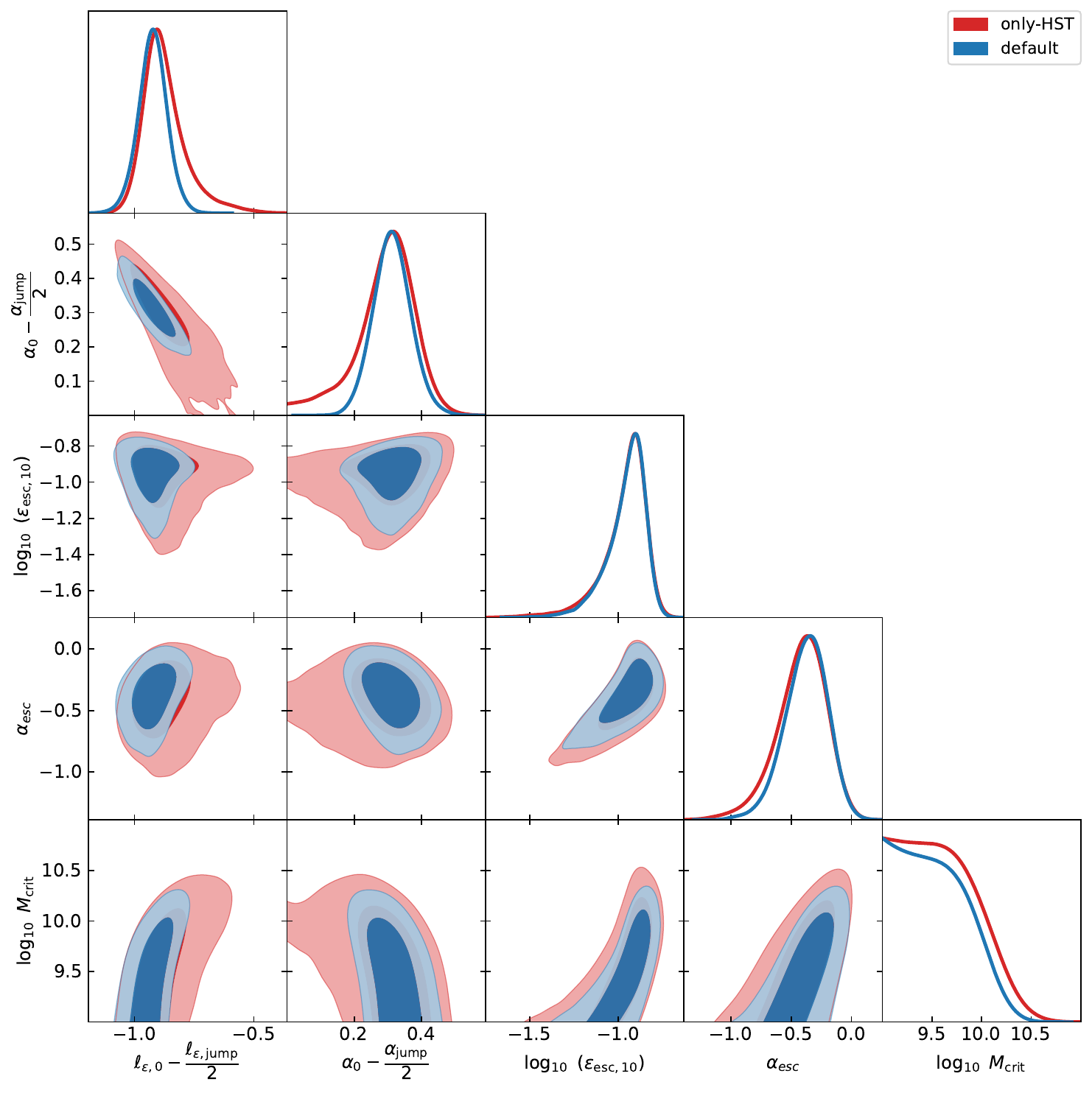}
\caption{Comparison of the posterior distributions of a subset of the free parameters for the \textbf{default} case (\textit{in blue}) and the \textbf{only-HST} case (\textit{in red}). The contours represent the 68\% and 95\% confidence intervals.}
\label{fig:compare_defaultHST}
\end{figure}

In the \textbf{default} case, we used UVLF measurements obtained from different galaxy surveys conducted with both the Hubble Space Telescope ($z \leq 9$) and the James Webb Space Telescope ($z \gtrsim 9$). To understand the significance of the inclusion of the JWST datasets, we ran a variant of the MCMC analysis, that we call the \textbf{only-HST} case,  by retaining only the observational data from HST studies - which are shown by black circles in \fig{fig:allQHI_UVLF}. The corresponding parameter constraints for this case are mentioned in the fourth column of \tab{tab:mcmc_results}. 

As one would expect, we find that the model parameters that contribute to the calculation of the evolving UVLF are the ones that are most affected by the exclusion of the JWST datasets. The individual free parameters like $\ell_{\varepsilon,0} + \ell_{\varepsilon, \mathrm{jump}}/2$ ,  $\alpha_0 + \alpha_\mathrm{jump}/2$ , $z_\varepsilon$, $\Delta z_\varepsilon$, $z_\alpha$ and  $\Delta z_\alpha$, which determine the UV efficiency of galaxies at higher $z$ and its overall evolution with redshift, are barely constrained. 

We find that the normalization of the UV efficiency $\log_{10}(\varepsilon_{\mathrm{*10,UV}})$  does not require any evolution with redshift to explain the HST-determined UVLFs and the various reionization observables since our marginalized constraints imply $\ell_{\varepsilon, \mathrm{jump}} \approx 0$. The same can also be said for the power-law index $\alpha_*$ as the $1\sigma$ constraint on $\alpha_{\mathrm{jump}}$ is consistent with zero.

In \fig{fig:compare_defaultHST}, we show a comparison of the posterior distributions between the \textbf{default} case \textit{(in blue)} and \textbf{only-HST} case \textit{(in red)} for a subset of the free parameters. We notice that although the constraints of $\ell_{\varepsilon,0} - \ell_{\varepsilon, \mathrm{jump}}/2$, $\alpha_0 - \alpha_\mathrm{jump}/2$, $\log_{10}(\varepsilon_{\mathrm{esc,10}})$ and $\alpha_{esc}$ agree with that obtained for the \textbf{default} case, they are somewhat weakened - possibly because the behavior of these evolving parameters at the high-redshift end becomes entirely unconstrained when the JWST UVLF measurements are excluded. We also find that the $1\sigma$ upper limit on $\log_{10} (M_\mathrm{crit}/M_\odot)$ has slightly relaxed from $< 9.70$ to $< 9.75$.

\subsection{Utilizing model-independent $Q_{\mathrm{HI}}$ constraints for inferences}
\label{subsec:onlyJ23_case}

\begin{figure}[htbp]
\centering
\includegraphics[scale=0.55]{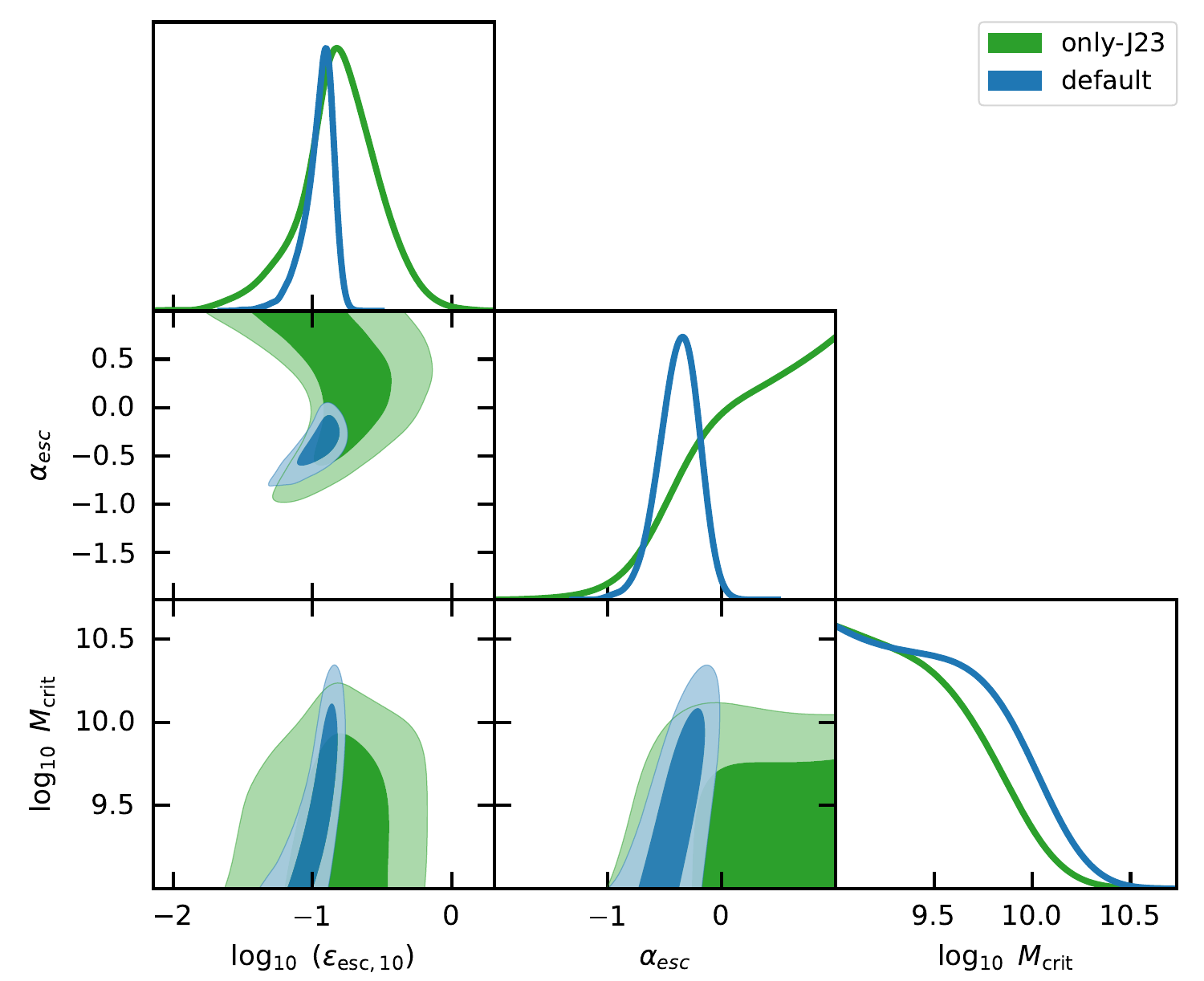}
\caption{Comparison of the posterior distributions of a subset of the free parameters for the \textbf{default} case (\textit{in blue}) and the \textbf{only-J23} case (\textit{in green}). The contours represent the 68\% and 95\% confidence intervals.}
\label{fig:compare_defaultJ23}
\end{figure}

In the \textbf{default} case, we used measurements of the neutral fraction $Q_\mathrm{HI}$ in the IGM obtained from absorption studies of Lyman-$\alpha$ emission originating from high-redshift sources. These estimates are usually model-dependent because they depend on several assumptions related to the intrinsic Ly$\alpha$ emission of the high-$z$ quasars and galaxies or the transmission of Ly$\alpha$ radiation in the intergalactic and circumgalactic medium. On the contrary, the analysis of dark pixel fraction \cite{McGreer2011, McGreer2015, Jin2023} provides us with nearly model-independent upper limits on  $Q_{\mathrm{HI}}$. Therefore, we ran another variant of the MCMC analysis, which we call the \textbf{only-J23} case, by replacing the earlier $Q_\mathrm{HI}$ measurements with model-independent but somewhat less stringent $Q_\mathrm{HI}$ upper limits \cite{Jin2023}.
In such cases, where we need to work with upper limits, the likelihood \cite{Ghara2020} for the  $Q_{\mathrm{HI}}$ data is given by
\begin{eqnarray*}
\mathcal {L}(\mathcal {D} \vert \boldsymbol \theta) = \prod _i \frac{1}{2}\left[1+\mathrm{erf}\left(\pm \frac{\mathcal {M}_i(\mathcal {\boldsymbol \theta})-\mathcal {D}_i}{\sqrt{2}~\sigma _i}\right)\right],
\end{eqnarray*}
As per the above definition, the likelihood of a model approaches unity when its prediction is less than $\mathcal {D}_i - \sigma _i$  for \textit{all} data points, and the likelihood approaches zero when the model prediction is greater than $\mathcal {D}_i + \sigma _i$ for \textit{any} data point $i$.

\begin{figure}[htbp]
\centering
\includegraphics[width=\columnwidth]{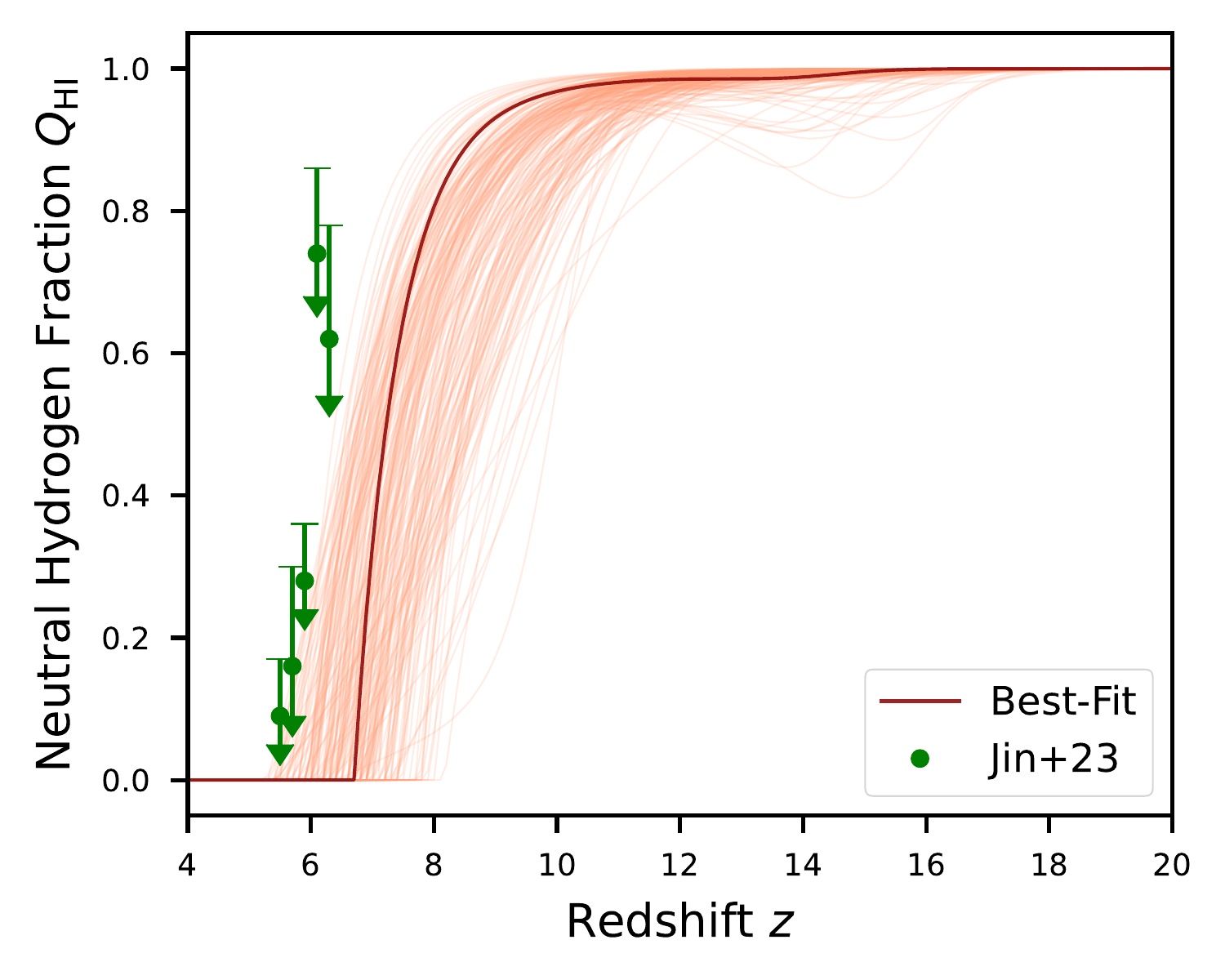}
\caption{The reionization histories for 200 random samples drawn from the MCMC chains for the \textbf{only-J23} case. The data points denote the observational constraints for $Q_{\rm HI}$($z$) used.}
\label{fig:reionJ23}
\end{figure}

For this case, the corresponding constraints on the free and derived parameters of the model are mentioned in the fifth column of \tab{tab:mcmc_results}. We compare the posterior distributions between the \textbf{default} case \textit{(in blue)} and \textbf{only-J23} case \textit{(in green)} for a subset of the free parameters in \fig{fig:compare_defaultJ23}. We find that the constraints on the astrophysical parameters controlling the star-formation rate and the UV emission show very marginal change compared to the \text{default} case, implying that the SFR-to-halo mass relation is already well constrained by the UVLF measurements. However, the power-law scaling index $\alpha_{esc}$ for the \textbf{only-J23} case now has a stronger preference for positive values, leading to a higher escape of LyC photons from galaxies residing inside heavier halos. The normalization value $\log_{10}(\varepsilon_{\mathrm{esc,10}})$ of the ionizing escape fraction also turns out to be higher in comparison to the \textbf{default} case. Consequently, in this scenario, the onus for providing ionizing photons is primarily delegated to the relatively heavier mass halos due to their higher escape fraction and efficiency in forming stars. Although massive halos are rarer at higher redshifts, they assemble rapidly at $z < 9$ and for this case, are also somewhat more immune to photosuppression in their star formation arising from radiative feedback (see the $M_{\rm{crit}}$ panel of \fig{fig:compare_defaultJ23}). Therefore, cosmic reionization progresses very rapidly in such models and completes comparatively early ($z \geq$ 6), as can also be seen from \fig{fig:reionJ23} where we show the reionization history for 200 random samples drawn from the MCMC chains of the \textbf{only-J23} case.

\section{Summary and Conclusions}
\label{sec:conclusion}

In this work, we have developed a semi-analytical model of galaxy formation and evolution that is self-consistently coupled to a physically motivated model for reionization. We use this model to constrain the astrophysical properties of high-redshift galaxies, which are believed to be the main drivers of reionization, by simultaneously comparing the model predictions with a variety of available observables, under an MCMC-based Bayesian framework. In particular, we use the measurement of optical depth $\tau_{el}$ from the Planck Collaboration, some of the recent estimates of the globally averaged neutral hydrogen fraction in the IGM from Lyman-$\alpha$ absorption studies, and the UV luminosity functions at $z \sim 6 - 15$ to obtain the constraints. Our model has eleven free parameters in total -  eight of which determine the redshift evolution of the star-formation efficiency (and consequently, the production efficiency of UV radiation) of high-$z$ galaxies, one to characterize the typical halo masses ($M_{\rm{crit}}$) below which effects of radiative feedback are dominant, while the remaining two describe the mass-dependent escape fraction $f_{\rm{esc}} (M_h) $ of ionizing UV photons emitted by galaxies.\\
\newline
The main findings of our analysis are :
\begin{itemize}
\item  Our model UVLFs are in agreement with observations at all redshifts where data are currently available. We notice that an enhancement in the star-formation rate efficiency and/or UV luminosity per stellar mass formed is required to reconcile with the recent JWST UVLF estimates at  $z \geq 10 $, as also found by several other studies.

\item We find that the UV emission from galaxies residing inside DM halos with masses $M_{\rm crit} < 10^{9.7}~M_\odot$ (68\% confidence limit) is strongly affected by feedback mechanisms arising due to photo-ionization heating as cosmic reionization progresses.

\item Our models are also consistent with the currently available constraints on reionization from studies of the cosmic microwave background and quasar absorption spectra if a $\textit{power-law}$ mass dependence is adopted for the escape fraction $f_{esc}$ of ionizing UV photons. In our model, $f_{esc}$ decreases with increasing halo mass, having a value of $\sim 11\%$ for 10$^{10} M_\odot$ halos based on the assumed properties ($\log_{10} \big[\xi_{\rm ion,fid}/({\rm ergs}^{-1}\ {\rm Hz}) \big] \approx 25.23$) of the stellar population, and the bulk of the ionizing photons for cosmic reionization seem to be produced by the fainter population of galaxies.
\end{itemize}

We emphasize that our galaxy model, which is entirely built on the dark matter HMFs and a semi-analytical prescription to model galaxy properties within these halos, is somewhat simplistic but very flexible. As a result, certain aspects of our present modeling can be easily improved in the future. For example, the treatment of radiative feedback in our model is rather agnostic to physical processes. Ideally, the effects of radiative feedback should be self-consistently calculated from the evolution of the thermal properties of the IGM \cite{Choudhury2005}. Secondly, we have modelled the inhomogeneities in the IGM through a clumping factor $\mathcal{C}$ that controls the \textit{effective} recombination rate, neglecting the density distribution of the IGM. However, since the recombination rate is typically higher in high-density regions, one would expect such regions to remain neutral for a longer time while regions of lower densities to be ionized earlier. Incorporating the density structure of the IGM \cite{Miralda2000, Wyithe2003, Choudhury2005, Choudhury2007, Mitra2011, Chatterjee2021} in recombination calculations would enable tracking the progress of reionization in a clumpy Universe more reliably. Furthermore, we have considered only a single class of stellar population, whose nature is captured through two numbers ($\kappa_{UV}$ and $\eta_{\gamma \ast}$) in our calculations. In other words, we have simply accounted for the total contribution from star-forming galaxies, without trying to segregate it into contributions from Pop-II or Pop-III stars. Therefore, from the perspective of the global observables considered here, any changes in these numbers ($\kappa_{UV}$ and $\eta_{\gamma \ast}$), arising as a result of the change in the nature of the stellar population, are completely degenerate with changes in other astrophysical parameters (such as $f_\ast$ and $f_{esc}$ respectively). As observations with the JWST have already started to reveal tantalizing insights into the complex nature of star formation in the early Universe, a self-consistent treatment of the properties of the pristine Population III stars and how they transition into the metal-enriched Population II stars in semi-analytical models like ours will be extremely beneficial for unambiguously understanding their impact on observables like the UV luminosity functions at $z > 12$ \cite{Ventura2024}, the 21-cm signal from Cosmic Dawn \cite{Ventura2023} and the chemical enrichment of the IGM \cite{Yamaguchi2023}. Furthermore, some studies have also invoked possibilities like contributions from non-stellar sources like an accreting supermassive black hole (AGN) \cite{Pacucci2022, Harikane2023} and/or decrement of dust attenuation at $z > 10$ \cite{Ferrara2023, Ziparo2023, Fiore2023, Ferrara2023_Paper2} for explaining the recent JWST observations. It would be interesting to include the effects of dust obscuration and AGN activity in our model and revisit the present analysis. We postpone these extensions for future work.

Interestingly, the nature and emission properties of high-redshift galaxies and their redshift evolution also have important consequences for the strength and fluctuations of the cosmological 21-cm signal during reionization and cosmic dawn \cite{Park++2019, Mirocha2019, Maity2022, Mittal2022, Shimabukuro2023, Chatterjee2023, Hassan2023}. Over the next decade, the extensive datasets acquired from high-$z$ galaxy surveys will be ably complemented by information on the redshifted 21-cm cosmological signal from interferometric as well as global experiments such as the Low Frequency Array (LOFAR) \cite{Gehlot2019_LOFAR, Mertens2020_LOFAR}, Murchison Widefield Array (MWA) \cite{Barry2019_MWA, Trott2020_MWA},  Precision Array for Probing the Epoch of Reionization (PAPER) \cite{Parsons2010_PAPER, Abdurashidova2022_HERA}, Hydrogen Epoch of Reionization Array (HERA) \cite{Kolopanis2019_PAPER}, Square Kilometre Array (SKA) \cite{Koopmans2015_SKA}, Shaped Antenna measurement of the background RAdio Spectrum (SARAS) \cite{Singh2022_SARAS}, Experiment to Detect the Global EoR Signature (EDGES) \cite{Bowman2018_EDGES}, Radio Experiment for the Analysis of Cosmic Hydrogen (REACH) \cite{Acedo2022_REACH}. In this context, expanding our current framework to incorporate the calculation of the 21-cm brightness temperature will offer a promising avenue to reliably deduce the astrophysical properties of galaxies in the early Universe.

\section*{Acknowledgments}

The authors acknowledge support of the Department of Atomic Energy, Government of India, under project no. 12-R\&D-TFR-5.02-0700.

\section*{Data Availability}

The data generated and presented in this paper will be made available upon reasonable request to the corresponding author.

\appendix
\section{Motivating the parameterization for $\varepsilon_{\ast 10,UV}(z)$ and $\alpha_\ast(z)$ }
\label{appendix:tanhParameterisation}

\begin{figure}[htbp]
\centering
\includegraphics[width=\columnwidth]{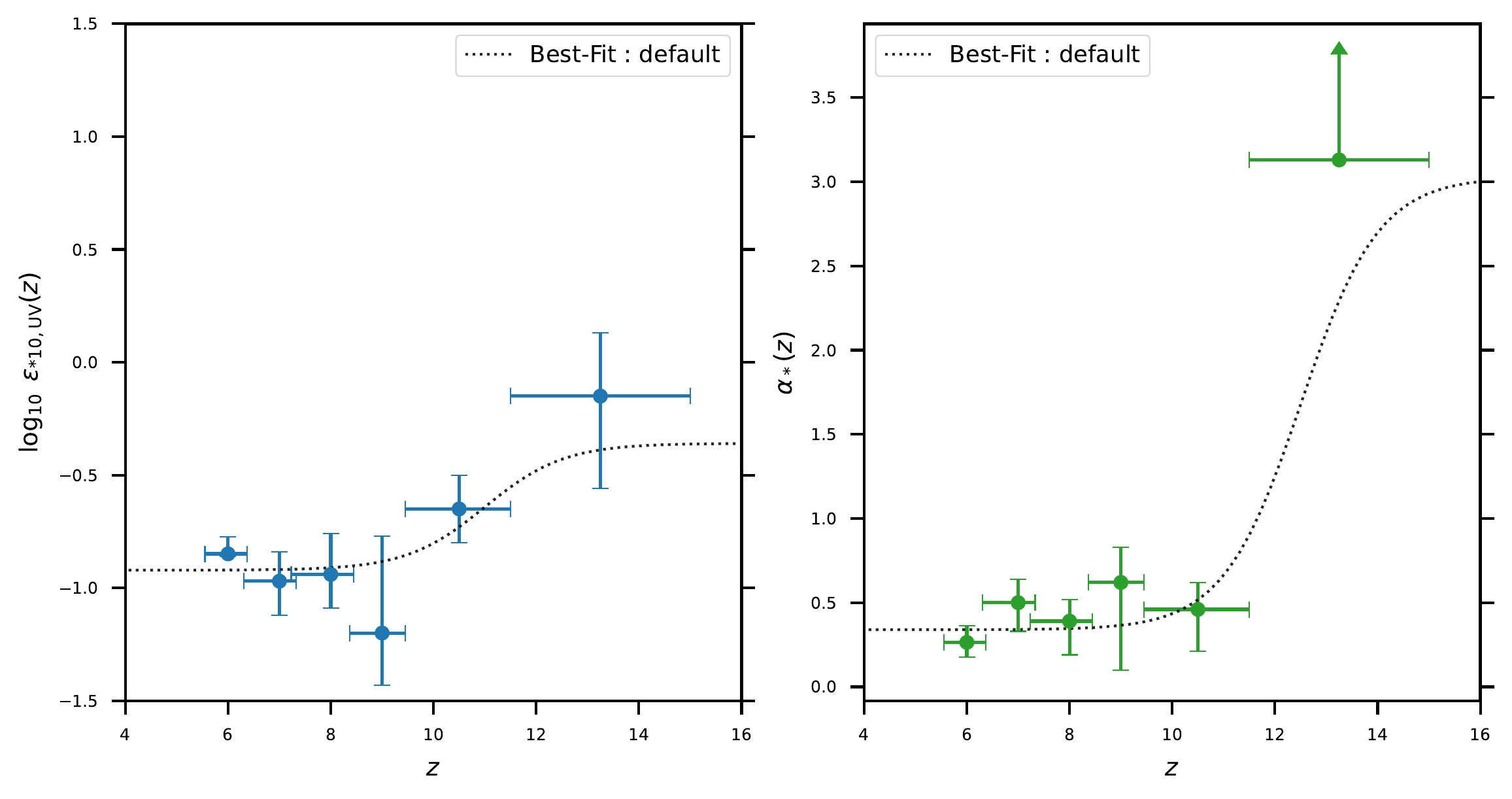}
\caption{The marginalized values of $\varepsilon_{\ast 10,UV}$ and $\alpha_\ast$ obtained by fitting the UVLF model to the corresponding measurements at each redshift bin \textit{individually} (see text for details). For illustration, we also show the $\textit{tanh}$ evolution corresponding to the best-fit model of the \textbf{default} case (\secn{subsec:results_default_case}) using a black dotted line.}
\label{fig:appendix_params}
\end{figure}

In this work, we used a \textit{tanh} parameterization to represent the smooth redshift-evolution of the parameters $\varepsilon_{\ast 10,UV}(z)$ and $\alpha_\ast(z)$. These two parameters affect the model-predicted UVLFs in a somewhat unambiguous manner while being completely degenerate with other parameters like $\varepsilon_{\mathrm esc}$ and $\alpha_{\mathrm esc}$ in determining the reionization history. 

As detailed in \secn{sec:galform_model}, the luminosity functions from our model at any given redshift depend on four parameters - $\varepsilon_{\ast 10, UV}$, $\alpha_\ast$, $M_{\rm crit}$ and $Q_{II}$.  In general, the shape of the UVLF is contingent on the $M_h$-$M_{\mathrm UV}$ relationship, which is decided by parameters such as $\varepsilon_{\ast 10, UV}$, $\alpha_\ast$ and $M_{\rm crit}$ and the shape of the dark matter halo mass function. At the very faint end, the UVLF is predominantly shaped by the impact of radiative feedback (determined by $M_{\mathrm crit}$) on low-mass galaxies that reside in ionized regions, which occupy a fraction $Q_{II}$ of the Universe by volume. However, for massive halos with masses $M_{\mathrm h} >> M_{\mathrm crit}$, the effect of UV feedback becomes negligible, and therefore, the UVLF towards the brighter end of the UV magnitude range becomes independent of $Q_{II}$ or $M_{\mathrm crit}$ and is entirely determined by the free parameters - $\varepsilon_{\ast 10,UV}$ and $\alpha_\ast$. With this motivation, we fit our model UVLFs to the observed measurements at \emph{each} redshift bin \textbf{\emph{individually}} by treating  $\log_{10}(\varepsilon_{\ast 10, UV})$, $\alpha_\ast$, $\log_{10}~M_{\rm crit}$ and $Q_{II}$ as free parameters. From this exercise, we find no sensible constraints on $\log_{10}~M_{\rm crit}$ and $Q_{II}$ at any of the redshift bins, while the values of $\log_{10}(\varepsilon_{\ast 10, UV})$ and $\alpha_\ast$ obtained at each $z$ from this exercise is shown in \fig{fig:appendix_params}. We find that both $\varepsilon_{\ast 10, UV}$ and $\alpha_{\ast}$ steeply increase for $z \geq 10$, while remaining roughly constant (within error bars) below $z \sim 10$. This motivated us to assume a \textit{tanh} parameterization for their evolution as it not only captures the behaviour shown in \fig{fig:appendix_params} but also offers reasonable flexibility in their evolution.


\bibliographystyle{JHEP}
\bibliography{manuscript_final_ver}

\providecommand{\href}[2]{#2}\begingroup\raggedright\begin{thebibliography}{100}

\bibitem{Sommerville2015}
R.S.~{Somerville} and R.~{Dav{\'e}}, \emph{{Physical Models of Galaxy Formation in a Cosmological Framework}}, \href{https://doi.org/10.1146/annurev-astro-082812-140951}{\emph{\araa} {\bfseries 53} (2015) 51} [\href{https://arxiv.org/abs/1412.2712}{{\ttfamily 1412.2712}}].

\bibitem{Dayal&Ferrara2018}
P.~{Dayal} and A.~{Ferrara}, \emph{{Early galaxy formation and its large-scale effects}}, \href{https://doi.org/10.1016/j.physrep.2018.10.002}{\emph{\physrep} {\bfseries 780} (2018) 1} [\href{https://arxiv.org/abs/1809.09136}{{\ttfamily 1809.09136}}].

\bibitem{Springel2005}
V.~{Springel}, \emph{{The cosmological simulation code GADGET-2}}, \href{https://doi.org/10.1111/j.1365-2966.2005.09655.x}{\emph{\mnras} {\bfseries 364} (2005) 1105} [\href{https://arxiv.org/abs/astro-ph/0505010}{{\ttfamily astro-ph/0505010}}].

\bibitem{BoylanKolchin2009}
M.~{Boylan-Kolchin}, V.~{Springel}, S.D.M.~{White}, A.~{Jenkins} and G.~{Lemson}, \emph{{Resolving cosmic structure formation with the Millennium-II Simulation}}, \href{https://doi.org/10.1111/j.1365-2966.2009.15191.x}{\emph{\mnras} {\bfseries 398} (2009) 1150} [\href{https://arxiv.org/abs/0903.3041}{{\ttfamily 0903.3041}}].

\bibitem{Klypin2011}
A.A.~{Klypin}, S.~{Trujillo-Gomez} and J.~{Primack}, \emph{{Dark Matter Halos in the Standard Cosmological Model: Results from the Bolshoi Simulation}}, \href{https://doi.org/10.1088/0004-637X/740/2/102}{\emph{\apj} {\bfseries 740} (2011) 102} [\href{https://arxiv.org/abs/1002.3660}{{\ttfamily 1002.3660}}].

\bibitem{Klypin2016}
A.~{Klypin}, G.~{Yepes}, S.~{Gottl{\"o}ber}, F.~{Prada} and S.~{He{\ss}}, \emph{{MultiDark simulations: the story of dark matter halo concentrations and density profiles}}, \href{https://doi.org/10.1093/mnras/stw248}{\emph{\mnras} {\bfseries 457} (2016) 4340} [\href{https://arxiv.org/abs/1411.4001}{{\ttfamily 1411.4001}}].

\bibitem{ST99}
R.K.~{Sheth} and G.~{Tormen}, \emph{{Large-scale bias and the peak background split}}, \href{https://doi.org/10.1046/j.1365-8711.1999.02692.x}{\emph{\mnras} {\bfseries 308} (1999) 119} [\href{https://arxiv.org/abs/astro-ph/9901122}{{\ttfamily astro-ph/9901122}}].

\bibitem{SMT2001}
R.K.~{Sheth}, H.J.~{Mo} and G.~{Tormen}, \emph{{Ellipsoidal collapse and an improved model for the number and spatial distribution of dark matter haloes}}, \href{https://doi.org/10.1046/j.1365-8711.2001.04006.x}{\emph{\mnras} {\bfseries 323} (2001) 1} [\href{https://arxiv.org/abs/astro-ph/9907024}{{\ttfamily astro-ph/9907024}}].

\bibitem{Vogelsberger2014}
M.~{Vogelsberger}, S.~{Genel}, V.~{Springel}, P.~{Torrey}, D.~{Sijacki}, D.~{Xu} et~al., \emph{{Introducing the Illustris Project: simulating the coevolution of dark and visible matter in the Universe}}, \href{https://doi.org/10.1093/mnras/stu1536}{\emph{\mnras} {\bfseries 444} (2014) 1518} [\href{https://arxiv.org/abs/1405.2921}{{\ttfamily 1405.2921}}].

\bibitem{Schaye2015}
J.~{Schaye}, R.A.~{Crain}, R.G.~{Bower}, M.~{Furlong}, M.~{Schaller}, T.~{Theuns} et~al., \emph{{The EAGLE project: simulating the evolution and assembly of galaxies and their environments}}, \href{https://doi.org/10.1093/mnras/stu2058}{\emph{\mnras} {\bfseries 446} (2015) 521} [\href{https://arxiv.org/abs/1407.7040}{{\ttfamily 1407.7040}}].

\bibitem{Dave2019}
R.~{Dav{\'e}}, D.~{Angl{\'e}s-Alc{\'a}zar}, D.~{Narayanan}, Q.~{Li}, M.H.~{Rafieferantsoa} and S.~{Appleby}, \emph{{SIMBA: Cosmological simulations with black hole growth and feedback}}, \href{https://doi.org/10.1093/mnras/stz937}{\emph{\mnras} {\bfseries 486} (2019) 2827} [\href{https://arxiv.org/abs/1901.10203}{{\ttfamily 1901.10203}}].

\bibitem{Choudhury2005}
T.R.~{Choudhury} and A.~{Ferrara}, \emph{{Experimental constraints on self-consistent reionization models}}, \href{https://doi.org/10.1111/j.1365-2966.2005.09196.x}{\emph{\mnras} {\bfseries 361} (2005) 577} [\href{https://arxiv.org/abs/astro-ph/0411027}{{\ttfamily astro-ph/0411027}}].

\bibitem{Samui2007}
S.~{Samui}, R.~{Srianand} and K.~{Subramanian}, \emph{{Probing the star formation history using the redshift evolution of luminosity functions}}, \href{https://doi.org/10.1111/j.1365-2966.2007.11603.x}{\emph{\mnras} {\bfseries 377} (2007) 285} [\href{https://arxiv.org/abs/astro-ph/0612271}{{\ttfamily astro-ph/0612271}}].

\bibitem{Mitra2011}
S.~{Mitra}, T.R.~{Choudhury} and A.~{Ferrara}, \emph{{Reionization constraints using principal component analysis}}, \href{https://doi.org/10.1111/j.1365-2966.2011.18234.x}{\emph{\mnras} {\bfseries 413} (2011) 1569} [\href{https://arxiv.org/abs/1011.2213}{{\ttfamily 1011.2213}}].

\bibitem{Dayal14}
P.~{Dayal}, A.~{Ferrara}, J.S.~{Dunlop} and F.~{Pacucci}, \emph{{Essential physics of early galaxy formation}}, \href{https://doi.org/10.1093/mnras/stu1848}{\emph{\mnras} {\bfseries 445} (2014) 2545} [\href{https://arxiv.org/abs/1405.4862}{{\ttfamily 1405.4862}}].

\bibitem{Mason2015}
C.A.~{Mason}, M.~{Trenti} and T.~{Treu}, \emph{{The Galaxy UV Luminosity Function before the Epoch of Reionization}}, \href{https://doi.org/10.1088/0004-637X/813/1/21}{\emph{\apj} {\bfseries 813} (2015) 21} [\href{https://arxiv.org/abs/1508.01204}{{\ttfamily 1508.01204}}].

\bibitem{Lacey2016}
C.G.~{Lacey}, C.M.~{Baugh}, C.S.~{Frenk}, A.J.~{Benson}, R.G.~{Bower}, S.~{Cole} et~al., \emph{{A unified multiwavelength model of galaxy formation}}, \href{https://doi.org/10.1093/mnras/stw1888}{\emph{\mnras} {\bfseries 462} (2016) 3854} [\href{https://arxiv.org/abs/1509.08473}{{\ttfamily 1509.08473}}].

\bibitem{Mutch2016}
S.J.~{Mutch}, P.M.~{Geil}, G.B.~{Poole}, P.W.~{Angel}, A.R.~{Duffy}, A.~{Mesinger} et~al., \emph{{Dark-ages reionization and galaxy formation simulation - III. Modelling galaxy formation and the epoch of reionization}}, \href{https://doi.org/10.1093/mnras/stw1506}{\emph{\mnras} {\bfseries 462} (2016) 250} [\href{https://arxiv.org/abs/1512.00562}{{\ttfamily 1512.00562}}].

\bibitem{Furlanetto2017}
S.R.~{Furlanetto}, J.~{Mirocha}, R.H.~{Mebane} and G.~{Sun}, \emph{{A minimalist feedback-regulated model for galaxy formation during the epoch of reionization}}, \href{https://doi.org/10.1093/mnras/stx2132}{\emph{\mnras} {\bfseries 472} (2017) 1576} [\href{https://arxiv.org/abs/1611.01169}{{\ttfamily 1611.01169}}].

\bibitem{Mirocha2017}
J.~{Mirocha}, S.R.~{Furlanetto} and G.~{Sun}, \emph{{The global 21-cm signal in the context of the high- z galaxy luminosity function}}, \href{https://doi.org/10.1093/mnras/stw2412}{\emph{\mnras} {\bfseries 464} (2017) 1365} [\href{https://arxiv.org/abs/1607.00386}{{\ttfamily 1607.00386}}].

\bibitem{Behroozi19}
P.~{Behroozi}, C.~{Conroy}, R.H.~{Wechsler}, A.~{Hearin}, C.C.~{Williams}, B.P.~{Moster} et~al., \emph{{The Universe at z > 10: predictions for JWST from the UNIVERSEMACHINE DR1}}, \href{https://doi.org/10.1093/mnras/staa3164}{\emph{\mnras} {\bfseries 499} (2020) 5702} [\href{https://arxiv.org/abs/2007.04988}{{\ttfamily 2007.04988}}].

\bibitem{Dayal2017}
P.~{Dayal}, T.R.~{Choudhury}, V.~{Bromm} and F.~{Pacucci}, \emph{{Reionization and Galaxy Formation in Warm Dark Matter Cosmologies}}, \href{https://doi.org/10.3847/1538-4357/836/1/16}{\emph{\apj} {\bfseries 836} (2017) 16} [\href{https://arxiv.org/abs/1501.02823}{{\ttfamily 1501.02823}}].

\bibitem{Park++2019}
J.~{Park}, A.~{Mesinger}, B.~{Greig} and N.~{Gillet}, \emph{{Inferring the astrophysics of reionization and cosmic dawn from galaxy luminosity functions and the 21-cm signal}}, \href{https://doi.org/10.1093/mnras/stz032}{\emph{\mnras} {\bfseries 484} (2019) 933} [\href{https://arxiv.org/abs/1809.08995}{{\ttfamily 1809.08995}}].

\bibitem{Maity2022}
B.~{Maity} and T.R.~{Choudhury}, \emph{{Probing the thermal history during reionization using a seminumerical photon-conserving code SCRIPT}}, \href{https://doi.org/10.1093/mnras/stac182}{\emph{\mnras} {\bfseries 511} (2022) 2239} [\href{https://arxiv.org/abs/2110.14231}{{\ttfamily 2110.14231}}].

\bibitem{Munoz2022_21cmFAST}
J.B.~{Mu{\~n}oz}, Y.~{Qin}, A.~{Mesinger}, S.G.~{Murray}, B.~{Greig} and C.~{Mason}, \emph{{The impact of the first galaxies on cosmic dawn and reionization}}, \href{https://doi.org/10.1093/mnras/stac185}{\emph{\mnras} {\bfseries 511} (2022) 3657} [\href{https://arxiv.org/abs/2110.13919}{{\ttfamily 2110.13919}}].

\bibitem{Barkana&Loeb2001}
R.~{Barkana} and A.~{Loeb}, \emph{{In the beginning: the first sources of light and the reionization of the universe}}, \href{https://doi.org/10.1016/S0370-1573(01)00019-9}{\emph{\physrep} {\bfseries 349} (2001) 125} [\href{https://arxiv.org/abs/astro-ph/0010468}{{\ttfamily astro-ph/0010468}}].

\bibitem{Wise2019}
J.H.~Wise, \emph{Cosmic reionisation}, \href{https://doi.org/10.1080/00107514.2019.1631548}{\emph{Contemporary Physics} {\bfseries 60} (2019) 145} [\href{https://arxiv.org/abs/https://doi.org/10.1080/00107514.2019.1631548}{{\ttfamily https://doi.org/10.1080/00107514.2019.1631548}}].

\bibitem{TRC2022Review}
T.R.~{Choudhury}, \emph{{A short introduction to reionization physics}}, \href{https://doi.org/10.1007/s10714-022-02987-4}{\emph{General Relativity and Gravitation} {\bfseries 54} (2022) 102} [\href{https://arxiv.org/abs/2209.08558}{{\ttfamily 2209.08558}}].

\bibitem{JWST_highzgalaxies}
C.L.~{Steinhardt}, C.K.~{Jespersen} and N.B.~{Linzer}, \emph{{Finding High-redshift Galaxies with JWST}}, \href{https://doi.org/10.3847/1538-4357/ac2a2f}{\emph{\apj} {\bfseries 923} (2021) 8} [\href{https://arxiv.org/abs/2111.14865}{{\ttfamily 2111.14865}}].

\bibitem{JWST_DeepSurveys}
M.~{Rieke}, S.~{Arribas}, A.~{Bunker}, S.~{Charlot}, S.~{Finkelstein}, R.~{Maiolino} et~al., \emph{{JWST GTO/ERS Deep Surveys}}, {\emph{\baas} {\bfseries 51} (2019) 45}.

\bibitem{TMTScienceCase_2015}
W.~{Skidmore}, {TMT International Science Development Teams} and T.~{Science Advisory Committee}, \emph{{Thirty Meter Telescope Detailed Science Case: 2015}}, \href{https://doi.org/10.1088/1674-4527/15/12/001}{\emph{Research in Astronomy and Astrophysics} {\bfseries 15} (2015) 1945} [\href{https://arxiv.org/abs/1505.01195}{{\ttfamily 1505.01195}}].

\bibitem{GMT_TMT_WhitePaper2019}
S.~{Finkelstein}, M.~{Bradac}, C.~{Casey}, M.~{Dickinson}, R.~{Endsley}, S.~{Furlanetto} et~al., \emph{{Unveiling the Phase Transition of the Universe During the Reionization Epoch with Lyman-alpha}}, \href{https://doi.org/10.48550/arXiv.1903.04518}{\emph{\baas} {\bfseries 51} (2019) 221} [\href{https://arxiv.org/abs/1903.04518}{{\ttfamily 1903.04518}}].

\bibitem{Bouwens2015}
R.J.~{Bouwens}, G.D.~{Illingworth}, P.A.~{Oesch}, M.~{Trenti}, I.~{Labb{\'e}}, L.~{Bradley} et~al., \emph{{UV Luminosity Functions at Redshifts z {\ensuremath{\sim}} 4 to z {\ensuremath{\sim}} 10: 10,000 Galaxies from HST Legacy Fields}}, \href{https://doi.org/10.1088/0004-637X/803/1/34}{\emph{\apj} {\bfseries 803} (2015) 34} [\href{https://arxiv.org/abs/1403.4295}{{\ttfamily 1403.4295}}].

\bibitem{Bouwens2017}
R.J.~{Bouwens}, P.A.~{Oesch}, G.D.~{Illingworth}, R.S.~{Ellis} and M.~{Stefanon}, \emph{{The z {\ensuremath{\sim}} 6 Luminosity Function Fainter than -15 mag from the Hubble Frontier Fields: The Impact of Magnification Uncertainties}}, \href{https://doi.org/10.3847/1538-4357/aa70a4}{\emph{\apj} {\bfseries 843} (2017) 129} [\href{https://arxiv.org/abs/1610.00283}{{\ttfamily 1610.00283}}].

\bibitem{Atek2018}
H.~{Atek}, J.~{Richard}, J.-P.~{Kneib} and D.~{Schaerer}, \emph{{The extreme faint end of the UV luminosity function at z {\ensuremath{\sim}} 6 through gravitational telescopes: a comprehensive assessment of strong lensing uncertainties}}, \href{https://doi.org/10.1093/mnras/sty1820}{\emph{\mnras} {\bfseries 479} (2018) 5184} [\href{https://arxiv.org/abs/1803.09747}{{\ttfamily 1803.09747}}].

\bibitem{Ono2018}
Y.~{Ono}, M.~{Ouchi}, Y.~{Harikane}, J.~{Toshikawa}, M.~{Rauch}, S.~{Yuma} et~al., \emph{{Great Optically Luminous Dropout Research Using Subaru HSC (GOLDRUSH). I. UV luminosity functions at z {\ensuremath{\sim}} 4-7 derived with the half-million dropouts on the 100 deg$^{2}$ sky}}, \href{https://doi.org/10.1093/pasj/psx103}{\emph{\pasj} {\bfseries 70} (2018) S10} [\href{https://arxiv.org/abs/1704.06004}{{\ttfamily 1704.06004}}].

\bibitem{Bowler2020}
R.A.A.~{Bowler}, M.J.~{Jarvis}, J.S.~{Dunlop}, R.J.~{McLure}, D.J.~{McLeod}, N.J.~{Adams} et~al., \emph{{A lack of evolution in the very bright end of the galaxy luminosity function from z = 8 to 10}}, \href{https://doi.org/10.1093/mnras/staa313}{\emph{\mnras} {\bfseries 493} (2020) 2059} [\href{https://arxiv.org/abs/1911.12832}{{\ttfamily 1911.12832}}].

\bibitem{harikane2022}
Y.~{Harikane}, Y.~{Ono}, M.~{Ouchi}, C.~{Liu}, M.~{Sawicki}, T.~{Shibuya} et~al., \emph{{GOLDRUSH. IV. Luminosity Functions and Clustering Revealed with 4,000,000 Galaxies at z 2-7: Galaxy-AGN Transition, Star Formation Efficiency, and Implication for Evolution at z > 10}}, \href{https://doi.org/10.3847/1538-4365/ac3dfc}{\emph{\apjs} {\bfseries 259} (2022) 20} [\href{https://arxiv.org/abs/2108.01090}{{\ttfamily 2108.01090}}].

\bibitem{Naidu2022}
R.P.~{Naidu}, P.A.~{Oesch}, P.~{van Dokkum}, E.J.~{Nelson}, K.A.~{Suess}, G.~{Brammer} et~al., \emph{{Two Remarkably Luminous Galaxy Candidates at z {\ensuremath{\approx}} 10-12 Revealed by JWST}}, \href{https://doi.org/10.3847/2041-8213/ac9b22}{\emph{\apjl} {\bfseries 940} (2022) L14} [\href{https://arxiv.org/abs/2207.09434}{{\ttfamily 2207.09434}}].

\bibitem{Castellano2022}
M.~{Castellano}, A.~{Fontana}, T.~{Treu}, P.~{Santini}, E.~{Merlin}, N.~{Leethochawalit} et~al., \emph{{Early Results from GLASS-JWST. III. Galaxy Candidates at z 9-15}}, \href{https://doi.org/10.3847/2041-8213/ac94d0}{\emph{\apjl} {\bfseries 938} (2022) L15} [\href{https://arxiv.org/abs/2207.09436}{{\ttfamily 2207.09436}}].

\bibitem{Finkelstein2022}
S.L.~{Finkelstein}, M.B.~{Bagley}, P.~{Arrabal Haro}, M.~{Dickinson}, H.C.~{Ferguson}, J.S.~{Kartaltepe} et~al., \emph{{A Long Time Ago in a Galaxy Far, Far Away: A Candidate z {\ensuremath{\sim}} 12 Galaxy in Early JWST CEERS Imaging}}, \href{https://doi.org/10.3847/2041-8213/ac966e}{\emph{\apjl} {\bfseries 940} (2022) L55} [\href{https://arxiv.org/abs/2207.12474}{{\ttfamily 2207.12474}}].

\bibitem{Atek2023}
H.~{Atek}, M.~{Shuntov}, L.J.~{Furtak}, J.~{Richard}, J.-P.~{Kneib}, G.~{Mahler} et~al., \emph{{Revealing galaxy candidates out to z 16 with JWST observations of the lensing cluster SMACS0723}}, \href{https://doi.org/10.1093/mnras/stac3144}{\emph{\mnras} {\bfseries 519} (2023) 1201} [\href{https://arxiv.org/abs/2207.12338}{{\ttfamily 2207.12338}}].

\bibitem{Adams2023}
N.J.~{Adams}, C.J.~{Conselice}, L.~{Ferreira}, D.~{Austin}, J.A.A.~{Trussler}, I.~{Juod{\v{z}}balis} et~al., \emph{{Discovery and properties of ultra-high redshift galaxies (9 < z < 12) in the JWST ERO SMACS 0723 Field}}, \href{https://doi.org/10.1093/mnras/stac3347}{\emph{\mnras} {\bfseries 518} (2023) 4755} [\href{https://arxiv.org/abs/2207.11217}{{\ttfamily 2207.11217}}].

\bibitem{Bradley2023}
L.D.~{Bradley}, D.~{Coe}, G.~{Brammer}, L.J.~{Furtak}, R.L.~{Larson}, V.~{Kokorev} et~al., \emph{{High-redshift Galaxy Candidates at z = 9-10 as Revealed by JWST Observations of WHL0137-08}}, \href{https://doi.org/10.3847/1538-4357/acecfe}{\emph{\apj} {\bfseries 955} (2023) 13} [\href{https://arxiv.org/abs/2210.01777}{{\ttfamily 2210.01777}}].

\bibitem{Whitler2023}
L.~{Whitler}, R.~{Endsley}, D.P.~{Stark}, M.~{Topping}, Z.~{Chen} and S.~{Charlot}, \emph{{On the ages of bright galaxies 500 Myr after the big bang: insights into star formation activity at z {\ensuremath{\gtrsim}} 15 with JWST}}, \href{https://doi.org/10.1093/mnras/stac3535}{\emph{\mnras} {\bfseries 519} (2023) 157} [\href{https://arxiv.org/abs/2208.01599}{{\ttfamily 2208.01599}}].

\bibitem{Castellano2023_UVLF}
M.~{Castellano}, A.~{Fontana}, T.~{Treu}, E.~{Merlin}, P.~{Santini}, P.~{Bergamini} et~al., \emph{{Early Results from GLASS-JWST. XIX. A High Density of Bright Galaxies at z {\ensuremath{\approx}} 10 in the A2744 Region}}, \href{https://doi.org/10.3847/2041-8213/accea5}{\emph{\apjl} {\bfseries 948} (2023) L14} [\href{https://arxiv.org/abs/2212.06666}{{\ttfamily 2212.06666}}].

\bibitem{Finkelstein2023_UVLF}
S.L.~{Finkelstein}, M.B.~{Bagley}, H.C.~{Ferguson}, S.M.~{Wilkins}, J.S.~{Kartaltepe}, C.~{Papovich} et~al., \emph{{CEERS Key Paper. I. An Early Look into the First 500 Myr of Galaxy Formation with JWST}}, \href{https://doi.org/10.3847/2041-8213/acade4}{\emph{\apjl} {\bfseries 946} (2023) L13} [\href{https://arxiv.org/abs/2211.05792}{{\ttfamily 2211.05792}}].

\bibitem{Donnan2023}
C.T.~{Donnan}, D.J.~{McLeod}, J.S.~{Dunlop}, R.J.~{McLure}, A.C.~{Carnall}, R.~{Begley} et~al., \emph{{The evolution of the galaxy UV luminosity function at redshifts z = 8 - 15 from deep JWST and ground-based near-infrared imaging}}, \href{https://doi.org/10.1093/mnras/stac3472}{\emph{\mnras} {\bfseries 518} (2023) 6011} [\href{https://arxiv.org/abs/2207.12356}{{\ttfamily 2207.12356}}].

\bibitem{Harikane2023}
Y.~{Harikane}, M.~{Ouchi}, M.~{Oguri}, Y.~{Ono}, K.~{Nakajima}, Y.~{Isobe} et~al., \emph{{A Comprehensive Study of Galaxies at z 9-16 Found in the Early JWST Data: Ultraviolet Luminosity Functions and Cosmic Star Formation History at the Pre-reionization Epoch}}, \href{https://doi.org/10.3847/1538-4365/acaaa9}{\emph{\apjs} {\bfseries 265} (2023) 5} [\href{https://arxiv.org/abs/2208.01612}{{\ttfamily 2208.01612}}].

\bibitem{Bouwens2023}
R.~{Bouwens}, G.~{Illingworth}, P.~{Oesch}, M.~{Stefanon}, R.~{Naidu}, I.~{van Leeuwen} et~al., \emph{{UV luminosity density results at z > 8 from the first JWST/NIRCam fields: limitations of early data sets and the need for spectroscopy}}, \href{https://doi.org/10.1093/mnras/stad1014}{\emph{\mnras} {\bfseries 523} (2023) 1009} [\href{https://arxiv.org/abs/2212.06683}{{\ttfamily 2212.06683}}].

\bibitem{McLeod2024}
D.J.~{McLeod}, C.T.~{Donnan}, R.J.~{McLure}, J.S.~{Dunlop}, D.~{Magee}, R.~{Begley} et~al., \emph{{The galaxy UV luminosity function at z $\simeq$ 11 from a suite of public JWST ERS, ERO, and Cycle-1 programs}}, \href{https://doi.org/10.1093/mnras/stad3471}{\emph{\mnras} {\bfseries 527} (2024) 5004} [\href{https://arxiv.org/abs/2304.14469}{{\ttfamily 2304.14469}}].

\bibitem{Haslbauer2022}
M.~{Haslbauer}, P.~{Kroupa}, A.H.~{Zonoozi} and H.~{Haghi}, \emph{{Has JWST Already Falsified Dark-matter-driven Galaxy Formation?}}, \href{https://doi.org/10.3847/2041-8213/ac9a50}{\emph{\apjl} {\bfseries 939} (2022) L31} [\href{https://arxiv.org/abs/2210.14915}{{\ttfamily 2210.14915}}].

\bibitem{Lovell2023}
C.C.~{Lovell}, I.~{Harrison}, Y.~{Harikane}, S.~{Tacchella} and S.M.~{Wilkins}, \emph{{Extreme value statistics of the halo and stellar mass distributions at high redshift: are JWST results in tension with {\ensuremath{\Lambda}}CDM?}}, \href{https://doi.org/10.1093/mnras/stac3224}{\emph{\mnras} {\bfseries 518} (2023) 2511} [\href{https://arxiv.org/abs/2208.10479}{{\ttfamily 2208.10479}}].

\bibitem{Mirocha2023}
J.~{Mirocha} and S.R.~{Furlanetto}, \emph{{Balancing the efficiency and stochasticity of star formation with dust extinction in $z > 10$ galaxies observed by JWST}}, \href{https://doi.org/10.1093/mnras/stac3578}{\emph{\mnras} {\bfseries 519} (2023) 843} [\href{https://arxiv.org/abs/2208.12826}{{\ttfamily 2208.12826}}].

\bibitem{Munoz2023}
J.B.~{Mu{\~n}oz}, J.~{Mirocha}, S.~{Furlanetto} and N.~{Sabti}, \emph{{Breaking degeneracies in the first galaxies with clustering}}, \href{https://doi.org/10.1093/mnrasl/slad115}{\emph{\mnras} {\bfseries 526} (2023) L47} [\href{https://arxiv.org/abs/2306.09403}{{\ttfamily 2306.09403}}].

\bibitem{Mason2023}
C.A.~{Mason}, M.~{Trenti} and T.~{Treu}, \emph{{The brightest galaxies at cosmic dawn}}, \href{https://doi.org/10.1093/mnras/stad035}{\emph{\mnras} {\bfseries 521} (2023) 497} [\href{https://arxiv.org/abs/2207.14808}{{\ttfamily 2207.14808}}].

\bibitem{Boylan2023}
M.~{Boylan-Kolchin}, \emph{{Stress testing {\ensuremath{\Lambda}}CDM with high-redshift galaxy candidates}}, \href{https://doi.org/10.1038/s41550-023-01937-7}{\emph{Nature Astronomy} {\bfseries 7} (2023) 731} [\href{https://arxiv.org/abs/2208.01611}{{\ttfamily 2208.01611}}].

\bibitem{Labbe2023}
I.~{Labb{\'e}}, P.~{van Dokkum}, E.~{Nelson}, R.~{Bezanson}, K.A.~{Suess}, J.~{Leja} et~al., \emph{{A population of red candidate massive galaxies 600 Myr after the Big Bang}}, \href{https://doi.org/10.1038/s41586-023-05786-2}{\emph{\nat} {\bfseries 616} (2023) 266} [\href{https://arxiv.org/abs/2207.12446}{{\ttfamily 2207.12446}}].

\bibitem{Liu2022}
B.~{Liu} and V.~{Bromm}, \emph{{Accelerating Early Massive Galaxy Formation with Primordial Black Holes}}, \href{https://doi.org/10.3847/2041-8213/ac927f}{\emph{\apjl} {\bfseries 937} (2022) L30} [\href{https://arxiv.org/abs/2208.13178}{{\ttfamily 2208.13178}}].

\bibitem{Padmanabhan2023}
H.~{Padmanabhan} and A.~{Loeb}, \emph{{Alleviating the Need for Exponential Evolution of JWST Galaxies in {}10$^{10}$ M $_{{\ensuremath{\odot}}}$ Haloes at z > 10 by a Modified {\ensuremath{\Lambda}}CDM Power Spectrum}}, \href{https://doi.org/10.3847/2041-8213/acea7a}{\emph{\apjl} {\bfseries 953} (2023) L4} [\href{https://arxiv.org/abs/2306.04684}{{\ttfamily 2306.04684}}].

\bibitem{Biagetti2023}
M.~{Biagetti}, G.~{Franciolini} and A.~{Riotto}, \emph{{High-redshift JWST Observations and Primordial Non-Gaussianity}}, \href{https://doi.org/10.3847/1538-4357/acb5ea}{\emph{\apj} {\bfseries 944} (2023) 113} [\href{https://arxiv.org/abs/2210.04812}{{\ttfamily 2210.04812}}].

\bibitem{Hutsi2023}
G.~{H{\"u}tsi}, M.~{Raidal}, J.~{Urrutia}, V.~{Vaskonen} and H.~{Veerm{\"a}e}, \emph{{Did JWST observe imprints of axion miniclusters or primordial black holes?}}, \href{https://doi.org/10.1103/PhysRevD.107.043502}{\emph{\prd} {\bfseries 107} (2023) 043502} [\href{https://arxiv.org/abs/2211.02651}{{\ttfamily 2211.02651}}].

\bibitem{Parashari2023}
P.~{Parashari} and R.~{Laha}, \emph{{Primordial power spectrum in light of JWST observations of high redshift galaxies}}, \href{https://doi.org/10.1093/mnrasl/slad107}{\emph{\mnras} {\bfseries 526} (2023) L63} [\href{https://arxiv.org/abs/2305.00999}{{\ttfamily 2305.00999}}].

\bibitem{Gong2023}
Y.~{Gong}, B.~{Yue}, Y.~{Cao} and X.~{Chen}, \emph{{Fuzzy Dark Matter as a Solution to Reconcile the Stellar Mass Density of High-z Massive Galaxies and Reionization History}}, \href{https://doi.org/10.3847/1538-4357/acc109}{\emph{\apj} {\bfseries 947} (2023) 28} [\href{https://arxiv.org/abs/2209.13757}{{\ttfamily 2209.13757}}].

\bibitem{Maio2023}
U.~{Maio} and M.~{Viel}, \emph{{JWST high-redshift galaxy constraints on warm and cold dark matter models}}, \href{https://doi.org/10.1051/0004-6361/202345851}{\emph{\aap} {\bfseries 672} (2023) A71} [\href{https://arxiv.org/abs/2211.03620}{{\ttfamily 2211.03620}}].

\bibitem{Dekel2023}
A.~{Dekel}, K.S.~{Sarkar}, Y.~{Birnboim}, N.~{Mandelker} and Z.~{Li}, \emph{{Efficient Formation of Massive Galaxies at Cosmic Dawn by Feedback-Free Starbursts}}, \href{https://doi.org/10.48550/arXiv.2303.04827}{\emph{arXiv e-prints} (2023) arXiv:2303.04827} [\href{https://arxiv.org/abs/2303.04827}{{\ttfamily 2303.04827}}].

\bibitem{Renzini2023}
A.~{Renzini}, \emph{{A transient overcooling in the early Universe? Clues from globular clusters formation}}, \href{https://doi.org/10.1093/mnrasl/slad091}{\emph{\mnras} {\bfseries 525} (2023) L117} [\href{https://arxiv.org/abs/2305.14476}{{\ttfamily 2305.14476}}].

\bibitem{Sipple2023}
J.~{Sipple} and A.~{Lidz}, \emph{{The Star Formation Efficiency During Reionization as Inferred from the Hubble Frontier Fields}}, \href{https://doi.org/10.48550/arXiv.2306.12087}{\emph{arXiv e-prints} (2023) arXiv:2306.12087} [\href{https://arxiv.org/abs/2306.12087}{{\ttfamily 2306.12087}}].

\bibitem{Shen2023}
X.~{Shen}, M.~{Vogelsberger}, M.~{Boylan-Kolchin}, S.~{Tacchella} and R.~{Kannan}, \emph{{The impact of UV variability on the abundance of bright galaxies at $z \geq 9$}}, \href{https://doi.org/10.48550/arXiv.2305.05679}{\emph{arXiv e-prints} (2023) arXiv:2305.05679} [\href{https://arxiv.org/abs/2305.05679}{{\ttfamily 2305.05679}}].

\bibitem{Pallottini2023}
A.~{Pallottini} and A.~{Ferrara}, \emph{{Stochastic star formation in early galaxies: JWST implications}}, \href{https://doi.org/10.48550/arXiv.2307.03219}{\emph{arXiv e-prints} (2023) arXiv:2307.03219} [\href{https://arxiv.org/abs/2307.03219}{{\ttfamily 2307.03219}}].

\bibitem{Inayoshi2022}
K.~{Inayoshi}, Y.~{Harikane}, A.K.~{Inoue}, W.~{Li} and L.C.~{Ho}, \emph{{A Lower Bound of Star Formation Activity in Ultra-high-redshift Galaxies Detected with JWST: Implications for Stellar Populations and Radiation Sources}}, \href{https://doi.org/10.3847/2041-8213/ac9310}{\emph{\apjl} {\bfseries 938} (2022) L10} [\href{https://arxiv.org/abs/2208.06872}{{\ttfamily 2208.06872}}].

\bibitem{Ferrara2023}
A.~{Ferrara}, A.~{Pallottini} and P.~{Dayal}, \emph{{On the stunning abundance of super-early, luminous galaxies revealed by JWST}}, \href{https://doi.org/10.1093/mnras/stad1095}{\emph{\mnras} {\bfseries 522} (2023) 3986} [\href{https://arxiv.org/abs/2208.00720}{{\ttfamily 2208.00720}}].

\bibitem{Ucci2021}
G.~{Ucci}, P.~{Dayal}, A.~{Hutter}, G.~{Yepes}, S.~{Gottl{\"o}ber}, L.~{Legrand} et~al., \emph{{Astraeus - II. Quantifying the impact of cosmic variance during the Epoch of Reionization}}, \href{https://doi.org/10.1093/mnras/stab1229}{\emph{\mnras} {\bfseries 506} (2021) 202} [\href{https://arxiv.org/abs/2004.11096}{{\ttfamily 2004.11096}}].

\bibitem{Jespersen2024}
C.~{Kragh Jespersen}, C.L.~{Steinhardt}, R.S.~{Somerville} and C.C.~{Lovell}, \emph{{On the Significance of Rare Objects at High Redshift: The Impact of Cosmic Variance}}, \href{https://doi.org/10.48550/arXiv.2403.00050}{\emph{arXiv e-prints} (2024) arXiv:2403.00050} [\href{https://arxiv.org/abs/2403.00050}{{\ttfamily 2403.00050}}].

\bibitem{Wise2014}
J.H.~{Wise}, V.G.~{Demchenko}, M.T.~{Halicek}, M.L.~{Norman}, M.J.~{Turk}, T.~{Abel} et~al., \emph{{The birth of a galaxy - III. Propelling reionization with the faintest galaxies}}, \href{https://doi.org/10.1093/mnras/stu979}{\emph{\mnras} {\bfseries 442} (2014) 2560} [\href{https://arxiv.org/abs/1403.6123}{{\ttfamily 1403.6123}}].

\bibitem{Bouwens2015_sources}
R.J.~{Bouwens}, G.D.~{Illingworth}, P.A.~{Oesch}, J.~{Caruana}, B.~{Holwerda}, R.~{Smit} et~al., \emph{{Reionization After Planck: The Derived Growth of the Cosmic Ionizing Emissivity Now Matches the Growth of the Galaxy UV Luminosity Density}}, \href{https://doi.org/10.1088/0004-637X/811/2/140}{\emph{\apj} {\bfseries 811} (2015) 140} [\href{https://arxiv.org/abs/1503.08228}{{\ttfamily 1503.08228}}].

\bibitem{Dayal2020}
P.~{Dayal}, M.~{Volonteri}, T.R.~{Choudhury}, R.~{Schneider}, M.~{Trebitsch}, N.Y.~{Gnedin} et~al., \emph{{Reionization with galaxies and active galactic nuclei}}, \href{https://doi.org/10.1093/mnras/staa1138}{\emph{\mnras} {\bfseries 495} (2020) 3065} [\href{https://arxiv.org/abs/2001.06021}{{\ttfamily 2001.06021}}].

\bibitem{Atek2024_Spectroscopy}
H.~{Atek}, I.~{Labb{\'e}}, L.J.~{Furtak}, I.~{Chemerynska}, S.~{Fujimoto}, D.J.~{Setton} et~al., \emph{{Most of the photons that reionized the Universe came from dwarf galaxies}}, \href{https://doi.org/10.1038/s41586-024-07043-6}{\emph{\nat} {\bfseries 626} (2024) 975}.

\bibitem{Miralda1994}
J.~{Miralda-Escud{\'e}} and M.J.~{Rees}, \emph{{Reionization and thermal evolution of a photoionized intergalactic medium.}}, \href{https://doi.org/10.1093/mnras/266.2.343}{\emph{\mnras} {\bfseries 266} (1994) 343}.

\bibitem{Finlator2011}
K.~{Finlator}, R.~{Dav{\'e}} and F.~{{\"O}zel}, \emph{{Galactic Outflows and Photoionization Heating in the Reionization Epoch}}, \href{https://doi.org/10.1088/0004-637X/743/2/169}{\emph{\apj} {\bfseries 743} (2011) 169} [\href{https://arxiv.org/abs/1106.4321}{{\ttfamily 1106.4321}}].

\bibitem{Finlator2012}
K.~{Finlator}, S.P.~{Oh}, F.~{{\"O}zel} and R.~{Dav{\'e}}, \emph{{Gas clumping in self-consistent reionization models}}, \href{https://doi.org/10.1111/j.1365-2966.2012.22114.x}{\emph{\mnras} {\bfseries 427} (2012) 2464} [\href{https://arxiv.org/abs/1209.2489}{{\ttfamily 1209.2489}}].

\bibitem{Hutter2021}
A.~{Hutter}, P.~{Dayal}, G.~{Yepes}, S.~{Gottl{\"o}ber}, L.~{Legrand} and G.~{Ucci}, \emph{{Astraeus I: the interplay between galaxy formation and reionization}}, \href{https://doi.org/10.1093/mnras/stab602}{\emph{\mnras} {\bfseries 503} (2021) 3698} [\href{https://arxiv.org/abs/2004.08401}{{\ttfamily 2004.08401}}].

\bibitem{Mitra2015}
S.~{Mitra}, T.R.~{Choudhury} and A.~{Ferrara}, \emph{{Cosmic reionization after Planck.}}, \href{https://doi.org/10.1093/mnrasl/slv134}{\emph{\mnras} {\bfseries 454} (2015) L76} [\href{https://arxiv.org/abs/1505.05507}{{\ttfamily 1505.05507}}].

\bibitem{Maity2022_ParameterConstraints}
B.~{Maity} and T.R.~{Choudhury}, \emph{{Constraining the reionization and thermal history of the Universe using a seminumerical photon-conserving code SCRIPT}}, \href{https://doi.org/10.1093/mnras/stac1847}{\emph{\mnras} {\bfseries 515} (2022) 617} [\href{https://arxiv.org/abs/2204.05268}{{\ttfamily 2204.05268}}].

\bibitem{Planck2014}
{Planck Collaboration}, P.A.R.~{Ade}, N.~{Aghanim}, C.~{Armitage-Caplan}, M.~{Arnaud}, M.~{Ashdown} et~al., \emph{{Planck 2013 results. XVI. Cosmological parameters}}, \href{https://doi.org/10.1051/0004-6361/201321591}{\emph{\aap} {\bfseries 571} (2014) A16} [\href{https://arxiv.org/abs/1303.5076}{{\ttfamily 1303.5076}}].

\bibitem{Behroozi&Silk2015}
P.S.~{Behroozi} and J.~{Silk}, \emph{{A Simple Technique for Predicting High-redshift Galaxy Evolution}}, \href{https://doi.org/10.1088/0004-637X/799/1/32}{\emph{\apj} {\bfseries 799} (2015) 32} [\href{https://arxiv.org/abs/1404.5299}{{\ttfamily 1404.5299}}].

\bibitem{Sun&Furlanetto2016}
G.~{Sun} and S.R.~{Furlanetto}, \emph{{Constraints on the star formation efficiency of galaxies during the epoch of reionization}}, \href{https://doi.org/10.1093/mnras/stw980}{\emph{\mnras} {\bfseries 460} (2016) 417} [\href{https://arxiv.org/abs/1512.06219}{{\ttfamily 1512.06219}}].

\bibitem{Madau&Dickinson2014}
P.~{Madau} and M.~{Dickinson}, \emph{{Cosmic Star-Formation History}}, \href{https://doi.org/10.1146/annurev-astro-081811-125615}{\emph{\araa} {\bfseries 52} (2014) 415} [\href{https://arxiv.org/abs/1403.0007}{{\ttfamily 1403.0007}}].

\bibitem{Starburst99}
C.~{Leitherer}, D.~{Schaerer}, J.D.~{Goldader}, R.M.G.~{Delgado}, C.~{Robert}, D.F.~{Kune} et~al., \emph{{Starburst99: Synthesis Models for Galaxies with Active Star Formation}}, \href{https://doi.org/10.1086/313233}{\emph{\apjs} {\bfseries 123} (1999) 3} [\href{https://arxiv.org/abs/astro-ph/9902334}{{\ttfamily astro-ph/9902334}}].

\bibitem{Choudhury&Dayal2019}
T.R.~{Choudhury} and P.~{Dayal}, \emph{{Probing the fluctuating ultraviolet background using the Hubble Frontier Fields}}, \href{https://doi.org/10.1093/mnrasl/sly186}{\emph{\mnras} {\bfseries 482} (2019) L19} [\href{https://arxiv.org/abs/1809.01798}{{\ttfamily 1809.01798}}].

\bibitem{Sobacchi&Mesinger2013}
E.~{Sobacchi} and A.~{Mesinger}, \emph{{How does radiative feedback from an ultraviolet background impact reionization?}}, \href{https://doi.org/10.1093/mnras/stt693}{\emph{\mnras} {\bfseries 432} (2013) 3340} [\href{https://arxiv.org/abs/1301.6781}{{\ttfamily 1301.6781}}].

\bibitem{Dayal2015}
P.~{Dayal}, A.~{Mesinger} and F.~{Pacucci}, \emph{{Early Galaxy Formation in Warm Dark Matter Cosmologies}}, \href{https://doi.org/10.1088/0004-637X/806/1/67}{\emph{\apj} {\bfseries 806} (2015) 67} [\href{https://arxiv.org/abs/1408.1102}{{\ttfamily 1408.1102}}].

\bibitem{Oke_ABmag}
J.B.~{Oke}, \emph{{Absolute Spectral Energy Distributions for White Dwarfs}}, \href{https://doi.org/10.1086/190287}{\emph{\apjs} {\bfseries 27} (1974) 21}.

\bibitem{Oke&Gunn_ABmag}
J.B.~{Oke} and J.E.~{Gunn}, \emph{{Secondary standard stars for absolute spectrophotometry.}}, \href{https://doi.org/10.1086/160817}{\emph{\apj} {\bfseries 266} (1983) 713}.

\bibitem{Jenkins2001}
A.~{Jenkins}, C.S.~{Frenk}, S.D.M.~{White}, J.M.~{Colberg}, S.~{Cole}, A.E.~{Evrard} et~al., \emph{{The mass function of dark matter haloes}}, \href{https://doi.org/10.1046/j.1365-8711.2001.04029.x}{\emph{\mnras} {\bfseries 321} (2001) 372} [\href{https://arxiv.org/abs/astro-ph/0005260}{{\ttfamily astro-ph/0005260}}].

\bibitem{HMFcodePaper}
S.G.~{Murray}, C.~{Power} and A.S.G.~{Robotham}, \emph{{HMFcalc: An online tool for calculating dark matter halo mass functions}}, \href{https://doi.org/10.1016/j.ascom.2013.11.001}{\emph{Astronomy and Computing} {\bfseries 3} (2013) 23} [\href{https://arxiv.org/abs/1306.6721}{{\ttfamily 1306.6721}}].

\bibitem{Shapiro1987}
P.R.~{Shapiro} and M.L.~{Giroux}, \emph{{Cosmological H II Regions and the Photoionization of the Intergalactic Medium}}, \href{https://doi.org/10.1086/185015}{\emph{\apjl} {\bfseries 321} (1987) L107}.

\bibitem{Madau1999}
P.~{Madau}, F.~{Haardt} and M.J.~{Rees}, \emph{{Radiative Transfer in a Clumpy Universe. III. The Nature of Cosmological Ionizing Sources}}, \href{https://doi.org/10.1086/306975}{\emph{\apj} {\bfseries 514} (1999) 648} [\href{https://arxiv.org/abs/astro-ph/9809058}{{\ttfamily astro-ph/9809058}}].

\bibitem{McQuinn2011}
M.~{McQuinn}, S.P.~{Oh} and C.-A.~{Faucher-Gigu{\`e}re}, \emph{{On Lyman-limit Systems and the Evolution of the Intergalactic Ionizing Background}}, \href{https://doi.org/10.1088/0004-637X/743/1/82}{\emph{\apj} {\bfseries 743} (2011) 82} [\href{https://arxiv.org/abs/1101.1964}{{\ttfamily 1101.1964}}].

\bibitem{Shull2012}
J.M.~{Shull}, A.~{Harness}, M.~{Trenti} and B.D.~{Smith}, \emph{{Critical Star Formation Rates for Reionization: Full Reionization Occurs at Redshift z {\ensuremath{\approx}} 7}}, \href{https://doi.org/10.1088/0004-637X/747/2/100}{\emph{\apj} {\bfseries 747} (2012) 100}.

\bibitem{Pawlik2015}
A.H.~{Pawlik}, J.~{Schaye} and C.~{Dalla Vecchia}, \emph{{Spatially adaptive radiation-hydrodynamical simulations of galaxy formation during cosmological reionization}}, \href{https://doi.org/10.1093/mnras/stv976}{\emph{\mnras} {\bfseries 451} (2015) 1586} [\href{https://arxiv.org/abs/1501.01980}{{\ttfamily 1501.01980}}].

\bibitem{DAloisio2020}
A.~{D'Aloisio}, M.~{McQuinn}, H.~{Trac}, C.~{Cain} and A.~{Mesinger}, \emph{{Hydrodynamic Response of the Intergalactic Medium to Reionization}}, \href{https://doi.org/10.3847/1538-4357/ab9f2f}{\emph{\apj} {\bfseries 898} (2020) 149} [\href{https://arxiv.org/abs/2002.02467}{{\ttfamily 2002.02467}}].

\bibitem{Qin2021}
Y.~{Qin}, A.~{Mesinger}, S.E.I.~{Bosman} and M.~{Viel}, \emph{{Reionization and galaxy inference from the high-redshift Ly {\ensuremath{\alpha}} forest}}, \href{https://doi.org/10.1093/mnras/stab1833}{\emph{\mnras} {\bfseries 506} (2021) 2390} [\href{https://arxiv.org/abs/2101.09033}{{\ttfamily 2101.09033}}].

\bibitem{Robertson2015}
B.E.~{Robertson}, R.S.~{Ellis}, S.R.~{Furlanetto} and J.S.~{Dunlop}, \emph{{Cosmic Reionization and Early Star-forming Galaxies: A Joint Analysis of New Constraints from Planck and the Hubble Space Telescope}}, \href{https://doi.org/10.1088/2041-8205/802/2/L19}{\emph{\apjl} {\bfseries 802} (2015) L19} [\href{https://arxiv.org/abs/1502.02024}{{\ttfamily 1502.02024}}].

\bibitem{Planck2020}
{Planck Collaboration}, N.~{Aghanim}, Y.~{Akrami}, M.~{Ashdown}, J.~{Aumont}, C.~{Baccigalupi} et~al., \emph{{Planck 2018 results. VI. Cosmological parameters}}, \href{https://doi.org/10.1051/0004-6361/201833910}{\emph{\aap} {\bfseries 641} (2020) A6} [\href{https://arxiv.org/abs/1807.06209}{{\ttfamily 1807.06209}}].

\bibitem{Bouwens2021}
R.J.~{Bouwens}, P.A.~{Oesch}, M.~{Stefanon}, G.~{Illingworth}, I.~{Labb{\'e}}, N.~{Reddy} et~al., \emph{{New Determinations of the UV Luminosity Functions from z 9 to 2 Show a Remarkable Consistency with Halo Growth and a Constant Star Formation Efficiency}}, \href{https://doi.org/10.3847/1538-3881/abf83e}{\emph{\aj} {\bfseries 162} (2021) 47} [\href{https://arxiv.org/abs/2102.07775}{{\ttfamily 2102.07775}}].

\bibitem{Mauerhofer&Dayal2023}
V.~{Mauerhofer} and P.~{Dayal}, \emph{{The dust enrichment of early galaxies in the JWST and ALMA era}}, \href{https://doi.org/10.1093/mnras/stad2734}{\emph{\mnras} {\bfseries 526} (2023) 2196} [\href{https://arxiv.org/abs/2305.01681}{{\ttfamily 2305.01681}}].

\bibitem{Gaikwad2023}
P.~{Gaikwad}, M.G.~{Haehnelt}, F.B.~{Davies}, S.E.I.~{Bosman}, M.~{Molaro}, G.~{Kulkarni} et~al., \emph{{Measuring the photoionization rate, neutral fraction, and mean free path of H I ionizing photons at 4.9 {\ensuremath{\leq}} z {\ensuremath{\leq}} 6.0 from a large sample of XShooter and ESI spectra}}, \href{https://doi.org/10.1093/mnras/stad2566}{\emph{\mnras} {\bfseries 525} (2023) 4093} [\href{https://arxiv.org/abs/2304.02038}{{\ttfamily 2304.02038}}].

\bibitem{Davies2018}
F.B.~{Davies}, J.F.~{Hennawi}, E.~{Ba{\~n}ados}, Z.~{Luki{\'c}}, R.~{Decarli}, X.~{Fan} et~al., \emph{{Quantitative Constraints on the Reionization History from the IGM Damping Wing Signature in Two Quasars at z > 7}}, \href{https://doi.org/10.3847/1538-4357/aad6dc}{\emph{\apj} {\bfseries 864} (2018) 142} [\href{https://arxiv.org/abs/1802.06066}{{\ttfamily 1802.06066}}].

\bibitem{Greig2022}
B.~{Greig}, A.~{Mesinger}, F.B.~{Davies}, F.~{Wang}, J.~{Yang} and J.F.~{Hennawi}, \emph{{IGM damping wing constraints on reionization from covariance reconstruction of two z {\ensuremath{\gtrsim}} 7 QSOs}}, \href{https://doi.org/10.1093/mnras/stac825}{\emph{\mnras} {\bfseries 512} (2022) 5390} [\href{https://arxiv.org/abs/2112.04091}{{\ttfamily 2112.04091}}].

\bibitem{Durovcikova2024}
D.~{{\v{D}}urov{\v{c}}{\'\i}kov{\'a}}, A.-C.~{Eilers}, H.~{Chen}, S.~{Satyavolu}, G.~{Kulkarni}, R.A.~{Simcoe} et~al., \emph{{Chronicling the reionization history at $6\lesssim z \lesssim 7$ with emergent quasar damping wings}}, \href{https://doi.org/10.48550/arXiv.2401.10328}{\emph{arXiv e-prints} (2024) arXiv:2401.10328} [\href{https://arxiv.org/abs/2401.10328}{{\ttfamily 2401.10328}}].

\bibitem{Umeda2023}
H.~{Umeda}, M.~{Ouchi}, K.~{Nakajima}, Y.~{Harikane}, Y.~{Ono}, Y.~{Xu} et~al., \emph{{JWST Measurements of Neutral Hydrogen Fractions and Ionized Bubble Sizes at $z=7-12$ Obtained with Ly$\alpha$ Damping Wing Absorptions in 26 Bright Continuum Galaxies}}, \href{https://doi.org/10.48550/arXiv.2306.00487}{\emph{arXiv e-prints} (2023) arXiv:2306.00487} [\href{https://arxiv.org/abs/2306.00487}{{\ttfamily 2306.00487}}].

\bibitem{cobaya}
J.~{Torrado} and A.~{Lewis}, \emph{{Cobaya: code for Bayesian analysis of hierarchical physical models}}, \href{https://doi.org/10.1088/1475-7516/2021/05/057}{\emph{\jcap} {\bfseries 2021} (2021) 057} [\href{https://arxiv.org/abs/2005.05290}{{\ttfamily 2005.05290}}].

\bibitem{Jin2023}
X.~{Jin}, J.~{Yang}, X.~{Fan}, F.~{Wang}, E.~{Ba{\~n}ados}, F.~{Bian} et~al., \emph{{(Nearly) Model-independent Constraints on the Neutral Hydrogen Fraction in the Intergalactic Medium at z 5-7 Using Dark Pixel Fractions in Ly{\ensuremath{\alpha}} and Ly{\ensuremath{\beta}} Forests}}, \href{https://doi.org/10.3847/1538-4357/aca678}{\emph{\apj} {\bfseries 942} (2023) 59} [\href{https://arxiv.org/abs/2211.12613}{{\ttfamily 2211.12613}}].

\bibitem{Becker2015}
G.D.~{Becker}, J.S.~{Bolton}, P.~{Madau}, M.~{Pettini}, E.V.~{Ryan-Weber} and B.P.~{Venemans}, \emph{{Evidence of patchy hydrogen reionization from an extreme Ly{\ensuremath{\alpha}} trough below redshift six}}, \href{https://doi.org/10.1093/mnras/stu2646}{\emph{\mnras} {\bfseries 447} (2015) 3402} [\href{https://arxiv.org/abs/1407.4850}{{\ttfamily 1407.4850}}].

\bibitem{Bosman2018}
S.E.I.~{Bosman}, X.~{Fan}, L.~{Jiang}, S.~{Reed}, Y.~{Matsuoka}, G.~{Becker} et~al., \emph{{New constraints on Lyman-{\ensuremath{\alpha}} opacity with a sample of 62 quasarsat z > 5.7}}, \href{https://doi.org/10.1093/mnras/sty1344}{\emph{\mnras} {\bfseries 479} (2018) 1055} [\href{https://arxiv.org/abs/1802.08177}{{\ttfamily 1802.08177}}].

\bibitem{Kulkarni2019}
G.~{Kulkarni}, L.C.~{Keating}, M.G.~{Haehnelt}, S.E.I.~{Bosman}, E.~{Puchwein}, J.~{Chardin} et~al., \emph{{Large Ly {\ensuremath{\alpha}} opacity fluctuations and low CMB {\ensuremath{\tau}} in models of late reionization with large islands of neutral hydrogen extending to z < 5.5}}, \href{https://doi.org/10.1093/mnrasl/slz025}{\emph{\mnras} {\bfseries 485} (2019) L24} [\href{https://arxiv.org/abs/1809.06374}{{\ttfamily 1809.06374}}].

\bibitem{Choudhury2021}
T.R.~{Choudhury}, A.~{Paranjape} and S.E.I.~{Bosman}, \emph{{Studying the Lyman {\ensuremath{\alpha}} optical depth fluctuations at z {\ensuremath{\sim}} 5.5 using fast semi-numerical methods}}, \href{https://doi.org/10.1093/mnras/stab045}{\emph{\mnras} {\bfseries 501} (2021) 5782} [\href{https://arxiv.org/abs/2003.08958}{{\ttfamily 2003.08958}}].

\bibitem{Bosman2022}
S.E.I.~{Bosman}, F.B.~{Davies}, G.D.~{Becker}, L.C.~{Keating}, R.L.~{Davies}, Y.~{Zhu} et~al., \emph{{Hydrogen reionization ends by z = 5.3: Lyman-{\ensuremath{\alpha}} optical depth measured by the XQR-30 sample}}, \href{https://doi.org/10.1093/mnras/stac1046}{\emph{\mnras} {\bfseries 514} (2022) 55} [\href{https://arxiv.org/abs/2108.03699}{{\ttfamily 2108.03699}}].

\bibitem{Qin2023}
Y.~{Qin}, S.~{Balu} and J.S.B.~{Wyithe}, \emph{{Implications of z {\ensuremath{\gtrsim}} 12 JWST galaxies for galaxy formation at high redshift}}, \href{https://doi.org/10.1093/mnras/stad2448}{\emph{\mnras} {\bfseries 526} (2023) 1324} [\href{https://arxiv.org/abs/2305.17959}{{\ttfamily 2305.17959}}].

\bibitem{Yung2024}
L.Y.A.~{Yung}, R.S.~{Somerville}, S.L.~{Finkelstein}, S.M.~{Wilkins} and J.P.~{Gardner}, \emph{{Are the ultra-high-redshift galaxies at z > 10 surprising in the context of standard galaxy formation models?}}, \href{https://doi.org/10.1093/mnras/stad3484}{\emph{\mnras} {\bfseries 527} (2024) 5929} [\href{https://arxiv.org/abs/2304.04348}{{\ttfamily 2304.04348}}].

\bibitem{Sun2023}
G.~{Sun}, C.-A.~{Faucher-Gigu{\`e}re}, C.C.~{Hayward}, X.~{Shen}, A.~{Wetzel} and R.K.~{Cochrane}, \emph{{Bursty Star Formation Naturally Explains the Abundance of Bright Galaxies at Cosmic Dawn}}, \href{https://doi.org/10.3847/2041-8213/acf85a}{\emph{\apjl} {\bfseries 955} (2023) L35} [\href{https://arxiv.org/abs/2307.15305}{{\ttfamily 2307.15305}}].

\bibitem{Trinca2023}
A.~{Trinca}, R.~{Schneider}, R.~{Valiante}, L.~{Graziani}, A.~{Ferrotti}, K.~{Omukai} et~al., \emph{{Exploring the nature of UV-bright $z rsim 10$ galaxies detected by JWST: star formation, black hole accretion, or a non-universal IMF?}}, \href{https://doi.org/10.48550/arXiv.2305.04944}{\emph{arXiv e-prints} (2023) arXiv:2305.04944} [\href{https://arxiv.org/abs/2305.04944}{{\ttfamily 2305.04944}}].

\bibitem{Jermyn2018}
A.S.~{Jermyn}, C.L.~{Steinhardt} and C.A.~{Tout}, \emph{{The cosmic microwave background and the stellar initial mass function}}, \href{https://doi.org/10.1093/mnras/sty2123}{\emph{\mnras} {\bfseries 480} (2018) 4265} [\href{https://arxiv.org/abs/1809.03502}{{\ttfamily 1809.03502}}].

\bibitem{Chon2022}
S.~{Chon}, H.~{Ono}, K.~{Omukai} and R.~{Schneider}, \emph{{Impact of the cosmic background radiation on the initial mass function of metal-poor stars}}, \href{https://doi.org/10.1093/mnras/stac1549}{\emph{\mnras} {\bfseries 514} (2022) 4639} [\href{https://arxiv.org/abs/2205.15328}{{\ttfamily 2205.15328}}].

\bibitem{Chon2021}
S.~{Chon}, K.~{Omukai} and R.~{Schneider}, \emph{{Transition of the initial mass function in the metal-poor environments}}, \href{https://doi.org/10.1093/mnras/stab2497}{\emph{\mnras} {\bfseries 508} (2021) 4175} [\href{https://arxiv.org/abs/2103.04997}{{\ttfamily 2103.04997}}].

\bibitem{Susa2014}
H.~{Susa}, K.~{Hasegawa} and N.~{Tominaga}, \emph{{The Mass Spectrum of the First Stars}}, \href{https://doi.org/10.1088/0004-637X/792/1/32}{\emph{\apj} {\bfseries 792} (2014) 32} [\href{https://arxiv.org/abs/1407.1374}{{\ttfamily 1407.1374}}].

\bibitem{Hirano2015}
S.~{Hirano}, T.~{Hosokawa}, N.~{Yoshida}, K.~{Omukai} and H.W.~{Yorke}, \emph{{Primordial star formation under the influence of far ultraviolet radiation: 1540 cosmological haloes and the stellar mass distribution}}, \href{https://doi.org/10.1093/mnras/stv044}{\emph{\mnras} {\bfseries 448} (2015) 568} [\href{https://arxiv.org/abs/1501.01630}{{\ttfamily 1501.01630}}].

\bibitem{Stacy2016}
A.~{Stacy}, V.~{Bromm} and A.T.~{Lee}, \emph{{Building up the Population III initial mass function from cosmological initial conditions}}, \href{https://doi.org/10.1093/mnras/stw1728}{\emph{\mnras} {\bfseries 462} (2016) 1307} [\href{https://arxiv.org/abs/1603.09475}{{\ttfamily 1603.09475}}].

\bibitem{Cueto2023}
E.~{Rasmussen Cueto}, A.~{Hutter}, P.~{Dayal}, S.~{Gottl{\"o}ber}, K.E.~{Heintz}, C.~{Mason} et~al., \emph{{The impact of an evolving stellar initial mass function on early galaxies and reionisation}}, \href{https://doi.org/10.48550/arXiv.2312.12109}{\emph{arXiv e-prints} (2023) arXiv:2312.12109} [\href{https://arxiv.org/abs/2312.12109}{{\ttfamily 2312.12109}}].

\bibitem{Qin2020_CMB}
Y.~{Qin}, V.~{Poulin}, A.~{Mesinger}, B.~{Greig}, S.~{Murray} and J.~{Park}, \emph{{Reionization inference from the CMB optical depth and E-mode polarization power spectra}}, \href{https://doi.org/10.1093/mnras/staa2797}{\emph{\mnras} {\bfseries 499} (2020) 550} [\href{https://arxiv.org/abs/2006.16828}{{\ttfamily 2006.16828}}].

\bibitem{CurtisLake2023}
E.~{Curtis-Lake}, S.~{Carniani}, A.~{Cameron}, S.~{Charlot}, P.~{Jakobsen}, R.~{Maiolino} et~al., \emph{{Spectroscopic confirmation of four metal-poor galaxies at z = 10.3-13.2}}, \href{https://doi.org/10.1038/s41550-023-01918-w}{\emph{Nature Astronomy} {\bfseries 7} (2023) 622} [\href{https://arxiv.org/abs/2212.04568}{{\ttfamily 2212.04568}}].

\bibitem{Rinaldi2023}
P.~{Rinaldi}, K.I.~{Caputi}, E.~{Iani}, L.~{Costantin}, S.~{Gillman}, P.G.~{Perez-Gonzalez} et~al., \emph{{MIDIS: Unveiling the Role of Strong Ha-emitters during the Epoch of Reionization with JWST}}, \href{https://doi.org/10.48550/arXiv.2309.15671}{\emph{arXiv e-prints} (2023) arXiv:2309.15671} [\href{https://arxiv.org/abs/2309.15671}{{\ttfamily 2309.15671}}].

\bibitem{Endsley2023}
R.~{Endsley}, D.P.~{Stark}, L.~{Whitler}, M.W.~{Topping}, B.D.~{Johnson}, B.~{Robertson} et~al., \emph{{The Star-forming and Ionizing Properties of Dwarf z\raisebox{-0.5ex}\textasciitilde6-9 Galaxies in JADES: Insights on Bursty Star Formation and Ionized Bubble Growth}}, \href{https://doi.org/10.48550/arXiv.2306.05295}{\emph{arXiv e-prints} (2023) arXiv:2306.05295} [\href{https://arxiv.org/abs/2306.05295}{{\ttfamily 2306.05295}}].

\bibitem{Simmonds2024}
C.~{Simmonds}, S.~{Tacchella}, K.~{Hainline}, B.D.~{Johnson}, W.~{McClymont}, B.~{Robertson} et~al., \emph{{Low-mass bursty galaxies in JADES efficiently produce ionizing photons and could represent the main drivers of reionization}}, \href{https://doi.org/10.1093/mnras/stad3605}{\emph{\mnras} {\bfseries 527} (2024) 6139} [\href{https://arxiv.org/abs/2310.01112}{{\ttfamily 2310.01112}}].

\bibitem{AlvarezMarquez2024}
J.~{{\'A}lvarez-M{\'a}rquez}, L.~{Colina}, A.~{Crespo G{\'o}mez}, P.~{Rinaldi}, J.~{Melinder}, G.~{{\"O}stlin} et~al., \emph{{Spatially resolved H{\ensuremath{\alpha}} and ionizing photon production efficiency in the lensed galaxy MACS1149-JD1 at a redshift of 9.11}}, \href{https://doi.org/10.1051/0004-6361/202347946}{\emph{\aap} {\bfseries 686} (2024) A85} [\href{https://arxiv.org/abs/2309.06319}{{\ttfamily 2309.06319}}].

\bibitem{Calabro2024}
A.~{Calabro}, M.~{Castellano}, J.A.~{Zavala}, L.~{Pentericci}, P.~{Arrabal Haro}, T.J.L.C.~{Bakx} et~al., \emph{{Evidence of extreme ionization conditions and low metallicity in GHZ2/GLASS-z12 from a combined analysis of NIRSpec and MIRI observations}}, \href{https://doi.org/10.48550/arXiv.2403.12683}{\emph{arXiv e-prints} (2024) arXiv:2403.12683} [\href{https://arxiv.org/abs/2403.12683}{{\ttfamily 2403.12683}}].

\bibitem{Paardekooper2015}
J.-P.~{Paardekooper}, S.~{Khochfar} and C.~{Dalla Vecchia}, \emph{{The First Billion Years project: the escape fraction of ionizing photons in the epoch of reionization}}, \href{https://doi.org/10.1093/mnras/stv1114}{\emph{\mnras} {\bfseries 451} (2015) 2544} [\href{https://arxiv.org/abs/1501.01967}{{\ttfamily 1501.01967}}].

\bibitem{Xu2016}
H.~{Xu}, J.H.~{Wise}, M.L.~{Norman}, K.~{Ahn} and B.W.~{O'Shea}, \emph{{Galaxy Properties and UV Escape Fractions during the Epoch of Reionization: Results from the Renaissance Simulations}}, \href{https://doi.org/10.3847/1538-4357/833/1/84}{\emph{\apj} {\bfseries 833} (2016) 84} [\href{https://arxiv.org/abs/1604.07842}{{\ttfamily 1604.07842}}].

\bibitem{Lewis2020}
J.S.W.~{Lewis}, P.~{Ocvirk}, D.~{Aubert}, J.G.~{Sorce}, P.R.~{Shapiro}, N.~{Deparis} et~al., \emph{{Galactic ionizing photon budget during the epoch of reionization in the Cosmic Dawn II simulation}}, \href{https://doi.org/10.1093/mnras/staa1748}{\emph{\mnras} {\bfseries 496} (2020) 4342} [\href{https://arxiv.org/abs/2001.07785}{{\ttfamily 2001.07785}}].

\bibitem{Kimm2014}
T.~{Kimm} and R.~{Cen}, \emph{{Escape Fraction of Ionizing Photons during Reionization: Effects due to Supernova Feedback and Runaway OB Stars}}, \href{https://doi.org/10.1088/0004-637X/788/2/121}{\emph{\apj} {\bfseries 788} (2014) 121} [\href{https://arxiv.org/abs/1405.0552}{{\ttfamily 1405.0552}}].

\bibitem{Mutch2023}
S.J.~{Mutch}, B.~{Greig}, Y.~{Qin}, G.B.~{Poole} and J.S.B.~{Wyithe}, \emph{{Dark-ages reionization and galaxy formation simulation -- XXI. Constraining the evolution of the ionizing escape fraction}}, \href{https://doi.org/10.48550/arXiv.2303.07378}{\emph{arXiv e-prints} (2023) arXiv:2303.07378} [\href{https://arxiv.org/abs/2303.07378}{{\ttfamily 2303.07378}}].

\bibitem{Harikane2024}
Y.~{Harikane}, K.~{Nakajima}, M.~{Ouchi}, H.~{Umeda}, Y.~{Isobe}, Y.~{Ono} et~al., \emph{{Pure Spectroscopic Constraints on UV Luminosity Functions and Cosmic Star Formation History from 25 Galaxies at z $_{spec}$ = 8.61-13.20 Confirmed with JWST/NIRSpec}}, \href{https://doi.org/10.3847/1538-4357/ad0b7e}{\emph{\apj} {\bfseries 960} (2024) 56} [\href{https://arxiv.org/abs/2304.06658}{{\ttfamily 2304.06658}}].

\bibitem{Bolton2007}
J.S.~{Bolton} and M.G.~{Haehnelt}, \emph{{The observed ionization rate of the intergalactic medium and the ionizing emissivity at z >= 5: evidence for a photon-starved and extended epoch of reionization}}, \href{https://doi.org/10.1111/j.1365-2966.2007.12372.x}{\emph{\mnras} {\bfseries 382} (2007) 325} [\href{https://arxiv.org/abs/astro-ph/0703306}{{\ttfamily astro-ph/0703306}}].

\bibitem{Becker2013}
G.D.~{Becker} and J.S.~{Bolton}, \emph{{New measurements of the ionizing ultraviolet background over 2 < z < 5 and implications for hydrogen reionization}}, \href{https://doi.org/10.1093/mnras/stt1610}{\emph{\mnras} {\bfseries 436} (2013) 1023} [\href{https://arxiv.org/abs/1307.2259}{{\ttfamily 1307.2259}}].

\bibitem{DAloisio2018}
A.~{D'Aloisio}, M.~{McQuinn}, F.B.~{Davies} and S.R.~{Furlanetto}, \emph{{Large fluctuations in the high-redshift metagalactic ionizing background}}, \href{https://doi.org/10.1093/mnras/stx2341}{\emph{\mnras} {\bfseries 473} (2018) 560} [\href{https://arxiv.org/abs/1611.02711}{{\ttfamily 1611.02711}}].

\bibitem{Becker2021}
G.D.~{Becker}, A.~{D'Aloisio}, H.M.~{Christenson}, Y.~{Zhu}, G.~{Worseck} and J.S.~{Bolton}, \emph{{The mean free path of ionizing photons at 5 < z < 6: evidence for rapid evolution near reionization}}, \href{https://doi.org/10.1093/mnras/stab2696}{\emph{\mnras} {\bfseries 508} (2021) 1853} [\href{https://arxiv.org/abs/2103.16610}{{\ttfamily 2103.16610}}].

\bibitem{Keating2020_ATON0}
L.C.~{Keating}, L.H.~{Weinberger}, G.~{Kulkarni}, M.G.~{Haehnelt}, J.~{Chardin} and D.~{Aubert}, \emph{{Long troughs in the Lyman-{\ensuremath{\alpha}} forest below redshift 6 due to islands of neutral hydrogen}}, \href{https://doi.org/10.1093/mnras/stz3083}{\emph{\mnras} {\bfseries 491} (2020) 1736} [\href{https://arxiv.org/abs/1905.12640}{{\ttfamily 1905.12640}}].

\bibitem{Keating2020_ATON1}
L.C.~{Keating}, G.~{Kulkarni}, M.G.~{Haehnelt}, J.~{Chardin} and D.~{Aubert}, \emph{{Constraining the second half of reionization with the Ly {\ensuremath{\beta}} forest}}, \href{https://doi.org/10.1093/mnras/staa1909}{\emph{\mnras} {\bfseries 497} (2020) 906} [\href{https://arxiv.org/abs/1912.05582}{{\ttfamily 1912.05582}}].

\bibitem{Sharma2016}
M.~{Sharma}, T.~{Theuns}, C.~{Frenk}, R.~{Bower}, R.~{Crain}, M.~{Schaller} et~al., \emph{{The brighter galaxies reionized the Universe}}, \href{https://doi.org/10.1093/mnrasl/slw021}{\emph{\mnras} {\bfseries 458} (2016) L94} [\href{https://arxiv.org/abs/1512.04537}{{\ttfamily 1512.04537}}].

\bibitem{Naidu2020}
R.P.~{Naidu}, S.~{Tacchella}, C.A.~{Mason}, S.~{Bose}, P.A.~{Oesch} and C.~{Conroy}, \emph{{Rapid Reionization by the Oligarchs: The Case for Massive, UV-bright, Star-forming Galaxies with High Escape Fractions}}, \href{https://doi.org/10.3847/1538-4357/ab7cc9}{\emph{\apj} {\bfseries 892} (2020) 109} [\href{https://arxiv.org/abs/1907.13130}{{\ttfamily 1907.13130}}].

\bibitem{Joshi2024}
N.~{Joshi} and M.~{Sharma}, \emph{{The haloes that reionized the Universe}}, \href{https://doi.org/10.48550/arXiv.2403.12132}{\emph{arXiv e-prints} (2024) arXiv:2403.12132} [\href{https://arxiv.org/abs/2403.12132}{{\ttfamily 2403.12132}}].

\bibitem{Anderson2017}
L.~{Anderson}, F.~{Governato}, M.~{Karcher}, T.~{Quinn} and J.~{Wadsley}, \emph{{The little Galaxies that could (reionize the universe): predicting faint end slopes \& escape fractions at z>4}}, \href{https://doi.org/10.1093/mnras/stx709}{\emph{\mnras} {\bfseries 468} (2017) 4077} [\href{https://arxiv.org/abs/1606.05352}{{\ttfamily 1606.05352}}].

\bibitem{DalmassoHST}
N.~{Dalmasso}, M.~{Trenti} and N.~{Leethochawalit}, \emph{{Galaxy clustering measurements out to redshift z {\ensuremath{\sim}} 8 from Hubble Legacy Fields}}, \href{https://doi.org/10.1093/mnras/stad3901}{\emph{\mnras} {\bfseries 528} (2024) 898} [\href{https://arxiv.org/abs/2312.12329}{{\ttfamily 2312.12329}}].

\bibitem{DalmassoJWST}
N.~{Dalmasso}, N.~{Leethochawalit}, M.~{Trenti} and K.~{Boyett}, \emph{{Galaxy clustering at cosmic dawn from JWST/NIRCam observations to redshift z$\sim$11}}, \href{https://doi.org/10.48550/arXiv.2402.18052}{\emph{arXiv e-prints} (2024) arXiv:2402.18052} [\href{https://arxiv.org/abs/2402.18052}{{\ttfamily 2402.18052}}].

\bibitem{Barone-Nugent2014}
R.L.~{Barone-Nugent}, M.~{Trenti}, J.S.B.~{Wyithe}, R.J.~{Bouwens}, P.A.~{Oesch}, G.D.~{Illingworth} et~al., \emph{{Measurement of Galaxy Clustering at z \raisebox{-0.5ex}\textasciitilde 7.2 and the Evolution of Galaxy Bias from 3.8 < z < 8 in the XDF, GOODS-S, and GOODS-N}}, \href{https://doi.org/10.1088/0004-637X/793/1/17}{\emph{\apj} {\bfseries 793} (2014) 17} [\href{https://arxiv.org/abs/1407.7316}{{\ttfamily 1407.7316}}].

\bibitem{Harikane2016}
Y.~{Harikane}, M.~{Ouchi}, Y.~{Ono}, S.~{More}, S.~{Saito}, Y.-T.~{Lin} et~al., \emph{{Evolution of Stellar-to-Halo Mass Ratio at z = 0 - 7 Identified by Clustering Analysis with the Hubble Legacy Imaging and Early Subaru/Hyper Suprime-Cam Survey Data}}, \href{https://doi.org/10.3847/0004-637X/821/2/123}{\emph{\apj} {\bfseries 821} (2016) 123} [\href{https://arxiv.org/abs/1511.07873}{{\ttfamily 1511.07873}}].

\bibitem{Qiu2018}
Y.~{Qiu}, J.S.B.~{Wyithe}, P.A.~{Oesch}, S.J.~{Mutch}, Y.~{Qin}, I.~{Labb{\'e}} et~al., \emph{{Dependence of galaxy clustering on UV luminosity and stellar mass at z {\ensuremath{\sim}} 4-7}}, \href{https://doi.org/10.1093/mnras/sty2633}{\emph{\mnras} {\bfseries 481} (2018) 4885} [\href{https://arxiv.org/abs/1809.10161}{{\ttfamily 1809.10161}}].

\bibitem{McGreer2011}
I.D.~{McGreer}, A.~{Mesinger} and X.~{Fan}, \emph{{The first (nearly) model-independent constraint on the neutral hydrogen fraction at z {\ensuremath{\sim}} 6}}, \href{https://doi.org/10.1111/j.1365-2966.2011.18935.x}{\emph{\mnras} {\bfseries 415} (2011) 3237} [\href{https://arxiv.org/abs/1101.3314}{{\ttfamily 1101.3314}}].

\bibitem{McGreer2015}
I.D.~{McGreer}, A.~{Mesinger} and V.~{D'Odorico}, \emph{{Model-independent evidence in favour of an end to reionization by z {\ensuremath{\approx}} 6}}, \href{https://doi.org/10.1093/mnras/stu2449}{\emph{\mnras} {\bfseries 447} (2015) 499} [\href{https://arxiv.org/abs/1411.5375}{{\ttfamily 1411.5375}}].

\bibitem{Ghara2020}
R.~{Ghara}, S.K.~{Giri}, G.~{Mellema}, B.~{Ciardi}, S.~{Zaroubi}, I.T.~{Iliev} et~al., \emph{{Constraining the intergalactic medium at z {\ensuremath{\approx}} 9.1 using LOFAR Epoch of Reionization observations}}, \href{https://doi.org/10.1093/mnras/staa487}{\emph{\mnras} {\bfseries 493} (2020) 4728} [\href{https://arxiv.org/abs/2002.07195}{{\ttfamily 2002.07195}}].

\bibitem{Miralda2000}
J.~{Miralda-Escud{\'e}}, M.~{Haehnelt} and M.J.~{Rees}, \emph{{Reionization of the Inhomogeneous Universe}}, \href{https://doi.org/10.1086/308330}{\emph{\apj} {\bfseries 530} (2000) 1} [\href{https://arxiv.org/abs/astro-ph/9812306}{{\ttfamily astro-ph/9812306}}].

\bibitem{Wyithe2003}
J.S.B.~{Wyithe} and A.~{Loeb}, \emph{{Reionization of Hydrogen and Helium by Early Stars and Quasars}}, \href{https://doi.org/10.1086/367721}{\emph{\apj} {\bfseries 586} (2003) 693} [\href{https://arxiv.org/abs/astro-ph/0209056}{{\ttfamily astro-ph/0209056}}].

\bibitem{Choudhury2007}
T.R.~{Choudhury} and A.~{Ferrara}, \emph{{Searching for the reionization sources}}, \href{https://doi.org/10.1111/j.1745-3933.2007.00338.x}{\emph{\mnras} {\bfseries 380} (2007) L6} [\href{https://arxiv.org/abs/astro-ph/0703771}{{\ttfamily astro-ph/0703771}}].

\bibitem{Chatterjee2021}
A.~{Chatterjee}, T.R.~{Choudhury} and S.~{Mitra}, \emph{{CosmoReionMC: a package for estimating cosmological and astrophysical parameters using CMB, Lyman-{\ensuremath{\alpha}} absorption, and global 21 cm data}}, \href{https://doi.org/10.1093/mnras/stab2316}{\emph{\mnras} {\bfseries 507} (2021) 2405} [\href{https://arxiv.org/abs/2101.11088}{{\ttfamily 2101.11088}}].

\bibitem{Ventura2024}
E.M.~{Ventura}, Y.~{Qin}, S.~{Balu} and J.S.B.~{Wyithe}, \emph{{Semi-analytic modelling of Pop. III star formation and metallicity evolution - I. Impact on the UV luminosity functions at z = 9-16}}, \href{https://doi.org/10.1093/mnras/stae567}{\emph{\mnras} {\bfseries 529} (2024) 628} [\href{https://arxiv.org/abs/2401.07396}{{\ttfamily 2401.07396}}].

\bibitem{Ventura2023}
E.M.~{Ventura}, A.~{Trinca}, R.~{Schneider}, L.~{Graziani}, R.~{Valiante} and J.S.B.~{Wyithe}, \emph{{The role of Pop III stars and early black holes in the 21-cm signal from Cosmic Dawn}}, \href{https://doi.org/10.1093/mnras/stad237}{\emph{\mnras} {\bfseries 520} (2023) 3609} [\href{https://arxiv.org/abs/2210.10281}{{\ttfamily 2210.10281}}].

\bibitem{Yamaguchi2023}
N.~{Yamaguchi}, S.R.~{Furlanetto} and A.C.~{Trapp}, \emph{{The extent of intergalactic metal enrichment from galactic winds during the Cosmic Dawn}}, \href{https://doi.org/10.1093/mnras/stad315}{\emph{\mnras} {\bfseries 520} (2023) 2922} [\href{https://arxiv.org/abs/2209.09345}{{\ttfamily 2209.09345}}].

\bibitem{Pacucci2022}
F.~{Pacucci}, P.~{Dayal}, Y.~{Harikane}, A.K.~{Inoue} and A.~{Loeb}, \emph{{Are the newly-discovered z 13 drop-out sources starburst galaxies or quasars?}}, \href{https://doi.org/10.1093/mnrasl/slac035}{\emph{\mnras} {\bfseries 514} (2022) L6} [\href{https://arxiv.org/abs/2201.00823}{{\ttfamily 2201.00823}}].

\bibitem{Ziparo2023}
F.~{Ziparo}, A.~{Ferrara}, L.~{Sommovigo} and M.~{Kohandel}, \emph{{Blue monsters. Why are JWST super-early, massive galaxies so blue?}}, \href{https://doi.org/10.1093/mnras/stad125}{\emph{\mnras} {\bfseries 520} (2023) 2445} [\href{https://arxiv.org/abs/2209.06840}{{\ttfamily 2209.06840}}].

\bibitem{Fiore2023}
F.~{Fiore}, A.~{Ferrara}, M.~{Bischetti}, C.~{Feruglio} and A.~{Travascio}, \emph{{Dusty-wind-clear JWST Super-early Galaxies}}, \href{https://doi.org/10.3847/2041-8213/acb5f2}{\emph{\apjl} {\bfseries 943} (2023) L27} [\href{https://arxiv.org/abs/2211.08937}{{\ttfamily 2211.08937}}].

\bibitem{Ferrara2023_Paper2}
A.~{Ferrara}, \emph{{Super-early JWST galaxies, outflows and Lyman alpha visibility in the EoR}}, \href{https://doi.org/10.48550/arXiv.2310.12197}{\emph{arXiv e-prints} (2023) arXiv:2310.12197} [\href{https://arxiv.org/abs/2310.12197}{{\ttfamily 2310.12197}}].

\bibitem{Mirocha2019}
J.~{Mirocha} and S.R.~{Furlanetto}, \emph{{What does the first highly redshifted 21-cm detection tell us about early galaxies?}}, \href{https://doi.org/10.1093/mnras/sty3260}{\emph{\mnras} {\bfseries 483} (2019) 1980} [\href{https://arxiv.org/abs/1803.03272}{{\ttfamily 1803.03272}}].

\bibitem{Mittal2022}
S.~{Mittal} and G.~{Kulkarni}, \emph{{Implications of the cosmological 21-cm absorption profile for high-redshift star formation and deep JWST surveys}}, \href{https://doi.org/10.1093/mnras/stac1961}{\emph{\mnras} {\bfseries 515} (2022) 2901} [\href{https://arxiv.org/abs/2203.07733}{{\ttfamily 2203.07733}}].

\bibitem{Shimabukuro2023}
H.~{Shimabukuro}, K.~{Hasegawa}, A.~{Kuchinomachi}, H.~{Yajima} and S.~{Yoshiura}, \emph{{Exploring the cosmic dawn and epoch of reionization with the 21 cm line}}, \href{https://doi.org/10.1093/pasj/psac042}{\emph{\pasj} {\bfseries 75} (2023) S1} [\href{https://arxiv.org/abs/2303.07594}{{\ttfamily 2303.07594}}].

\bibitem{Chatterjee2023}
A.~{Chatterjee}, P.~{Dayal} and V.~{Mauerhofer}, \emph{{Predictions of the 21 cm global signal in the JWST and ALMA era}}, \href{https://doi.org/10.1093/mnras/stad2307}{\emph{\mnras} {\bfseries 525} (2023) 620} [\href{https://arxiv.org/abs/2306.03149}{{\ttfamily 2306.03149}}].

\bibitem{Hassan2023}
S.~{Hassan}, C.C.~{Lovell}, P.~{Madau}, M.~{Huertas-Company}, R.S.~{Somerville}, B.~{Burkhart} et~al., \emph{{JWST Constraints on the UV Luminosity Density at Cosmic Dawn: Implications for 21 cm Cosmology}}, \href{https://doi.org/10.3847/2041-8213/ad0239}{\emph{\apjl} {\bfseries 958} (2023) L3} [\href{https://arxiv.org/abs/2305.02703}{{\ttfamily 2305.02703}}].

\bibitem{Gehlot2019_LOFAR}
B.K.~{Gehlot}, F.G.~{Mertens}, L.V.E.~{Koopmans}, M.A.~{Brentjens}, S.~{Zaroubi}, B.~{Ciardi} et~al., \emph{{The first power spectrum limit on the 21-cm signal of neutral hydrogen during the Cosmic Dawn at z = 20-25 from LOFAR}}, \href{https://doi.org/10.1093/mnras/stz1937}{\emph{\mnras} {\bfseries 488} (2019) 4271} [\href{https://arxiv.org/abs/1809.06661}{{\ttfamily 1809.06661}}].

\bibitem{Mertens2020_LOFAR}
F.G.~{Mertens}, M.~{Mevius}, L.V.E.~{Koopmans}, A.R.~{Offringa}, G.~{Mellema}, S.~{Zaroubi} et~al., \emph{{Improved upper limits on the 21 cm signal power spectrum of neutral hydrogen at z {\ensuremath{\approx}} 9.1 from LOFAR}}, \href{https://doi.org/10.1093/mnras/staa327}{\emph{\mnras} {\bfseries 493} (2020) 1662} [\href{https://arxiv.org/abs/2002.07196}{{\ttfamily 2002.07196}}].

\bibitem{Barry2019_MWA}
N.~{Barry}, M.~{Wilensky}, C.M.~{Trott}, B.~{Pindor}, A.P.~{Beardsley}, B.J.~{Hazelton} et~al., \emph{{Improving the Epoch of Reionization Power Spectrum Results from Murchison Widefield Array Season 1 Observations}}, \href{https://doi.org/10.3847/1538-4357/ab40a8}{\emph{\apj} {\bfseries 884} (2019) 1} [\href{https://arxiv.org/abs/1909.00561}{{\ttfamily 1909.00561}}].

\bibitem{Trott2020_MWA}
C.M.~{Trott}, C.H.~{Jordan}, S.~{Midgley}, N.~{Barry}, B.~{Greig}, B.~{Pindor} et~al., \emph{{Deep multiredshift limits on Epoch of Reionization 21 cm power spectra from four seasons of Murchison Widefield Array observations}}, \href{https://doi.org/10.1093/mnras/staa414}{\emph{\mnras} {\bfseries 493} (2020) 4711} [\href{https://arxiv.org/abs/2002.02575}{{\ttfamily 2002.02575}}].

\bibitem{Parsons2010_PAPER}
A.R.~{Parsons}, D.C.~{Backer}, G.S.~{Foster}, M.C.H.~{Wright}, R.F.~{Bradley}, N.E.~{Gugliucci} et~al., \emph{{The Precision Array for Probing the Epoch of Re-ionization: Eight Station Results}}, \href{https://doi.org/10.1088/0004-6256/139/4/1468}{\emph{\aj} {\bfseries 139} (2010) 1468} [\href{https://arxiv.org/abs/0904.2334}{{\ttfamily 0904.2334}}].

\bibitem{Abdurashidova2022_HERA}
Z.~{Abdurashidova}, J.E.~{Aguirre}, P.~{Alexander}, Z.S.~{Ali}, Y.~{Balfour}, A.P.~{Beardsley} et~al., \emph{{First Results from HERA Phase I: Upper Limits on the Epoch of Reionization 21 cm Power Spectrum}}, \href{https://doi.org/10.3847/1538-4357/ac1c78}{\emph{\apj} {\bfseries 925} (2022) 221} [\href{https://arxiv.org/abs/2108.02263}{{\ttfamily 2108.02263}}].

\bibitem{Kolopanis2019_PAPER}
M.~{Kolopanis}, D.C.~{Jacobs}, C.~{Cheng}, A.R.~{Parsons}, S.A.~{Kohn}, J.C.~{Pober} et~al., \emph{{A Simplified, Lossless Reanalysis of PAPER-64}}, \href{https://doi.org/10.3847/1538-4357/ab3e3a}{\emph{\apj} {\bfseries 883} (2019) 133} [\href{https://arxiv.org/abs/1909.02085}{{\ttfamily 1909.02085}}].

\bibitem{Koopmans2015_SKA}
L.~{Koopmans}, J.~{Pritchard}, G.~{Mellema}, J.~{Aguirre}, K.~{Ahn}, R.~{Barkana} et~al., \emph{{The Cosmic Dawn and Epoch of Reionisation with SKA}},  in \emph{Advancing Astrophysics with the Square Kilometre Array (AASKA14)}, p.~1, April, 2015, \href{https://doi.org/10.22323/1.215.0001}{DOI} [\href{https://arxiv.org/abs/1505.07568}{{\ttfamily 1505.07568}}].

\bibitem{Singh2022_SARAS}
S.~{Singh}, N.T.~{Jishnu}, R.~{Subrahmanyan}, N.~{Udaya Shankar}, B.S.~{Girish}, A.~{Raghunathan} et~al., \emph{{On the detection of a cosmic dawn signal in the radio background}}, \href{https://doi.org/10.1038/s41550-022-01610-5}{\emph{Nature Astronomy} {\bfseries 6} (2022) 607} [\href{https://arxiv.org/abs/2112.06778}{{\ttfamily 2112.06778}}].

\bibitem{Bowman2018_EDGES}
J.D.~{Bowman}, A.E.E.~{Rogers}, R.A.~{Monsalve}, T.J.~{Mozdzen} and N.~{Mahesh}, \emph{{An absorption profile centred at 78 megahertz in the sky-averaged spectrum}}, \href{https://doi.org/10.1038/nature25792}{\emph{\nat} {\bfseries 555} (2018) 67} [\href{https://arxiv.org/abs/1810.05912}{{\ttfamily 1810.05912}}].

\bibitem{Acedo2022_REACH}
E.~{de Lera Acedo}, D.I.L.~{de Villiers}, N.~{Razavi-Ghods}, W.~{Handley}, A.~{Fialkov}, A.~{Magro} et~al., \emph{{The REACH radiometer for detecting the 21-cm hydrogen signal from redshift z {\ensuremath{\approx}} 7.5-28}}, \href{https://doi.org/10.1038/s41550-022-01709-9}{\emph{Nature Astronomy} {\bfseries 6} (2022) 984} [\href{https://arxiv.org/abs/2210.07409}{{\ttfamily 2210.07409}}].

\end{thebibliography}\endgroup

\end{document}